\documentclass[a4paper,11pt]{article}
\pdfoutput=1 

\usepackage{jheppub} 

\usepackage{subcaption}
\usepackage{verbatim}

\usepackage[T1]{fontenc} 

\newcommand\numberthis{\addtocounter{equation}{1}\tag{\theequation}}

\usepackage[dvipsnames]{xcolor}

\title{Dual Subtractions}

\author[a]{Renato Maria Prisco}
\author[a,b]{and Francesco Tramontano}

\affiliation[a]{Universit\`a Federico II di Napoli,
  Complesso Universitario di Monte Sant'Angelo,
  Via Cintia, 80126 Napoli, Italy} 
\affiliation[b]{INFN, Sezione di Napoli, 
  Complesso Universitario di Monte Sant'Angelo,
  Via Cintia, 80126 Napoli, Italy} 

\emailAdd{renatomaria.prisco@unina.it}
\emailAdd{francesco.tramontano@unina.it}

\abstract{
We propose a novel local subtraction scheme for the computation of Next-to-Leading
Order contributions to theoretical predictions for scattering processes in perturbative Quantum Field Theory.
With respect to well known schemes proposed since many years that build upon the
analysis of the real radiation matrix elements, our construction starts from the loop diagrams and exploits their dual representation.
Our scheme implements exact phase space factorization, handles final state as well as initial state singularities and is suitable for both massless and massive particles.
}

\begin{document} 
\maketitle
\flushbottom

\section{Introduction}
\label{sec:intro}

The success of the physics studies at the LHC, culminated with the discovery of the Higgs
boson~\cite{Englert:1964et, Higgs:1966ev} at CERN~\cite{Aad:2012tfa, Chatrchyan:2012ufa}, has its roots in the deep level of understanding reached
in both experimental and theoretical aspects of the physics of hadronic collisions.
On the theory side, the computation of scattering amplitudes including higher
order perturbative corrections plays the main role.
The calculation of tree--level and one-loop matrix elements is nowadays fully automated in very efficient ways. Furthermore, well known schemes for the treatment of their infrared and collinear behaviour have been formulated since decades, like
the Catani--Seymour and the Frixione-Kunszt-Signer local schemes~\cite{Catani:1996vz,Catani:2002hc,Frixione:1995ms}.
For these reasons the Next to Leading Order (NLO) computation is considered a completely solved problem.
On the other hand, new precise exclusive measurements will be
available soon, as the ongoing monumental campaign of data collection goes on, so that,
for a meaningful comparison among theory and experiment, Next-to-Next-to Leading Order (NNLO) calculations are required.
At the NNLO, one is faced with the construction of appropriate integration schemes that allows for the cancellation of non integrable infrared and collinear singularities across contributions that live on three different phase spaces.
Several techniques have been developed to address this problem, such as $q_T$~\cite{Catani:2007vq}
and $N$-jettiness~\cite{Boughezal:2015dva,Gaunt:2015pea} slicing methods, subtraction schemes like antenna~\cite{GehrmannDeRidder:2005cm,
Daleo:2006xa,GehrmannDeRidder:2005aw,GehrmannDeRidder:2005hi,Daleo:2009yj,
Gehrmann:2011wi,Boughezal:2010mc,GehrmannDeRidder:2012ja,Currie:2013vh},
CoLoRFulNNLO~\cite{Somogyi:2005xz,Somogyi:2006da,Somogyi:2006db,Somogyi:2008fc,
Aglietti:2008fe,Somogyi:2009ri,Bolzoni:2009ye,Bolzoni:2010bt,DelDuca:2013kw,
Somogyi:2013yk}, residue-improved~\cite{Czakon:2010td,Czakon:2011ve,
Czakon:2014oma,Czakon:2019tmo}, nested soft-collinear~\cite{Caola:2017dug,
Caola:2018pxp,Delto:2019asp,Caola:2019nzf,Caola:2019pfz} and 
projection-to-Born~\cite{Cacciari:2015jma}.
Also, other approaches are under development~\cite{Magnea:2018hab,Magnea:2018ebr,Magnea:2020trj,Herzog:2018ily}. 
Although there is still not the same level of automation as for the NLO, progress on the subtraction of the singularities is going
quite fast.

However, the bottleneck for NNLO computations is represented by the calculation
of multi-loop amplitudes with several massive internal and external particles. This field of research is very active, although it is still not clear if the procedures
proposed so far can really simplify the job. The most efficient paradigm followed for
decades keeps separated the two problems mentioned above. Reduction of tensor integrals
to a base of master integrals and computation of the master integrals from
one side, and the construction of a subtraction scheme based on the analysis
of the radiation matrix elements, on the other.

An interesting new path was proposed in~\cite{Catani:2008xa, Bierenbaum:2010cy}.
The computation of loop diagrams
is turned into phase space integrals thanks to a Loop--Tree duality (LTD) theorem
proved in the same papers. The possibility to follow such a strategy is of course
very interesting because the high mathematical complexity accompanying the analytic
computation of higher loop calculation and that of building a subtraction scheme, might simply
not be there, being replaced by the numerical integration of properly defined integrands.
Such integrations are of course non trivial, but one can think that computer science
has already developed a large set of tools to address the associated technicalities.
This path has been pursued in~\cite{Sborlini:2016gbr,Hernandez-Pinto:2015ysa,Sborlini:2016hat} at one--loop and in~\cite{Driencourt-Mangin:2019aix}
at two loop as examples. Another point of view offered by the LTD theorem is the possibility
to simplify the direct numerical computation of multi-loop amplitudes.
Non-trivial numerical applications of LTD have been performed in~\cite{Buchta:2014dfa,Buchta:2015wna,Buchta:2015xda}.
Furthermore, in~\cite{Ramirez-Uribe:2020hes,Aguilera-Verdugo:2020kzc,Verdugo:2020kzh} the better numerical behaviour of LTD due to causality has been conjectured and demonstrated.
Further steps in this direction have been done in~\cite{Anastasiou:2020sdt, Capatti:2019edf,Capatti:2020ytd}.
In particular, in~\cite{Anastasiou:2020sdt} a general strategy is derived for
the subtraction of the divergences by one and two loop QED amplitudes with the
effect that the finite remnants can be numerically evaluated with
relatively small effort.

In the present paper, we will consider a slightly different perspective from
the ones outlined above.
We investigate the possibility to extract the divergences from the loop
amplitude in a way that builds subtractions for the real radiation contribution.
This path is somewhat opposite to the usual strategy to build subtractions,
that is based on the analysis of the divergences in the real sector,
nevertheless universality and cancellation of course must hold in both directions.
The opportunity to follow such a path is offered by LTD, but we will not apply
it directly to one-loop Feynman diagrams. We will first consider the reduction of
the tensor integrals to scalar integrals and then apply LTD to a small class
of scalar integrals, made by namely just two integrals, one triangle and one
massless bubble (in massless QCD). Note that, to build a proper subtraction we
will also need an appropriate integrand expression for the wave--function renormalization.
We recall the available formulas for such wave--function contributions for the case of fermions and compute the one for the gluon.
We note that a procedure to build an expression for the gluon wave--function renormalization constant at the integrand--level has also been already presented in~\cite{Seth:2016hmv}.
In that work,
counterterms for the virtual contribution are built and shown to cancel the singularities of the counterterms for the real part.
On the other hand, our approach consists in building counterterms for the real contributions through their direct extraction from the virtual part.

The paper is organized as follows: in Section~\ref{sec:method}
we discuss the singular behaviour of one--loop matrix elements
while in Section~\ref{sec:algorithm} we report the basics of LTD and use them to
extract divergent counterterms for the real radiation from final state massless particles.
Then, in Sections~\ref{sec:masses} and~\ref{sec:initial_state} we follow the same path to build
subtractions for the real contribution in the cases of radiation from
massive final state particles and from initial state particles, respectively.
In Section~\ref{sec:applications} we show a small collection of applications and
in Section~\ref{sec:conclusions} we comment our construction analyzing some
consequences and give our conclusion.
Finally, we include two appendices, one with about
the counting of the dual counterterms (Appendix~\ref{sec:counting}) and a second one
(Appendix~\ref{sec:masslessDS}) collecting all the formulae for the numerical implementation of the dual subtraction scheme for NLO QCD corrections
to scattering processes involving any number of massless external partons and non coloured particles.

\section{Singular behavior of one-loop matrix elements}
\label{sec:method}
We start by considering the easiest situation that is the one of processes
with a colourless initial state and $m$ massless partons in the final state.
Although most of the discussion in this section already applies to the case
of final states with massive particles, and that of coloured particles in the initial state,
in the present section we limit the discussion to the simpler case mentioned above
and postpone the extension to the other cases to dedicated following sections.
Our first step is to consider the full reduction of a one-loop amplitude to
scalar integrals. As it is well known a reduced amplitude has the following
form in terms of scalar boxes, triangles, bubbles and tadpoles
\begin{equation}
\label{eq:redamp}
A_m^{(1)} = \sum_{i<j<k<l} c_{ijkl} D^{ijkl}_0 
+\sum_{i<j<k} c_{ijk} C^{ijk}_0 +\sum_{i<j} c_{ij} B^{ij}_0
+\sum_{i} c_{i} A^{i}_0
\end{equation}
where the indices $i,j,k,l$ run over all possible external momenta $\{p\}$ and their sums. It is worth to note that up to this integral transformation from
its expression in terms of Feynman diagrams, the amplitude has retained exactly the same
meaning and values at all orders in the dimensional parameter that we also
keep in both the scalar integrals and their coefficients.
The same is obviously true also if we use different sets or combinations
of the scalar integrals. A particularly convenient transformation consists
in expressing the scalar boxes in terms of its six dimensional version ($D^{6-2\varepsilon}_0$)
or, equivalently, in terms of the rank two form factor that multiplies the metric
tensor in a covariant decomposition (that we conventionally denote by $D_{00}$).
After this exchange, Eq.(\ref{eq:redamp}) becomes
\begin{equation}
\label{eq:nredamp}
A_m^{(1)} = \sum_{i<j<k<l} c'_{ijkl} D^{ijkl}_{00} 
+\sum_{i<j<k} c'_{ijk} C^{ijk}_0 +\sum_{i<j} c_{ij} B^{ij}_0
+\sum_{i} c_{i} A^{i}_0 \,\,.
\end{equation}
The $D_{00}$ function consists of a linear combination of the relative box
function and the four triangle functions that are obtained pinching one
denominator at the time. This is the reason why we have not changed the
coefficients of the scalar bubble and tadpole integrals.

Now we analyze the infrared and collinear behaviour of the reduced amplitude.
First, we remind that, irrespective from the value of both the internal and external
masses, the $D_{00}$ function is completely finite with respect to
the virtual loop integration, being in particular free of both Ultra-Violet (UV) and
Infra-Red and collinear (IR) poles\footnote{This property can be deduced by combining the scalar integrals
of the reduced $D_{00}$ function or by inspecting its direct calculation in terms of Feynman parameters.
An alternative proof based on the analysis of the soft and collinear behaviour of rank two tensor
box integrals can be found
in~\cite{Denner:2005nn}.}.

Moving to triangle functions $C^{ijk}_{0}$, there are two kinds of them which have IR divergences.
\begin{figure}[t]
  \begin{subfigure}{.48\textwidth}
    \centering
    \includegraphics[width=0.7\linewidth]{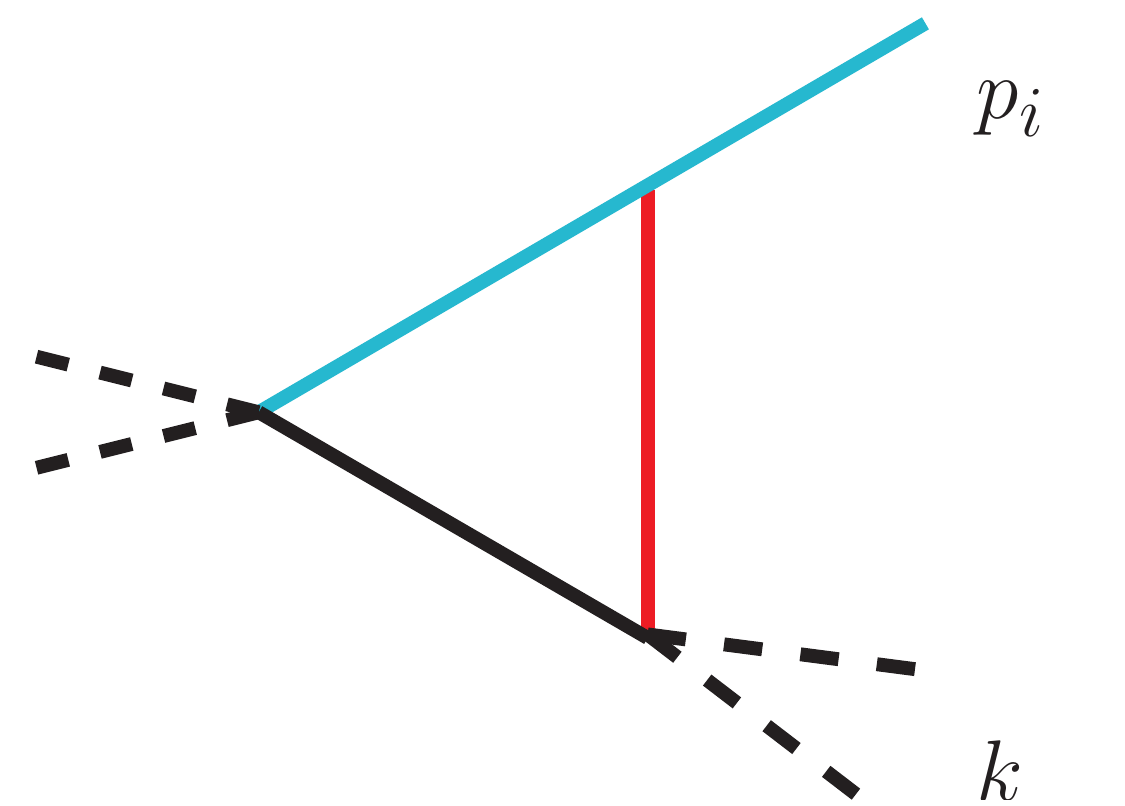}
    \caption{\label{fig:tr1} One external parton}
  \end{subfigure}
  \begin{subfigure}{.48\textwidth}
    \centering‰
    \includegraphics[width=0.7\linewidth]{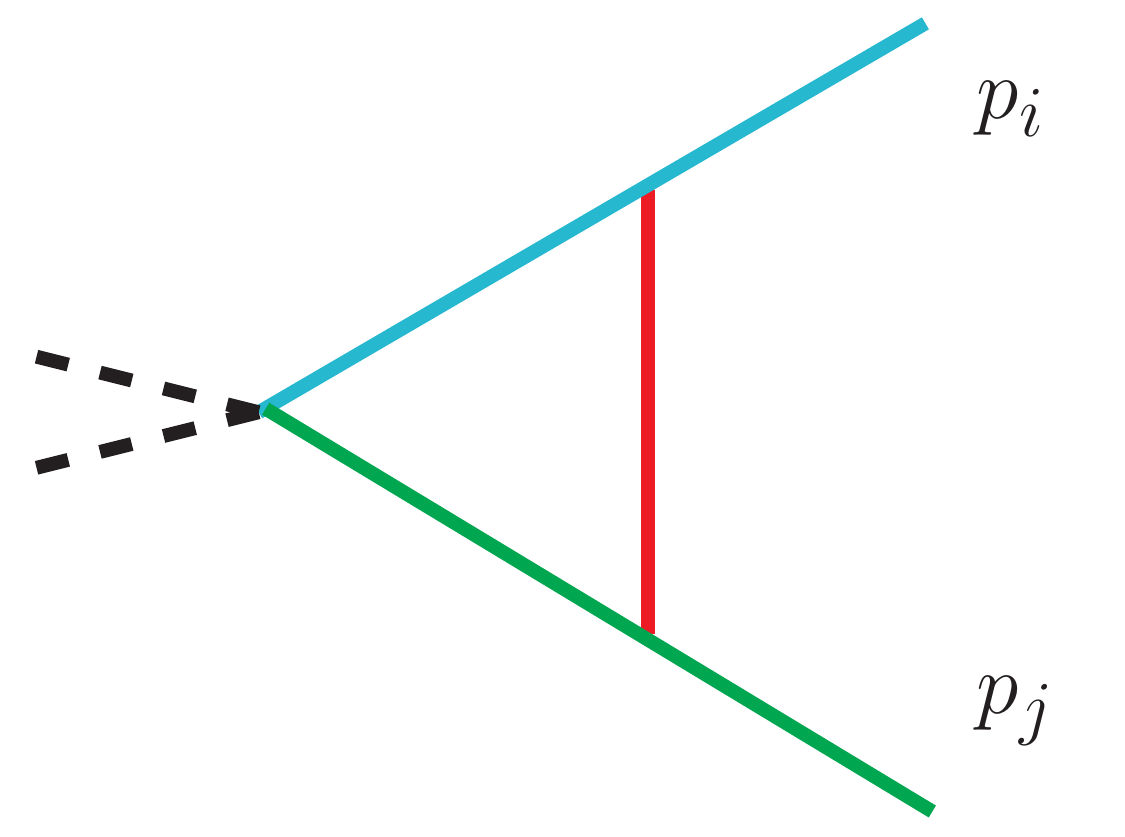}
    \caption{\label{fig:tr2} Two external partons}
  \end{subfigure}
  \caption{\label{fig:triangles} Divergent scalar triangle functions. Momentum flows
  from left to right. The red colour indicates a massless propagator while the blue and the green colour going from an internal to an external line means that the mass of the propagator is the same of the external (on-shell) particle with the same colour.}
\end{figure}
The first one, shown in Fig.\eqref{fig:tr1}, has an outgoing momentum $p_i$ connected, through a massless propagator, to a combination $k$ of at least two other external momenta; we can call it generically $C_i (k)$.
The second type is shown in Fig.\eqref{fig:tr2} and has two outgoing momenta, $p_i$ and $p_j$, connected by a massless propagator; for massless partons, it corresponds to the scalar three-point function $C_0(p_i, p_j)$.
Up to this point, the analysis equally applies also to the case of massive external partons.
The real part of the triangular functions is given, in the massless case, by
\begin{align}
{\rm Re}\{C_i(k)\} &= 
\frac{(4\pi)^{\varepsilon-2}}{\Gamma(1-\varepsilon)}
\frac{1}{(2\,p_i \cdot k - k^2)}
\frac{1}{\varepsilon^2} 
\left[ \left(\frac{\mu^2}{2\, p_i\cdot k}\right)^\varepsilon
-\left(\frac{\mu^2}{k^2}\right)^\varepsilon \right] \nonumber \\
&=
\frac{(4\pi)^{\varepsilon-2}}{\Gamma(1-\varepsilon)}
\frac{1}{(2\,p_i \cdot k - k^2)}
\frac{1}{\varepsilon} 
\left[
\log\left(\frac{\mu^2}{2\,p_i \cdot k}\right)
-\log\left(\frac{\mu^2}{k^2}\right) + \mathcal{O} (\varepsilon)
\right] \\
{\rm Re}\{C_0(p_i, p_j)\} &=
\frac{(4\pi)^{\varepsilon-2}}{\Gamma(1-\varepsilon)}
\frac{1}{2 \,p_i \cdot p_j} 
\left(\frac{\mu^2}{2 \,p_i\cdot p_j}\right)^\varepsilon
\left(\frac{1}{\varepsilon^2}-\frac{\pi^2}{2}\right)\,\,.
\label{eqn:triangle_Cij}
\end{align}

Finally, massless scalar two-point function $B_0^{ij}$
depending on a single massless external momentum $p_i$,
bring IR divergences
\begin{equation}
B_0 (p_i) \sim \frac{(4\pi)^{\varepsilon-2}}{\Gamma(1-\varepsilon)} \frac{1}{\varepsilon} 
\end{equation}
while we assume that massless tadpoles give no contributions to IR poles (i.e. we take them as vanishing).

The structure of the singularities of one-loop amplitudes is completed by adding the ones brought by the renormalization
procedure. While coupling renormalization involves only
UV poles, wave--function renormalization of the
external particles introduces both UV and IR poles.
These contributions can only bring single poles and are proportional to the whole leading order matrix element.

Now that we have analyzed the IR poles of Eq.\eqref{eq:nredamp}, we can compare the results with the general formula for the singular behaviour of the interference among the renormalized one--loop amplitude $A_m^{(1,R)}$ and the tree--level one $A_m^{(0)}$~\cite{Kunszt:1994np, Giele:1991vf}, that is
\begin{equation}
\label{eq:singular}
2\, \text{Re} \{A^{(0)\dagger}_m A^{(1,R)}_m \} \sim \frac{\alpha_S}{2\,\pi} \frac{(4\pi)^{\varepsilon}}{\Gamma(1-\varepsilon)}
\sum_i\sum_{j\ne i}\left[\frac{1}{\varepsilon^2}
\left(\frac{\mu^2}{2\,p_i\cdot p_j}\right)^\varepsilon
+\frac{1}{\varepsilon}\frac{\gamma_i}{{\bf T}_i^2}
\right] 
\langle 1,...,m | \mathbf{T}_i \mathbf{T}_j| 1,...,m \rangle
\end{equation}
where the usual definition of color charge operators and color correlated
matrix elements is implied and the constants $\gamma_i$ are given by
\begin{align*}
\gamma_{q} &= \gamma_{\bar{q}} = \frac{3}{2}\,C_F \\
\gamma_g  & = \frac{11}{6}\,C_A -\frac{2}{3}\,T_R \,N_f \numberthis \,\,.
\end{align*}
By inspecting Eq.(\ref{eq:singular}) we observe that there are no
single poles with a logarithm of an invariant formed with more than two external momenta ($2\,p_i \cdot k$).
These poles
are carried on individually and exclusively by the triangle integrals $C_i(k)$. Thus
we deduce that for full amplitudes the coefficients of the $C_i(k)$ scalar integrals
in Eq.(\ref{eq:nredamp}) must vanish\footnote{We remind here that the $C_i(k)$
triangle functions are still contained in the reduction of the $D_{00}$ functions,
and that their poles will anyway cancel with those of the other scalar integrals in
that reduction.}.
It is then clear that the double pole and the logarithmic part
of the single pole in Eq.\eqref{eq:singular} are produced by the triangle functions
$C_0 (p_i, p_j)$. Furthermore, the coefficients of these functions
(there is one for every pair of the external partons as for
the terms of the double sum in Eq.(\ref{eq:singular}))
obtained summing over all possible Feynman diagrams
contributing to a specific full one-loop process is
\begin{equation}
f_{ij} = 8 \pi \alpha_S \,(2\, p_i \cdot p_j )\,
\langle 1,...,m | {\mathbf{T}_i \cdot \mathbf{T}_j} | 1,...,m \rangle
\end{equation}
while the remaining single poles (the non-logarithmic ones) come from the bubbles $B_0(p_i)$ and the wave--function renormalization counterterms $\Delta Z(p_i)$.
The latters are necessary to obtain a fully local cancellation of IR singularities between the real cross section and the dual cross section that we are going to build.
The knowledge of the IR poles coming from $\Delta Z(p_i)$ and Eq.\eqref{eq:singular} allow to determine the coefficients of the bubble integrals $B_0(p_i)$, that turn out to be
\begin{equation}
f_i = 8 \pi \alpha_S \, C_i \, | A^{(0)}_m |^2 \left(1-\frac{\delta_{ig}}{2}\right) \,, \quad\quad i=q,\bar{q},g
\end{equation}
where $C_q=C_{\bar{q}}=C_F$ and $C_g=C_A$.

The triangles $C_0 (p_i, p_j)$, the bubbles $B_0(p_i)$ and the renormalization counterterms $\Delta Z(p_i)$ represent the key ingredients for the construction of the dual subtraction scheme.
In fact, our prescription to find all the necessary dual counterterms imposes to:
\begin{itemize}
\item collect triangles, bubbles and wave--function renormalization counterterms;
\item construct their dual representation by application of LTD;
\item extract their IR singular part by properly selecting
the loop integration domains.
\end{itemize}
Following these steps, in the next section we build a set
of dual subtractions.

\section{Loop--Tree duality and counterterms}
\label{sec:algorithm}
In this Section we show how to extract the singularities
from the virtual amplitude constructing counterterms. We first recall the
basics of LTD in section~\ref{sec:maps}. In particular, we list the main results that we need for our
construction and refer the interested reader to the seminal paper~\cite{Catani:2008xa} where they have
been derived for the first time, and also to the useful applications in~\cite{Sborlini:2016gbr, Sborlini:2016hat}, from which we borrow most of the discussion.
Then, in section~\ref{sec:real_beh} we will make a short summary of the relevant well know
limiting behaviours of tree--level amplitudes. In the subsequent section~\ref{sec:dual_sub} we build the dual subtractions and
we prove the local cancellation of the
singularities in the real matrix element in section~\ref{sec:count-limit}. In section~\ref{sec:int_dual_sub} we present the results of the integrated subtractions
and show the agreement with the general structure of the singularities expressed in Eq.(\ref{eq:singular}).

\subsection{Loop--Tree duality}
\label{sec:maps}
\begin{figure}[t]
\centering
\includegraphics[width=0.4\linewidth]{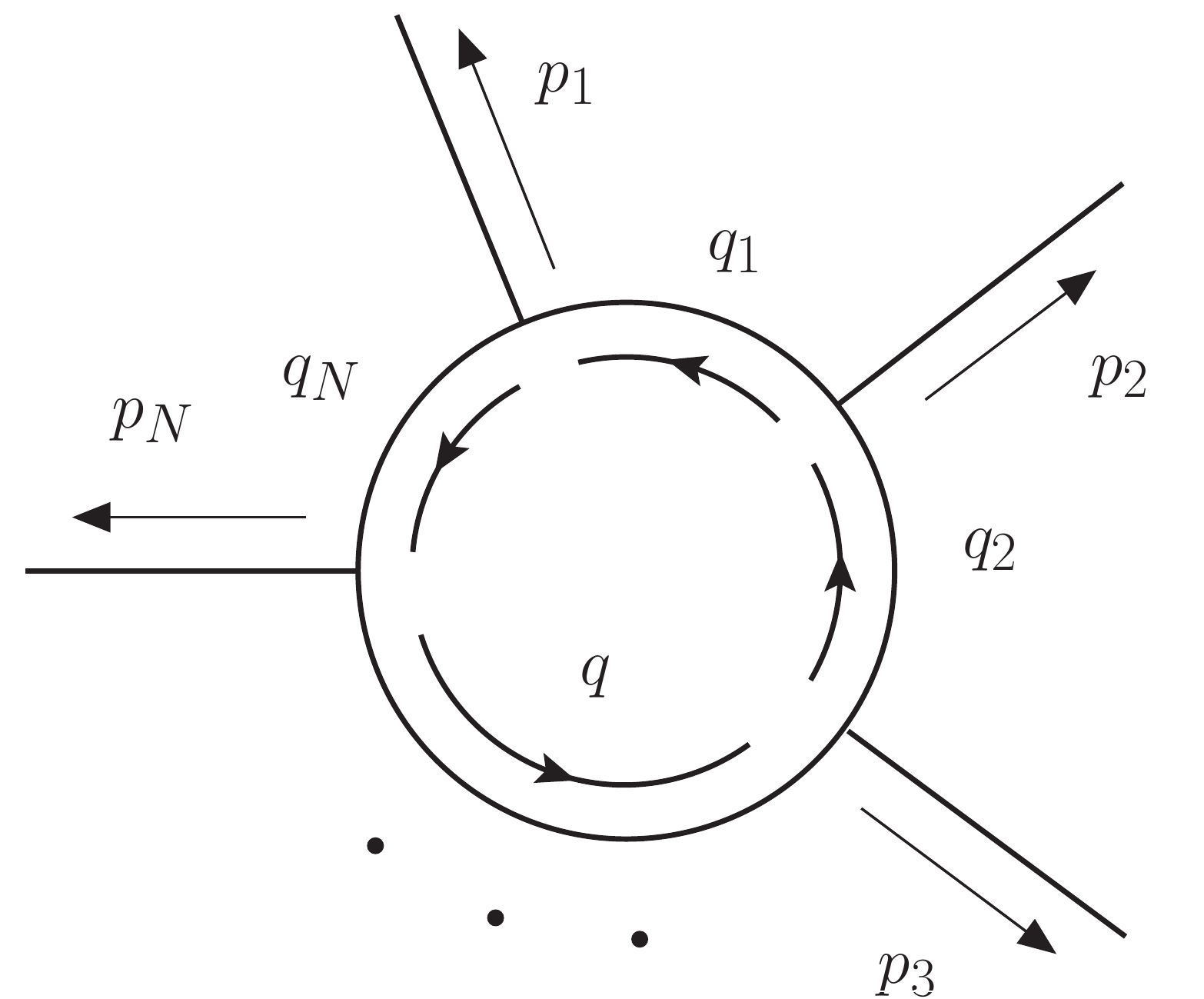}
\caption{scalar loop with $N$ external legs. The momenta $p_1, \ldots, p_N$ are taken as outgoing, while the loop momentum $q$ flows counter-clockwise.}
\label{fig: scalar_loop}
\end{figure}
Let us consider a generic $N$-particle scalar loop, which is shown in figure \ref{fig: scalar_loop}.  The external momenta are denoted as $p_1, \ldots, p_N$ and are taken as outgoing, $q$ is the counter-clockwise flowing loop momentum, and $q_i$ are the momenta of the internal lines.
Momentum conservation reads
\begin{equation}
\sum_{i = 1}^N p_i = 0
\end{equation}
while the internal momenta $q_i$ are related to $q$ and the external momenta by
\begin{equation}
\label{eqn:internal_momenta}
q_i = q + \sum_{k = 1}^i p_k
\end{equation}
which implies the choice $q_N=q$. With this notation, the $N$-particle scalar one-loop integral is given by\footnote{Adding a numerator to Eq.\eqref{def:loop integral} one can extend the discussion to any kind of integral.}
\begin{equation}
\label{def:loop integral}
L^{(N)} (p_1, \ldots, p_N) \equiv \int_{q} \prod^N_{i = 1} G_F (q_i)
\end{equation}
where $G_F (q_i) \equiv (q^2_i - m^2_i + i 0)^{- 1}$ are the Feynman propagators and we have used the shorthand notation
\begin{equation}
\int_{q} \cdot \,\, \equiv - i \mu^{4 - d} \int \frac{d^d q}{(2 \pi)^d} \,\,\, \cdot
\end{equation}
$\mu$ being an arbitrary energy scale used to restore the physical dimensions of the integral.

The LTD theorem states that the loop integral in Eq.\eqref{def:loop integral} has the following \textit{dual representation}
\begin{equation}
\label{th:LTD_theorem}
L^{(N)} (p_1, \ldots, p_N) = - \sum_{i = 1}^N \int_{q} \tilde{\delta} (q_i)  \prod^N_{j = 1, \, j \neq i} G_D (q_i; q_j)
\end{equation}
where $\tilde{\delta} (q_i) \equiv 2 \pi i \theta (q_i^0) \delta (q_i^2 - m_i^2)$ sets the internal momentum $q_i$ on-shell and 
\begin{equation}
\label{def:dual propagator}
G_D (q_i; q_j) \equiv \frac{1}{q_j^2 - m_j^2 - i 0\, \eta \cdot (q_j - q_i)}
\end{equation}
are \textit{dual propagators}. Here $\eta$ is an arbitrary time-like or light-like vector with positive definite energy which can be chosen as $\eta \equiv (1,\mathbf{0})$.
The only difference between the dual propagator and the Feynman propagator $G_F$ lies in the different $i 0$ prescription which regularizes the singularity of the right-hand side of Eq.\eqref{def:dual propagator}.
Using $\eta \equiv (1,\mathbf{0})$, we see that this \textit{dual $i 0$ prescription} depends on the sign of $q_j^0-q_i^0$ which in turn is related, through Eq.\eqref{eqn:internal_momenta}, to the energy components of the external momenta
\begin{equation}
q^0_j - q^0_i = q^0 + \sum_{k = 1}^j p^0_k - q^0 - \sum_{k = 1}^i p^0_k = \text{sign} (j - i) \sum_{k = \text{min} \{ i, j \}}^{\text{max} \{ i, j \}} p^0_k \,\,.
\end{equation}
To the right-hand side of Eq.\eqref{th:LTD_theorem} we find $N$ different integrals.
In each of them, one Feynman propagator is removed, the corresponding internal momentum is set on-shell and the remaining Feynman propagators are turned into dual propagators.
The role of the on-shell internal momentum is passed through the $N$ internal momenta $q_1, \ldots, q_N$ without repetition, so that each momentum $q_i$ is on-shell in one and only one integral.
Moreover, in each integral we can change the integration variable in order to integrate over the on-shell momentum itself. If $q_i$ is the on-shell momentum, this is done by mean of a simple translation
\begin{equation}
\label{eqn:translation}
q^{} \rightarrow q' = q + \sum_{k = 1}^i p^{}_k
\end{equation}
suggested by Eq.\eqref{eqn:internal_momenta}.
This means that the dual representation in Eq.\eqref{th:LTD_theorem} allows us to replace a one-loop integral over an off-shell virtual momentum by a linear combination of integrals over on-shell virtual momenta.

Let us now consider the dual representation of a virtual amplitude, and suppose we have built a dual subtraction from it. In the framework of the subtraction method, we should integrate the dual cross section together with the real one.
Since they live on different phase spaces, we need a mapping $(\Phi_m, q_i) \leftrightarrow \Phi_{m+1}$ connecting the virtual to the real phase space.
To be precise, we will use different mappings for different dual counterterms.
This means that, in a Monte Carlo integration, we can generate  a point in the real phase space and then obtain, through the application of each mapping, as many virtual configurations as the number of counterterms.
Moreover, in order to achieve a local cancellation of divergences, we require that IR singularities of the dual integrations are mapped into soft and collinear  singularities of the $(m+1)$-particle matrix element squared.
\begin{figure}[t]
\centering
\begin{subfigure}{.48\textwidth}
\centering
\includegraphics[width=0.6\linewidth]{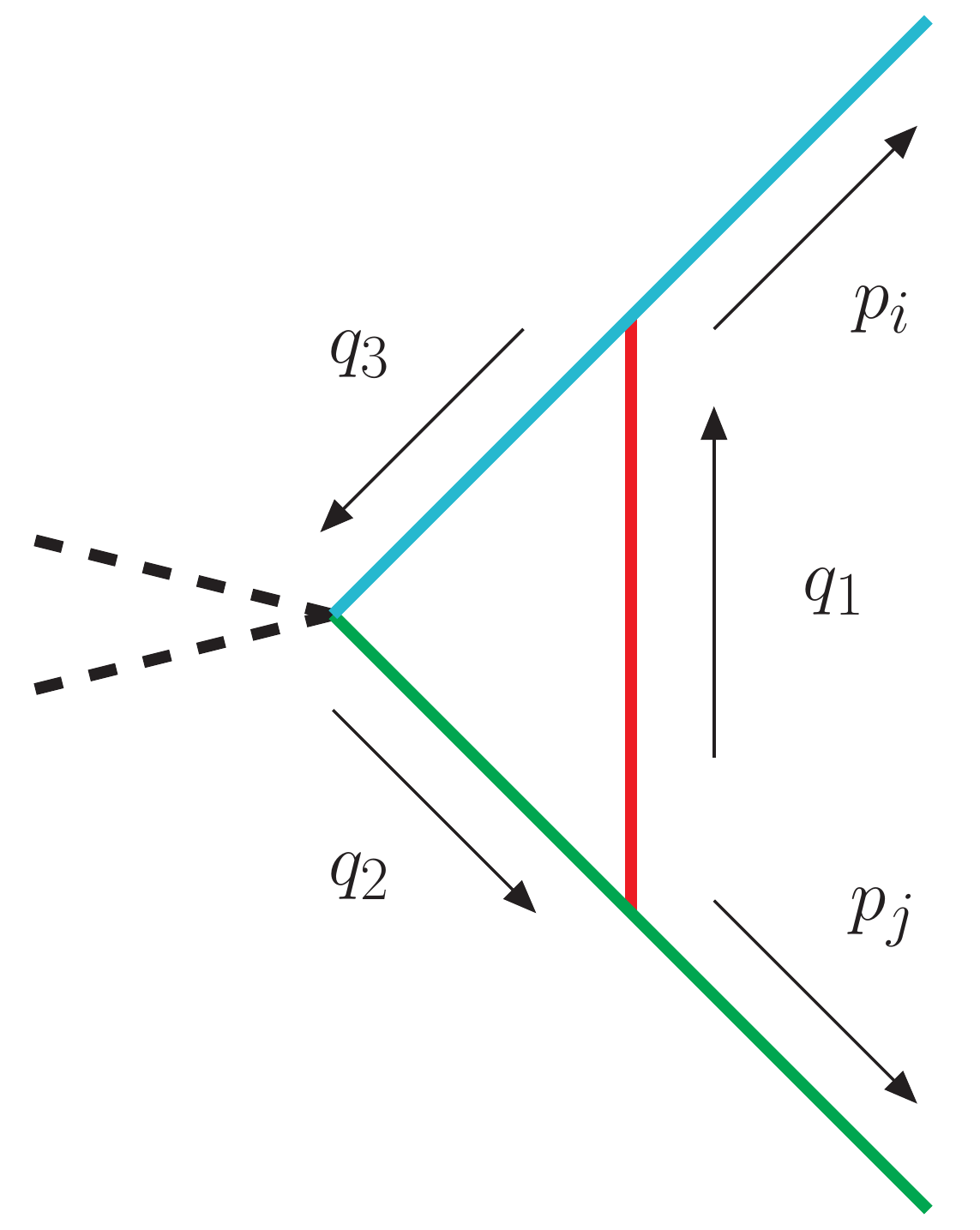}
\caption{\label{fig:vertex_ij1}}
\end{subfigure}
\begin{subfigure}{.48\textwidth}
\centering
\includegraphics[width=0.6\linewidth]{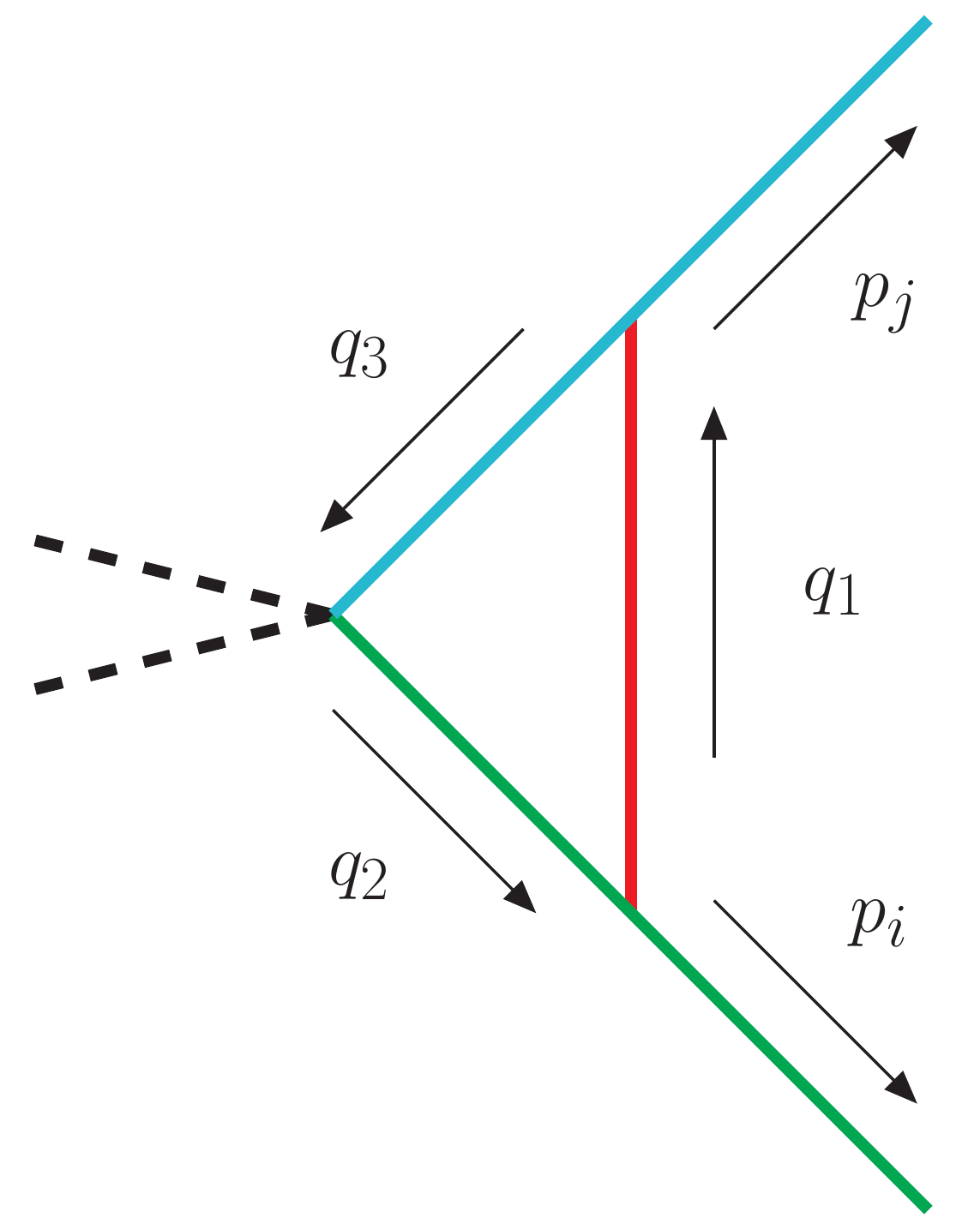}
\caption{\label{fig:vertex_ij2}}
\end{subfigure}
\caption{\label{fig:vertex_ij}Diagrams associated with three-point functions coming from the reduction of the virtual amplitude.
The meaning of the colours is the same of Fig.\eqref{fig:triangles}.
The triangle of panel (\subref{fig:vertex_ij2}) is obtained by exchanging $p_i \leftrightarrow p_j$ in the triangle of panel (\subref{fig:vertex_ij1}).}
\end{figure}
Considering the emitter-spectator pair $(i,j)$, we always denote the internal momenta of three-point function $C_0 (p_i, p_j)$ as in Fig.\eqref{fig:vertex_ij1}, namely $q_1$ is the virtual momentum connecting the external particles, while $q_2=q_1+p_j$ and $q_3=q_1-p_i$.
The dual contributions where $q_1$ is set on-shell by LTD bring IR singularities that can be associated with a real sector parton $p'_c$ being emitted from a parton of momentum $p'_a$ and absorbed by a spectator $p'_b$.
To this end, we make use of the following momentum mapping~\cite{Sborlini:2016gbr}
\begin{equation}
\label{map:map1inv}
\begin{split}
q_1 &= p'_c \\
p_i &= p'_a +p'_c - \frac{\alpha^{(1)}_{abc}}{1-\alpha^{(1)}_{abc}} p'_b \\
p_j &= \frac{1}{1-\alpha^{(1)}_{abc}} p'_b 
\end{split}
\end{equation}
where $\alpha^{(1)}_{abc} \equiv s_{ac}/s_{abc}$, $s_{ac}=(p_a+p_c)^2$ (that in the massless case reduces to $s_{ac}=2 p_a \cdot p_c$), $s_{abc}= (p_a+p_b+p_c)^2$, $q_1$ is the virtual on-shell loop momentum, and all the other momenta are unchanged.
The definition domain of Eq.\eqref{map:map1inv} is chosen to be the region where $s_{ac} < s_{bc}$.
This selects a compact subset of the loop momentum phase space, which will be parametrized later.
The momentum mapping in Eq.\eqref{map:map1inv} automatically verifies $p'_1+p'_2+p'_3=p_1+p_2$ (momentum conservation) and $p_1^2=p_2^2=0$ (on-shell conditions).
It is useful to consider also the inverse momentum mapping, which is given by
\begin{align}
\label{map:map1}
p'_a &= p_i -q_1 + \alpha^{(1)}_{ij}\, p_j & \nonumber \\
p'_b &= (1-\alpha^{(1)}_{ij}) p_j & \alpha^{(1)}_{ij} \equiv \frac{q_1 \cdot p_i}{p_j\cdot (p_i-q_1)} = \alpha^{(1)}_{abc} \\
p'_c &= q_1 \nonumber
\end{align}
and is the one explicitly used in~\cite{Sborlini:2016gbr}.
The soft emission  $p'_c \rightarrow 0$ is mapped into $q_1 \rightarrow 0$, while the collinear configuration $p'_a || p'_c$ is given by $q_1 || p_i$, as can be deduced by noticing that $\alpha^{(1)}_{ij} \rightarrow 0$ when $q_1 \cdot p_i \rightarrow 0$. Indeed one has that
\begin{equation}
p'_a \cdot p'_c = q_1 \cdot p_i + \alpha^{(1)}_{ij}\, q_1 \cdot p_j \rightarrow 0 \quad\text{if}\quad q_1 \cdot p_i \rightarrow 0 \,\,.
\end{equation}
Also, Eq.\eqref{map:map1inv} clearly shows that, when $p'_a$ and $p'_c$ become collinear, the momentum $p_i$ is mapped into the emitter momentum $p'_a+p'_c$.

In order to properly relate the integration over $\Phi_{m+1}$ to the one over $(\Phi_m, q_i)$, we need to compute the jacobian of the transformation in Eq.\eqref{map:map1inv}.
To do this, let us move to the center-of-mass frame of  $p_i$ and $p_j$, where the loop momenta $q_k$ can be assigned in a spherical coordinate system whose zenith is the direction of $p_i$
\begin{equation}
\label{def:param_loop_momenta}
q_k = \xi_k \frac{\sqrt{s_{ij}}}{2} \left( 1, 2 \sqrt{v_k (1 - v_k)} \cos \varphi_k, 2 \sqrt{v_k (1 - v_k)} \sin \varphi_k, 1 - 2 v_k \right) \,\,.
\end{equation}
Here, $\xi_k \sqrt{s_{ij}}/2$ is the energy of $q_k$, $v_k \in \left[ 0,1 \right]$ is related to the cosine of the polar angle by $\cos \theta_k \equiv 1- 2 v_k$ and $\varphi_k$ is the azimuthal angle.
We can now express the momentum mapping in Eq.\eqref{map:map1inv} as a relation between the dimensionless variables $(\xi_1,v_1)$ and the kinematic invariants $y'_{mn} \equiv s_{mn}/s_{abc}$, obtaining
\begin{equation}
\label{map:inv_kin_inva1}
\begin{split}
\xi_1 &= y'_{ac} + y'_{bc} \\
v_1 &=  \frac{y'_{ac} (1 - y'_{ac} - y'_{bc})}{(1-y'_{ac})(y'_{ac}+y'_{bc})} \,\,.
\end{split}
\end{equation}
The collinear limit $y'_{ac} \rightarrow 0$ is mapped into $v_1 \rightarrow 0$, while the soft one is approached when $\xi_1 \rightarrow 1$.
The jacobian related to Eq.\eqref{map:inv_kin_inva1} is given by
\begin{equation}
\label{map: J1}
J (y'_{ac},y'_{bc})= \frac{1-y'_{ac}-y'_{bc}}{(1-y'_{ac})^2 (y'_{ac}+y'_{bc})}\,\,.
\end{equation}
Also, by using the inverse of Eq.\eqref{map:inv_kin_inva1}
\begin{align}
\label{map:kin_inva1}
y'_{a c} &= \frac{v_1 \xi_1}{1 - (1 - v_1) \xi_1} \nonumber \\
y'_{b c} &= \frac{(1 - v_1) (1 - \xi_1) \xi_1}{1 - (1 - v_1) \xi_1}
\end{align}
we can find the boundaries of the loop integration domain selected by the
momentum mapping in Eq.\eqref{map:map1inv}
\begin{equation}
\label{def:R_1}
\theta (y'_{bc} - y'_{ac}) \equiv \mathcal{R}_1(\xi_1, v_1) = \theta (1 - 2 v_1) \,\, \theta\left( \frac{1 - 2 v_1}{1 - v_1} - \xi_1 \right) \,\,.
\end{equation}
We can now express the phase space factorization as
\begin{equation}
\label{eqn:phase_space}
\int d \Phi_m \int_{q_1} \tilde{\delta} (q_1) \, \mathcal{R}_1 (\xi_1, v_1)
= \mu^{2 \varepsilon} \int d \Phi_{m+1}
\theta (y'_{bc} - y'_{ac}) J(y'_{ac},y'_{bc}) \xi_1(y'_{ac},y'_{bc}) \,\,.
\end{equation}
where $d \Phi_m$ ($d \Phi_{m+1}$) is the customary differential phase space element for $m$ ($m+1$) partons in the final state.

In the region of the real phase space where $s'_{bc} < s'_{ac}$ (i.e. the one where $p'_b$ and $p'_c$ can be collinear), the dual contributions in which $q_2$ (green propagator in Fig.\eqref{fig:vertex_ij1}) is set on-shell can be used to match the soft and the collinear singularities~\cite{Sborlini:2016gbr}.
To this end, a second mapping can be used
\begin{align*}
\label{map:map2inv}
q_2 &= p'_b &  & \\
p_i &= \frac{1}{1-\alpha^{(2)}_{abc}} p'_a & \alpha^{(2)}_{abc} &=  \frac{s'_{bc}}{s'_{abc}} = y'_{bc} \\
p_j &= p'_b +p'_c - \frac{\alpha^{(2)}_{abc}}{1-\alpha^{(2)}_{abc}} p'_a & & \numberthis
\end{align*}
with all the other partons left unaltered.
In terms of $(y'_{ac},y'_{bc})$ and $(\xi_2, v_2)$, we have
\begin{align*}
\label{map:inv_kin_inva2}
\xi_2 &= 1-y'_{ac} \\
v_2 &=  \frac{1 - y'_{ac} - y'_{bc}}{(1-y'_{ac})(1-y'_{bc})} \numberthis
\end{align*}
and the associated jacobian is
\begin{equation}
\label{map:J2}
J_2(y'_{ac},y'_{bc})= \frac{y'_{ac}}{(1-y'_{ac}) (1-y'_{bc})^2} \,\,.
\end{equation}
The inverse mapping
\begin{align*}
\label{map:map2}
p'_a &= (1-\alpha^{(2)}_{ij}) p_i &&\\
p'_b &= q_2 & \alpha^{(2)}_{ij} &\equiv \frac{q_2 \cdot p_j}{p_i \cdot (p_j-q_2)} = \alpha^{(2)}_{abc}  \\
p'_c &= p_j - q_2 + \alpha^{(2)}_{ij} p_i && \numberthis
\end{align*}
used in~\cite{Sborlini:2016gbr} leads to
\begin{align*}
\label{map:kin_inva2}
y'_{a c} &= 1 - \xi_2 \\
y'_{b c} &= \frac{(1 - v_2) \xi_2}{1 - v_2 \xi_2} \numberthis
\end{align*}
which gives us the definition domain through
\begin{equation}
\label{def:R_2}
\theta (y'_{ac} - y'_{bc}) \equiv \mathcal{R}_2 (\xi_2, v_2) = \theta\left( \frac{1 }{1 + \sqrt{1-v_2}} - \xi_2 \right) \,\,.
\end{equation}
Finally, phase space factorization is given by
\begin{equation}
\label{eqn:phase_space2}
\int d \Phi_m \int_{q_2} \tilde{\delta} (q_2) \, \mathcal{R}_2 (\xi_2, v_2)
= \mu^{2 \varepsilon} \int d \Phi_{m+1}\,
\theta (y'_{bc} - y'_{ac})\, J_2(y'_{ac},y'_{bc})\, \xi_2(y'_{ac},y'_{bc}) \,\,.
\end{equation}

Given a pair $(i,j)$ of partons with momenta $p_i$ and $p_j$, LTD reconstructs the entire triangle in Fig.\eqref{fig:vertex_ij1} by cutting the propagators of $q_1$, $q_2$ and $q_3$.
Nonetheless, to match single soft and collinear singularities associated with the $(a,c,b)$ real parton phase space configurations, we just need to collect the dual contributions with $q_1$ and $q_2$ set on-shell, selecting the loop integration domains by using the functions $\mathcal{R}_1$ and $\mathcal{R}_2$ defined in Eqs.\eqref{def:R_1} and \eqref{def:R_2}~\cite{Sborlini:2016gbr}.
In the construction of a working subtraction scheme, we may follow a slightly different approach, though.
In fact, the region where $s'_{bc} < s'_{ac}$ can be covered by simply exchanging $p'_a \leftrightarrow p'_b$ and $p_i \leftrightarrow p_j$ in the momentum mapping of Eq.\eqref{map:map1inv}.
Then, to cancel the corresponding singularities, we just need to exchange $p_i \leftrightarrow p_j$ in the dual counterterms with $q_1$ on-shell.
This corresponds to cutting the propagator of $q_1$ in the triangle of Fig.\eqref{fig:vertex_ij2} instead of the propagator of $q_2$ in the triangle of Fig.\eqref{fig:vertex_ij1}.
In this way, we always put $q_1$ on-shell, avoiding to introduce different counterterms, and we use one mapping instead of two.
Moreover, if we chose to set on-shell $q_2$ and use the momentum mapping in Eq.\eqref{map:map2inv}, this would reflect into a
more involved counting of the counterterm
matching the singularities of the $(m+1)$-parton phase space. More details
on this subject are given in Appendix~\ref{sec:counting}.

\subsection{Singular behavior of real amplitudes}
\label{sec:real_beh}
In this Section we analyze the factorization properties of tree--level amplitudes with $m+1$ final state partons in the soft and collinear regions of the phase space~\cite{Altarelli:1977zs,Bassetto:1984ik,Catani:1996vz}.
The formulas of this section will be our reference point to check that the dual counterterms we are going to define have the same \textit{local} behavior of the real cross section in the IR singular regions.

The singular behaviour of an $(m+1)$-parton tree--level amplitude does not depend on the details of its structure.
Furthermore, it turns out that in the soft~\cite{Bassetto:1984ik} and collinear~\cite{Altarelli:1977zs} regions the $(m+1)$-parton amplitude is factorizable with respect to the $m$-parton one.

Let us first consider the limit where two momenta, say $p'_a$ and $p'_c$, become collinear.
By definition, this configuration is parametrized by
\begin{subequations}
\label{par:coll}
\begin{align}
\label{par:coll_pa}
{p'_a}^{\mu} &= z p^{\mu} + k_{\bot}^\mu - \frac{k_{\bot}^2}{z} \frac{n^{\mu}}{2 p \cdot n} \\
\label{par:coll_pc}
{p'_c}^{\mu} &= (1-z) p^{\mu} - k_{\bot}^\mu - \frac{k_{\bot}^2}{1-z} \frac{n^{\mu}}{2 p \cdot n} \\
p'_a \cdot p'_c &= - \frac{k_{\bot}^2}{2 z (1 - z)}
\end{align}
\end{subequations}
where $p$ and $n$ are light-like vector (with $p$ individuating the collinear direction), $k_\bot$ ($k_\bot^2<0$) is the transverse component ($k_\bot \cdot p = k_\bot \cdot n = 0$) and $z$ is the fraction of momentum $p$ carried by $p'_a$ in the collinear limit which is approached when $k_\bot \rightarrow 0$.

In the collinear limit, the $(m+1)$ tree--level matrix element squared has the following behaviour~\cite{Altarelli:1977zs}
\begin{multline}
\label{eqn:coll_beh}
_{m+1}\langle 1, \dots, m+1|1,\dots,m+1 \rangle_{m+1} \rightarrow \\
\frac{1}{p'_a \cdot p'_c} 4 \pi \alpha_S \mu^{2 \varepsilon} \,\, _m \langle 1, \dots, m + 1| \hat{P}_{(a c), c} (z, k_{\bot}, \varepsilon) |1, \dots, m + 1 \rangle_m
\end{multline}
where the $m$-parton matrix element on the right-hand side is obtained by replacing the partons $a$ and $c$ in the $(m+1)$-parton matrix element by a single parton denoted by $a c$ (the emitter), with momentum equal to $p'_a + p'_c$ and other quantum numbers (flavour, colour) given by the following rule: anything $+$ gluon gives anything, and quark $+$ antiquark gives gluon.
The matrix in color and spin $\hat{P}_{(a c), c}$ on the right-hand side of Eq.\eqref{eqn:coll_beh} is the tree--level $d$-dimensional Altarelli--Parisi splitting function. It is a matrix acting on the spin indices of the parton $ac$ in $_m \langle 1, \dots, m + 1|$ and $|1, \dots, m + 1 \rangle_m$. Its value depends on whether the partons $ac$ and $a$ are quarks, antiquarks or gluons, according to
\begin{subequations}
\label{def:splitting_functions}
\begin{align}
\langle s | \hat{P}_{q, q} (z, k_{\bot}, \varepsilon) | s' \rangle &\equiv \delta_{s s'} C_F \left[ \frac{1 + z^2}{1 - z} - \varepsilon (1 - z) \right] \\
\langle s | \hat{P}_{q, g} (z, k_{\bot}, \varepsilon) | s' \rangle &\equiv \delta_{s s'} C_F \left[ \frac{1 + (1 - z)^2}{z} - \varepsilon z \right] \\
\langle \mu | \hat{P}_{g, q} (z, k_{\bot}, \varepsilon) | \nu \rangle &\equiv T_R \left[ - g^{\mu \nu} + 4 z (1 - z) \frac{k^{\mu}_{\bot} k^{\nu}_{\bot}}{k^2_{\bot}} \right] \\
\label{def:P(g,g)}
\langle \mu | \hat{P}_{g, g} (z, k_{\bot}, \varepsilon) | \nu \rangle &\equiv 2 C_A \left[ - g^{\mu \nu} \left( \frac{z}{1 - z} + \frac{1 - z}{z} \right) - 2 (1 - \varepsilon) z (1 - z) \frac{k^{\mu}_{\bot} k^{\nu}_{\bot}}{k^2_{\bot}} \right]
\end{align}
\end{subequations}
where $\langle s | \cdot | s' \rangle$ and $\langle \mu | \cdot | \nu \rangle$ denote the components of the splitting function $\hat{P}_{(a c), a}$ in the spin space of the parton $ac$.
In Eqs.\eqref{def:splitting_functions}, the label $g$ stands for gluon, while $q$ stands both for quark and antiquark.

In case the parton $a$ belongs to the initial state, we use a quite different parametrization
\begin{subequations}
\label{par:coll_init}
\begin{align}
{p'_c}^{\mu} &= (1-x) {p'_a}^{\mu} + k_{\bot} - \frac{k_{\bot}^2}{1-x} \frac{n^{\mu}}{2 p'_a \cdot n} \\
p'_c \cdot p'_a &= - \frac{k_{\bot}^2}{2(1 - x)}
\end{align}
\end{subequations}
describing the splitting process $a \rightarrow ac + c$, where $c$ is a final state parton and $ac$ is the parton entering the hard scattering, whose quantum numbers are assigned according to the following rule: if $a$ and $c$ are partons of the same type, then $ac$ is a gluon, while if $a$ is a fermion (gluon) and $c$ is a gluon (fermion), then $ac$ is a fermion (antifermion).
For the case of emission from an initial state parton, in the limit $k_\perp \rightarrow 0$, the tree--level matrix element behaves as follows
\begin{multline}
\label{eqn:coll_beh_init}
_{m+1}\langle 1, \dots, m+1|1,\dots,m+1 \rangle_{m+1} \rightarrow \\
\frac{1}{x}\frac{1}{p'_a \cdot p'_c} 4 \pi \alpha_S \mu^{2 \varepsilon} \,\, _m \langle 1, \dots, m + 1| \bar{P}_{a, (ac)} (x, k_{\bot}, \varepsilon) |1, \dots, m + 1 \rangle_m
\end{multline}
where the $m$-parton matrix element on the right-hand side is obtained by replacing the partons $a$ and $c$ in the $(m+1)$-particle matrix element by the initial state parton $ac$, which takes the role of $a$.
The tree--level Altarelli--Parisi splitting functions for the case of initial state radiation can be obtained from the ones in Eqs.(\ref{def:splitting_functions}) by the crossing relation
\begin{equation}
\bar{P}_{a, (ac)} (x, k_{\bot}, \varepsilon)=
-(-1)^{F(a)+F(ac)}x \hat{P}_{(ac), c} (1/x, k_{\bot}, \varepsilon)
\end{equation}
with $F(q)=F(\bar{q})=1$, $F(g)=0$. They are given by
\begin{subequations}
\label{def:splitting_functions_initial}
\begin{align}
\langle s | \bar{P}_{q, q} (x, k_{\bot}, \varepsilon) | s' \rangle &\equiv \delta_{s s'} C_F \left[ \frac{1 + x^2}{1 - x} - \varepsilon (1 - x) \right] \\
\langle s | \bar{P}_{g, q} (x, k_{\bot}, \varepsilon) | s' \rangle &\equiv \delta_{s s'} C_F \left[1 - \varepsilon - 2 x( 1-x) \right] \\
\langle \mu | \bar{P}_{q, g} (x, k_{\bot}, \varepsilon) | \nu \rangle &\equiv T_R \left[ - g^{\mu \nu} x - 4 \frac{1 - x}{x} \frac{k^{\mu}_{\bot} k^{\nu}_{\bot}}{k^2_{\bot}} \right] \\
\langle \mu | \bar{P}_{g, g} (x, k_{\bot}, \varepsilon) | \nu \rangle &\equiv 2 C_A \left[ - g^{\mu \nu} \left( \frac{x}{1 - x} + x (1 - x) \right) - 2 (1 - \varepsilon) \frac{1 - x}{x} \frac{k^{\mu}_{\bot} k^{\nu}_{\bot}}{k^2_{\bot}} \right]\,.
\end{align}
\end{subequations}
The limit where a gluon, say $p'_c$, becomes soft, is parametrized by
\begin{equation}
\label{par:soft}
{p'_c}^\mu = \lambda p^\mu , \quad\quad \lambda \rightarrow 0
\end{equation}
where $p^\mu$ is an arbitrary vector that specifies the direction along which the parton $c$ approaches the soft limit.
When $\lambda \rightarrow 0$, we have the following behaviour for the $(m+1)$-parton matrix element squared~\cite{Bassetto:1984ik}
\begin{multline}
\label{eqn:soft_beh}
_{m+1}\langle 1, \dots, m+1|1,\dots,m+1 \rangle_{m+1} \rightarrow \\
- \frac{1}{\lambda^2} 4 \pi \alpha_S \mu^{2 \varepsilon}  \sum_{a, b} \frac{p'_a \cdot p'_b}{(p'_a \cdot p) (p'_b \cdot p)} \,_m\langle 1, \dots, m + 1| \mathbf{T}_a \mathbf{T}_b |1, \dots, m + 1 \rangle_m
\end{multline}
where
the $m$-parton matrix element on the right-hand side is obtained by removing the soft gluon $p'_c$ from the $(m+1)$-parton matrix element.
As we can see, the $m$-parton matrix element is not exactly factorized, because of \textit{colour correlations} induced by the colour-charge operator $\mathbf{T}_a \mathbf{T}_b$\,.
Nonetheless, we still have exact factorization in processes with $m=2$ and $m=3$ since, in these cases, the product of two colour-charge operators is always proportional to a linear combination of Casimir operators.

If there are partons in the initial state, Eq.\eqref{eqn:soft_beh} still holds:
we just need to extend the sum over more pairs, also including the initial state partons.

Let us now move to the case of massive partons in the final state~\cite{Catani:2000ef, keller:1998tf}.
If two momenta become collinear, and at least one of them is massive, there is no collinear singularity.
However, the matrix element squared is enhanced for very small values of the parton masses.
To take this effect into account, we parametrize the momenta $p'_a$ and $p'_c$ of the two collinear partons in the following way~\cite{Catani:2002hc}
\begin{subequations}
\label{par:coll_mass}
\begin{align}
{p'_a}^{\mu} &= z p^{\mu} + k_{\perp} - \frac{k_{\bot}^2+z^2 m^2_{ac} - m_a^2}{z} \frac{n^{\mu}}{2 p \cdot n} \\
{p'_c}^{\mu} &= (1-z) p^{\mu} - k_{\bot} - \frac{k_{\bot}^2+(1-z)^2 m^2_{ac} - m_c^2}{1-z} \frac{n^{\mu}}{2 p \cdot n} \\
(p'_a+ p'_c)^2 &= - \frac{k_{\bot}^2}{z(1 - z)}+\frac{m_a^2}{z}+\frac{m_c^2}{1-z}
\end{align}
\end{subequations}
where $m_a$, $m_c$ and $m_{ac}$ are the masses of the partons $a$, $c$ and the emitter parton $ac$, respectively, $p$ is a time-like vector which individuates the collinear direction, while $n$, $k_\perp$ and $z$ have the same meaning as in the massless case (see Eqs.\eqref{par:coll}).
We then consider the behavior of the tree--level matrix element under the following uniform rescaling
\begin{equation}
\label{rescaling}
k_\perp \rightarrow \lambda k_\perp, \quad m_a \rightarrow \lambda m_a, \quad m_c \rightarrow \lambda m_c \quad m_{ac} \rightarrow \lambda m_{ac}
\end{equation}
in the limit $\lambda \rightarrow 0$.
We have~\cite{Catani:2002hc}
\begin{multline}
\label{eqn:coll_beh_mass}
_{m+1}\langle 1, \dots, m+1|1,\dots,m+1 \rangle_{m+1} \rightarrow \\
\frac{1}{\lambda^2}\frac{ 8 \pi \alpha_S \mu^{2 \varepsilon}}{(p'_a+p'_c)^2-m^2_{ac}} \,\, _m \langle 1, \dots, m + 1| \hat{P}_{(a c), a} (z, k_{\bot}, \{m\},\varepsilon) |1, \dots, m + 1 \rangle_m
\end{multline}
where the $m$-parton matrix element on the right-hand side is obtained in the exact same way as for Eq.\eqref{eqn:coll_beh}.
The function $\hat{P}_{(a c), a}$ on the right-hand side of Eq.\eqref{eqn:coll_beh_mass} is the generalization of the $d$-dimensional Altarelli-Parisi splitting function to the massive case (the dependence on the masses $m_a$, $m_c$ and $m_{ac}$ being denoted by $\{m\}$).
As in the massless case, it is a matrix acting on the spin indices of the parton $a c$ in $_m \langle 1, \dots, m + 1|$ and $|1, \dots, m + 1 \rangle_m$, whose expression depends on the type of partons involved in the splitting process $ac \rightarrow a + c$.
Denoting quarks and antiquarks by $q$ and gluons by $g$, we have
\begin{subequations}
\label{def:splitting_functions_mass}
\begin{align}
\langle s | \hat{P}_{q, q} (z, k_{\bot}, \{m\}, \varepsilon) | s' \rangle &\equiv \delta_{s s'} C_F \left[ \frac{1 + z^2}{1 - z} - \varepsilon (1 - z) - \frac{m_q^2}{p'_q \cdot p'_g} \right] \\
\langle s | \hat{P}_{q, g} (z, k_{\bot}, \{m\}, \varepsilon) | s' \rangle &\equiv \delta_{s s'} C_F \left[ \frac{1 + (1 - z)^2}{z} - \varepsilon z -\frac{m_q^2}{p'_q \cdot p'_g} \right] \\
\langle \mu | \hat{P}_{g, q} (z, k_{\bot}, \{m\}, \varepsilon) | \nu \rangle &\equiv T_R \left[ - g_{\mu \nu} - 4 \frac{k^{\mu}_{\bot} k^{\nu}_{\bot}}{(p'_q+p'_{\bar{q}})^2}\right] 
\end{align}
\end{subequations}
where, as usual, $\langle s | \cdot | s' \rangle$ and $\langle \mu | \cdot | \nu \rangle$ label the components of the splitting function $\hat{P}_{(a c), a}$ in the spin space of the emitter parton $ac$.
The function $\hat{P}_{g, g}$ related to the splitting process $g \rightarrow g+g$ remains the same of Eq.\eqref{def:P(g,g)}, since gluons are massless in any case.

Unlike the collinear limit, the soft one leads to a real singularity in the tree--level matrix element squared, whatever value the emitter mass has.
The soft emission can be parametrized as in the massless case (see Eq.\eqref{par:soft}).
The $(m+1)$-parton matrix element squared then behaves as follows
\begin{multline}
\label{eqn:soft_beh_mass}
_{m+1}\langle 1, \dots, m+1|1,\dots,m+1 \rangle_{m+1} \rightarrow \\
- \frac{1}{\lambda^2} 4 \pi \alpha_S \mu^{2 \varepsilon}  \sum_{a \neq b} \left[ \frac{p'_a \cdot p'_b}{(p'_a \cdot p) (p'_b \cdot p)} - \frac{m_a^2}{(p'_a \cdot p)^2}  \right] \,_m\langle 1, \dots, m + 1| \mathbf{T}_a \mathbf{T}_b |1, \dots, m + 1 \rangle_m
\end{multline}
where the $m$-parton matrix element on the right-hand side is obtained, as in the massless case, by removing the soft gluon $p'_c$ from the $(m+1)$-parton matrix element.
The difference between Eq.\eqref{eqn:soft_beh_mass} and its massless analogue (Eq.\eqref{eqn:soft_beh}) lies in the presence of terms proportional to the mass of the particles that radiate the soft gluon.

\subsection{Dual Subtractions}
\label{sec:dual_sub}

Selecting a pair of hard partons $(i,j)$, we can apply LTD to extract the associated IR divergences by cutting the $q_1$ and $q_2$ propagators in Fig.\eqref{fig:vertex_ij1}.
As anticipated in section~\ref{sec:maps}, in place of the dual contributions with $q_2$ set on-shell, we can exchange $p_i \leftrightarrow p_j$ (see Fig.\eqref{fig:vertex_ij2}) and describe the rest of the divergences again by cutting $q_1$.
Nonetheless, for completeness in the following we will present dual subtraction formulas including also the cut over $q_2$.

\subsubsection*{\textit{Contributions of the cut over $q_1$}}

Let us start by considering the case in which a gluon with momentum $p'_c$ is radiated collinear to a quark of momentum $p'_a$ and absorbed by a spectator of momentum $p'_b$.
The corresponding dual subtraction term is constructed by summing the dual representation of two contributions:
\begin{itemize}
\item the triangle involving the emitter and the spectator in the final state, as well as the bubble depending on the emitter momentum, that we collectively call $V_{ac,b}^{(1)}$;
\item the wave--function renormalization of the emitter, that we denote by $G_{ac,b}^{(1)}$.
\end{itemize}
Let $p_i$, $p_j$ and $q_1$ be the virtual sector momenta of the emitter, the spectator and the virtual radiation, respectively, which are matched to the real sector momenta by the mapping in Eq.\eqref{map:map1inv}.
The dual counterterm is then given by
\begin{equation}
\label{def:fqg}
\sigma_{qg,b}^{D\, (1)} \equiv 8\pi\alpha_S \frac{\mathcal{N}_{in}}{S_{\{m\}}} \int d \Phi_m \,_m\langle 1, \dots, m| \mathbf{T}_{ac} \mathbf{T}_b |1, \dots, m \rangle_m  \left[  V^{(1)}_{qg,b} (p_i, p_j) + G^{(1)}_{qg,b} (p_i,p_j) \right] 
\end{equation}
where $\mathcal{N}_{in}$ includes all the non-QCD factors, $S_{\{m\}}$ is the Bose symmetry factor for identical partons in the final state, $|1, \dots, m \rangle_m$ is the $m$-parton Born amplitude and $\mathbf{T}_{ac} \mathbf{T}_b$ is a colour-charge operator acting on the colour indices of the partons $i$ and $j$, respectively.
We have used the labels $ac$ and $b$ for these operators because the flavours of the partons $i$ and $j$ are the same of the partons $ac$ and $b$ of the real sector, respectively.
The functions $V^{(1)}_{qg,b}$ and $G^{(1)}_{qg,b}$ in Eq.\eqref{def:fqg} are defined by
\begin{align*}
\label{def:count_qg_comp}
V_{qg,b}^{(1)} &\equiv \int_{q_1} \tilde{\delta} (q_1) \,\mathcal{R}_1 \left[- \frac{2 s_{ij}}{(- 2 q_1 \cdot p_i)(2 q_1 \cdot p_j)} +  \frac{2}{(-2 q_1 \cdot p_i)}  \right] \\
G_{qg,b}^{(1)} &\equiv - \frac{(1-\varepsilon)}{s_{ij}} \int_{q_1} \tilde{\delta} (q_1) \,\mathcal{R}_1  \frac{2 q_1 \cdot p_j}{(-2 q_1 \cdot p_i)}
\numberthis
\end{align*}
where $V_{qg,b}^{(1)}$ groups together the dual contributions from the reduction of the virtual amplitude.
In particular, in $V_{qg,b}^{(1)}$ we can identify the dual representation of the following scalar integrals
\begin{subequations}
\label{scheme1}
\begin{align}
C_0(p_i,p_j)  =& - \int_{q_1} \frac{\tilde{\delta}(q_1)}{(-2 q_1 \cdot p_i) (2 q_1 \cdot p_j) }
- \int_{q_2}  \frac{\tilde{\delta}(q_2)}{(- 2 q_2 \cdot p_j) (- 2 q_2 \cdot p_i - 2 q_2 \cdot p_j+ s_{ij} + i 0)} \nonumber \\
&- \int_{q_3}  \frac{\tilde{\delta}(q_3)}{(2 q_3 \cdot p_i) (2 q_3 \cdot p_i+ 2 q_3 \cdot p_j + s_{ij})}
\\
B_0(p_i) = &- \int_{q_1} 
\frac{\tilde{\delta}(q_1)}{(-2 q_1 \cdot p_i)} 
- \int_{q_3} \frac{\tilde{\delta}(q_3)}{(2 q_3 \cdot p_i)}
\end{align}
\end{subequations}
and we have extracted only terms where the internal momentum $q_1$ is set on-shell in the general formula for LTD in  Eq.\eqref{th:LTD_theorem}.
Moreover, we have not included the $B_0 (p_j)$ bubble integral neither the wave--function renormalization of the spectator parton because none of them 
develop a divergence cutting the $q_1$ internal propagator and integrating over the phase-space region selected by $\mathcal{R}_1$.

As stated above, $G_{qg,b}^{(1)}$ comes from the quark wave--function renormalization~\cite{Sborlini:2016gbr, Sborlini:2016hat}
\begin{equation}
\label{eqn:quark_ren}
\Delta Z_{\text{quark}} (p_i) = - g_s^2 C_F  \int_{q} G_F(q_1) G_F(q_3) \left[ 2(1 - \varepsilon) \frac{q_1 \cdot p}{p_i \cdot p} + 4 M^2 \left( 1 - \frac{q_1 \cdot p}{p_i \cdot p} \right) G_F(q_3) \right]
\end{equation}
with $M$ being the on-shell fermion mass that we assume to be zero for the time being, postponing to Section~\ref{sec:masses} the case of radiation off massive fermions.
Note that the integrand of Eq.\eqref{eqn:quark_ren} depends on an auxiliary momentum $p$, but the integral does not.
When contracting $\Delta Z_{\text{quark}}$ with the Born matrix elements, a dependence on the spectator colour-charge operator $\mathbf{T}_{\text{spec}} \equiv \mathbf{T}_b$ can be introduced by expressing the Casimir $C_F = - \mathbf{T}_{\text{emit}}^2 =- \mathbf{T}_{ac}^2$ according to the colour conservation relation
\begin{equation}
\label{eqn:colour_conserv}
\mathbf{T}_{\text{emit}}^2 = -\sum_{\substack{\text{spec},\\ \text{spec} \, \neq \, \text{emit}}} \mathbf{T}_{\text{emit}} \mathbf{T}_{\text{spec}} \,\,.
\end{equation}
Then, in each term of the sum over the different spectators we can choose the auxiliary momentum $p$ to coincide with the spectator momentum $p_j$.
In this way, the wave--function renormalization counterterm is decomposed as
\begin{multline}
\label{eqn:quark_ren2}
\,_m\langle 1, \dots, m| \frac{ \Delta Z_{\text{quark}}(p_i)}{2} |1, \dots, m \rangle_m  = \\ 
\frac{1}{2} g_s^2 \!\!\!\!\!\! \sum_{\substack{\text{spec},\\ \text{spec} \, \neq \, \text{emit}}} \!\!\!\!\!\! \,_m\langle 1, \dots, m| \mathbf{T}_{ac} \mathbf{T}_b |1, \dots, m \rangle_m \int_{q} G_F(q_1) G_F(q_3) \left[ 2(1 - \varepsilon) \frac{q_1 \cdot p_j}{p_i \cdot p_j} \right]\,\,.
\end{multline}
By applying LTD, selecting the dual contribution with $q_1$ on-shell and restricting the integration domain to the $\mathcal{R}_1$ region, the definition for the contribution $G^{(1)}$ in Eqs.\eqref{def:fqg} and \eqref{def:count_qg_comp} follows.
Note that, the presence of the spectator momentum and the restriction of the integration domain to $\mathcal{R}_1$, introduces a dependence on the $s_{ij}$ invariant.
However, such a dependence does not affect the singular behaviour and only reflects in the finite part of the integral.

In case a gluon is radiated collinear to another gluon, the corresponding dual counterterm $\sigma^{D\,(1)}_{gg,b}$ is given by
\begin{multline}
\label{def:fgg}
\sigma^{D\,(1)}_{gg,b} \equiv 8\pi\alpha_S \frac{\mathcal{N}_{in}}{S_{\{m\}}} \int \! d \Phi_m \,_m\langle 1, \dots, m| \mathbf{T}_{ac} \mathbf{T}_b  \left[  V^{(1) \mu\nu}_{gg,b} (p_i, p_j) + G^{(1) \mu\nu}_{gg,b} (p_i,p_j) \right]  |1, \dots, m \rangle_m \\
+ (p'_a \leftrightarrow p'_c)
\end{multline}
where the symmetrization $p'_a \leftrightarrow p'_c$ has to be performed in the right-hand side of the momentum mapping in Eq.\eqref{map:map1inv} and gives rise to a distinct counter event.
Once again, in Eq.\eqref{def:fgg} we have extracted only the dual contributions with $q_1$ on-shell, obtaining
\begin{subequations}
\label{def:count_gg_comp}
\begin{align}
\label{def:count_gg_comp_V}
V^{(1) \mu\nu}_{gg,b} \!\! &\equiv\! - \!\!\int_{q_1} \tilde{\delta} (q_1) \mathcal{R}_1 \left[- \frac{2 s_{ij}}{(- 2 q_1 \cdot p_i) (2 q_1 \cdot p_j)} + \frac{1}{(-2 q_1 \cdot p_i)} \right] g^{\mu \nu} \\
\label{def:count_gg_comp_G}
G^{(1) \mu\nu}_{gg,b} \!\! &\equiv\! - \!\!\int_{q_1} \!\!  \frac{\tilde{\delta} (q_1) \mathcal{R}_1}{(-2 q_1 \cdot p_i)}  \!\left[ g^{\mu \nu}\! + \frac{d-2}{(-2 q_1 \cdot p_i)}\!\! \left( \frac{q_1 \cdot p_j}{p_i \cdot p_j} -1 \right) \!\!
\left( q_1^{\mu} - \frac{q_1 \cdot p_i}{p_i \cdot p_j} p_j^{\mu} \right)\!\! \left( q_1^{\nu} - \frac{q_1 \cdot p_i}{p_i \cdot p_j} p_j^{\nu} \right) \! \right] \,\,.
\end{align}
\end{subequations}
In Eq.\eqref{def:fgg}, $\mu$ and $\nu$ correspond to the spin polarization indices of the gluon $i$ into the bra $\,_m\langle 1, \dots, m|$ and the ket $ |1, \dots, m \rangle_m$, respectively.
The presence of free Lorentz indices into $\,_m\langle 1, \dots, m|$ and $ |1, \dots, m \rangle_m$, which have to be contracted with $V^{(1) \mu\nu}_{gg,q}$ and $G^{(1) \mu\nu}_{gg,q}$, impedes the factorization of the Born amplitude squared.
Nonetheless, this is a good feature for the dual counterterm, since Eq.\eqref{def:P(g,g)} shows us that the same kind of Lorentz structure appears in the $(m+1)$-parton matrix element squared when we approach the limit of a gluon being emitted collinear to another gluon.
Note however that, in the Feynman gauge, all the terms in Eqs.\eqref{def:count_gg_comp} proportional to $-g^{\mu\nu}$ can be easily contracted by
\begin{equation}
\label{eqn:factorization}
\,_m\langle 1, \dots, m| (-g^{\mu\nu}) \mathbf{T}_{ac}\mathbf{T}_b|1, \dots, m \rangle_m = \,_m\langle 1, \dots, m| \mathbf{T}_{ac}\mathbf{T}_b|1, \dots, m \rangle_m \,\,.
\end{equation}
It is implicit that, on the right-hand side of this last equation, there are no free Lorentz indices in the matrix elements, the only free indices being the colour ones (which are then contracted with the ones of the colour-charge operators).

Looking at Eq.\eqref{def:fgg} in the framework of LTD, we have that $V^{(1) \mu\nu}_{gg,b}$ results from the same scalar integrals of Eqs.\eqref{scheme1}, once we have embedded $-g^{\mu\nu}$ into the Born matrix elements as in Eq.\eqref{eqn:factorization}.
Note that the coefficient of the bubble contribution in Eqs.\eqref{def:count_gg_comp} is decreased by a factor $2$ with respect to the case of a gluon emitted from a quark, as a result of the virtual amplitude reduction.
The counterterm $G^{(1) \mu\nu}_{gg,b}$, on the other hand, is obtained by application of LTD to an integrand representation of the gluon and ghost contributions to the gluon wave--function renormalization counterterm $G^{\mu\nu}_{gg,b}$
\begin{equation}
G^{\mu\nu}_{gg,b} \! \equiv\!  \!\!\int_{q_1} \!\!\! G_F(q_1) G_F(q_3)  \!\left[ g^{\mu \nu}\! + (d-2) G_F(q_3)\!\! \left( \frac{q_1 \cdot p_j}{p_i \cdot p_j} -1 \! \right) \!\!
\left( \! q_1^{\mu} - \frac{q_1 \cdot p_i}{p_i \cdot p_j} p_j^{\mu} \! \right)\!\! \left(\! q_1^{\nu} - \frac{q_1 \cdot p_i}{p_i \cdot p_j} p_j^{\nu} \! \right) \! \right].
\end{equation}
The above expression is also justified by both its $\varepsilon$-poles structure and its tensorial properties.
Given that the integrand representation $G^{\mu\nu}_{gg,b}$ of the gluon wave-function renormalization is scaleless, its unconstrained integration over the loop momentum vanishes.
The integration of $G^{(1) \mu\nu}_{gg,b}$
(performed in Section~\ref{sec:int_dual_sub}, Eq.\eqref{res:count_int_gg_G}) exhibits a single IR pole that coincides, up to a minus sign, with the correct UV pole for the gluon wave--function renormalization.
As for the tensorial structure of the counterterm, note that, once defined
\begin{equation}
N^{(1) \mu\nu} = \left( q_1^{\mu} - \frac{q_1 \cdot p_i}{p_i \cdot p_j} p_j^{\mu} \right) \left( q_1^{\nu} - \frac{q_1 \cdot p_i}{p_i \cdot p_j} p_j^{\nu} \right) \equiv u^\mu u^\nu
\end{equation}
we have $N^{(1) \mu\nu} {p_1}_\mu = N^{(1) \mu\nu} {p_1}_\nu  = 0$, so that 
the contraction between the integrand on the right-hand side of Eq.\eqref{def:count_gg_comp_G} and the tensor ${p_1}_\mu {p_1}_\nu$ is equal to zero.
Moreover, it is interesting to compare the tensor $N^{(1) \mu\nu}$ with the one used in~\cite{Catani:1996vz} for the dipole associated with the emission of a gluon from another gluon, that we report here for convenience
\begin{equation}
M^{\mu\nu} \equiv ( \tilde{z}_a {p'}_a^\mu - \tilde{z}_c {p'}_c^\mu ) ( \tilde{z}_a {p'}_a^\nu - \tilde{z}_c {p'}_c^\nu ) \equiv w^\mu w^\nu\,, \qquad \tilde{z}_a \equiv \frac{p_j \cdot p'_a}{p_i \cdot p_j}\,, \quad  \tilde{z}_c \equiv 1 - \tilde{z}_a \,\,.
\end{equation}
In fact, by using the mapping in Eq.\eqref{map:map1} to express the vector $w^\mu$ in terms of the virtual sector variables, we have
\begin{align*}
\label{eqn:tensor_comparison}
w^\mu &= \left(1-\frac{q \cdot p_j}{p_i \cdot p_j} \right) (p_i^\mu - q^\mu + \alpha^{(1)}_{ij} p_j^\mu) - \frac{q \cdot p_j}{p_i \cdot p_j} q^\mu \\
&= - q^\mu + \alpha^{(1)}_{ij} \left(1-\frac{q \cdot p_j}{p_i \cdot p_j} \right) p_j^\mu + \left(1-\frac{q \cdot p_j}{p_i \cdot p_j} \right) p_i^\mu \\
&= - q^\mu + \frac{q \cdot p_i}{p_i \cdot p_j} p_j^\mu + \left(1-\frac{q \cdot p_j}{p_i \cdot p_j} \right) p_i^\mu \\
&= -u^\mu + \left(1-\frac{q \cdot p_j}{p_i \cdot p_j} \right) p_i^\mu \numberthis \,\,.
\end{align*}
Since, by gauge invariance, we have
\begin{equation}
\label{eqn:gauge}
p_i^\mu |1, \dots, m \rangle_m = 0
\end{equation}
Eq.\eqref{eqn:tensor_comparison} tells us that the two tensors $N^{(1) \mu\nu}$ and $M^{ \mu\nu}$ differ only by gauge terms.

The dual subtraction for the remaining case of a collinear quark-antiquark pair includes the quark contribution to the gluon wave--function renormalization, and takes no contribution from the loop correction. We have
\begin{equation}
\label{def:count_qq}
\sigma^{D\,(1)}_{q\bar{q},b} \equiv 8\pi\alpha_S \frac{\mathcal{N}_{in}}{S_{\{m\}}} \int \! d \Phi_m \,_m\langle 1, \dots, m| \mathbf{T}_{ac} \mathbf{T}_b \, G^{(1) \mu\nu}_{q \bar{q},b} (p_i,p_j) |1, \dots, m \rangle_m
\end{equation}
with
\begin{equation}
\label{def:fgq}
G^{(1) \mu\nu}_{q\bar{q},b} \!\! \equiv\! \!\!\int_{q_1} \!\!  \frac{\tilde{\delta} (q_1) \mathcal{R}_1 T_R N_f}{C_A (-2 q_1 \cdot p_i)}  \!\left[ g^{\mu \nu}\! \!+\! \frac{4}{(-2 q_1 \cdot p_i)}\!\! \left( \frac{q_1 \cdot p_j}{p_i \cdot p_j} -1 \right) \!\!
\left( q_1^{\mu} - \frac{q_1 \cdot p_i}{p_i \cdot p_j} p_j^{\mu} \right)\!\! \left( q_1^{\nu} - \frac{q_1 \cdot p_i}{p_i \cdot p_j} p_j^{\nu} \right) \! \right].
\end{equation}
The definition of $G^{(1) \mu\nu}_{q\bar{q},b}$ is obtained through application of LTD to the expression
\begin{equation}
G^{\mu\nu}_{q\bar{q},b} \! \equiv\!-  \!\!\!\int_{q_1} \!\!\! \frac{T_R N_f G_F(q_1) G_F(q_3)
}{C_A}\!\left[ g^{\mu \nu}\! + 4 \, G_F(q_3)\!\! \left( \frac{q_1 \cdot p_j}{p_i \cdot p_j} -1 \! \right) \!\!
\left( \! q_1^{\mu} \!-\! \frac{q_1 \cdot p_i}{p_i \cdot p_j} p_j^{\mu} \! \right)\!\! \left(\! q_1^{\nu} \! -\! \frac{q_1 \cdot p_i}{p_i \cdot p_j} p_j^{\nu} \! \right) \! \right]
\end{equation}
which provides an integrand representation for the renormalization counterterm mentioned above.
In a way analogous to the gluon and ghost contribution, both the pole and the tensorial structure justify its expression.
Note that the insertion of the colour-charge operator $\mathbf{T}_{ac} \mathbf{T}_b$ has been forced, since the sum of Eq.\eqref{eqn:colour_conserv} simplify the Casimir $C_A$ at the denominator of Eq.\eqref{def:count_qq_comp}.
However, as it happens for the other renormalization counterterms, the dependence on the spectator plays no role in the singular behaviour.

For completeness, we show the integrand representation of the whole gluon wave--function renormalization counterterm, obtained by combining gluon, ghost and quark contributions
\begin{multline}
\,_m\langle 1, \dots, m|\Delta Z_{\text{gluon}} (p_i) |1, \dots, m \rangle_m  =  \,_m\langle 1, \dots, m| - 2 g_s^2 \!\!  \int_{q}\! G_F(q_1) G_F(q_3) \bigg[ (C_A - T_R N_f) g^{\mu\nu}\\
+ ((d-2) C_A - 4 T_R N_f) \, G_F(q_3)\!\! \left( \frac{q_1 \cdot p}{p_i \cdot p} -1 \! \right) \!\!
\left( \! q_1^{\mu} \!-\! \frac{q_1 \cdot p_i}{p_i \cdot p} p^{\mu} \! \right)\!\! \left(\! q_1^{\nu} \! -\! \frac{q_1 \cdot p_i}{p_i \cdot p} p^{\nu} \! \right) \! \bigg] |1, \dots, m \rangle_m 
\end{multline}
with $p$ an auxiliary momentum.

Finally, note that the roles of quarks and antiquarks in the dual counterterms defined in this Section are interchangeable: when a quark takes the place of an antiquark or vice versa, we just need to change the sign of their momenta in the wave--function renormalization counterterms.

\subsubsection*{\textit{Contributions of the cut over $q_2$}}

The dual counterterms obtained by selecting the terms where $q_2$ is set on-shell by LTD are
\begin{equation}
\label{def:count_qg2}
\sigma^{D\,(2)}_{ag,\bar{q}} \equiv 8\pi\alpha_S \frac{\mathcal{N}_{in}}{S_{\{m\}}} \int d \Phi_m \,_m\langle 1, \dots, m| \mathbf{T}_{bc} \mathbf{T}_a |1, \dots, m \rangle_m  \left[  V^{(2)}_{ag,\bar{q}} (p_i, p_j) + G^{(2)}_{ag,\bar{q}} (p_i,p_j) \right] 
\end{equation}
with
\begin{align*}
\label{def:count_qg2_comp}
V_{ag,\bar{q}}^{(2)} &\equiv \int_{q_2} \tilde{\delta} (q_2) \,\mathcal{R}_2 \left[- \text{Re} \frac{2 s_{ij}}{(- 2 q_2 \cdot p_j) (- 2 q_2 \cdot p_i - 2 q_2 \cdot p_j + s_{ij} + i 0)} - \frac{2}{(2 q_2 \cdot p_j)} \right] \\
G_{ag,\bar{q}}^{(2)} &\equiv (1-\varepsilon) \int_{q_2} \tilde{\delta} (q_2) \,\mathcal{R}_2  \frac{1}{(2 q_2 \cdot p_j)}  \left( 1 - \frac{q_2 \cdot p_i}{p_i \cdot p_j} \right) \numberthis
\end{align*}
for the emission of a gluon from an antiquark
as well as
\begin{multline}
\label{def:count_gg2}
\sigma^{D\,(2)}_{ag,g} \equiv 8\pi\alpha_S \frac{\mathcal{N}_{in}}{S_{\{m\}}} \int d \Phi_m \,_m\langle 1, \dots, m| \mathbf{T}_{bc} \mathbf{T}_a  \left[  V^{(2) \mu\nu}_{ag,g} (p_i, p_j) + G^{(2) \mu\nu}_{ag,g} (p_i,p_j) \right] \!|1, \dots, m \rangle_m \\
+ (p'_b \leftrightarrow p'_c)
\end{multline}
with
\begin{subequations}
\label{def:count_gg2_comp}
\begin{align}
V_{ag,g}^{(2) \mu\nu} &\equiv - \int_{q_2} \tilde{\delta} (q_2) \mathcal{R}_2 \left[- \text{Re} \frac{2 s_{ij}}{(- 2 q_2 \cdot p_j) (- 2 q_2 \cdot p_i - 2 q_2 \cdot p_j + s_{ij} + i 0)} - \frac{1}{(2 q_2 \cdot p_j)} \right] g^{\mu \nu} \\
\label{def:count_gg2_comp_G}
G_{ag,g}^{(2) \mu\nu} &\equiv \int_{q_2} \!\! \tilde{\delta} (q_2)  \frac{\mathcal{R}_2}{(2 q_2 \cdot p_j)} \! \left[ g^{\mu \nu} \! - \frac{d-2}{(2 q_2 \cdot p_j)} \! \left( \frac{q_2 \cdot p_i}{p_i \cdot p_j} - 1 \right)
\!\! \left( q_2^{\mu} - \frac{q_2 \cdot p_j }{p_i \cdot p_j} p_i^{\mu} \right) \!\! \left( q_2^{\nu} - \frac{q_2 \cdot p_j}{p_i \cdot p_j} p_i^{\nu} \right)
\right]
\end{align}
\end{subequations}
for the emission of a gluon from another gluon and
\begin{equation}
\label{def:count_qq2}
\sigma^{D\,(2)}_{a\bar{q},q} \equiv 8\pi\alpha_S \frac{\mathcal{N}_{in}}{S_{\{m\}}} \int \! d \Phi_m \,_m\langle 1, \dots, m| \mathbf{T}_{bc} \mathbf{T}_a \, G^{(2) \mu\nu}_{a \bar{q},q} (p_i,p_j) |1, \dots, m \rangle_m
\end{equation}
with
\begin{equation}
\label{def:count_qq2_comp}
G^{(2) \mu\nu}_{a\bar{q},q} \!\! \equiv\! \!\!\int_{q_2} \!\!  \frac{\tilde{\delta} (q_2) \mathcal{R}_2 T_R N_f}{C_A (-2 q_2 \cdot p_j)}  \!\left[ g^{\mu \nu}\! \!+\! \frac{4}{(-2 q_2 \cdot p_j)}\!\! \left( \frac{q_2 \cdot p_i}{p_i \cdot p_j} -1 \right) \!\!
\left( q_2^{\mu} - \frac{q_2 \cdot p_j}{p_i \cdot p_j} p_i^{\mu} \right)\!\! \left( q_2^{\nu} - \frac{q_2 \cdot p_j}{p_i \cdot p_j} p_i^{\nu} \right) \! \right].
\end{equation}
for a collinear quark-antiquark pair.
Note that, with the notation $\sigma^{D\,(k)}_{ac,b}$, we indicate that $q_k$ is set on-shell, $c$ is the radiated parton and parton $b$ is the spectator if $k=1$, while this  role is played by $a$ if $k=2$.
Analogous considerations hold also for $V_{ac,b}^{(k)}$, $G_{ac,b}^{(k)}$, $V_{ac,b}^{(k) \mu\nu}$ and $G_{ac,b}^{(k)\mu\nu}$.

\subsection{Singular behaviour of the dual counterterms}
\label{sec:count-limit}
In this Section we prove that the dual counterterms defined in Section~\ref{sec:dual_sub} locally cancel the corresponding singularities of the $(m+1)$-parton tree--level cross section.
We do this by applying the momentum mappings of Section~\ref{sec:maps} and analyzing the singular behaviour of the dual counterterms in the infrared and collinear limits of the $(m+1)$-parton phase space.
We will see that, in the singular regions, the local behaviour matches exactly the one of the real matrix element squared shown in Section~\ref{sec:real_beh}.

In the following, we denote by \textit{$q_1$-cut method} the strategy in which, to collect the singularities associated to the emission from the pair of hard partons $i$ and $j$, we cut only the propagator connecting them ($q_1$ in Fig.\eqref{fig:vertex_ij}).
Furthermore, we denote by \textit{$q_1$-$q_2$-cut method} the method in which both the cut contributions over $q_1$ and $q_2$ are included to collect the above mentioned singularities.
Nevertheless, we anticipate here that the \textit{$q_1$-cut method} will turn out to be more convenient.

\subsubsection*{\textit{$q_1$-cut method}}

Let us start by analyzing the region where the momentum of a gluon, say $p'_c$, becomes soft.
If we substitute the parametrization of Eq.\eqref{par:soft} into the right-hand side of Eq.\eqref{map:map1inv}, we obtain the following behaviour in the limit $\lambda \rightarrow 0$
\begin{align*}
\label{beh:soft_dots}
q_1 \cdot p_i &\overset{\lambda \rightarrow 0}{\sim} \lambda\, p'_a \cdot p + \mathcal{O}(\lambda^2) \\
q_1 \cdot p_j &\overset{\lambda \rightarrow 0}{\sim} \lambda\, p'_b \cdot p + \mathcal{O}(\lambda^2)\numberthis \,\,.
\end{align*}
Since in what follows we are going to use the phase space factorization of Eq.\eqref{eqn:phase_space}, it is useful to evaluate the behaviour of the product $J_1 \xi_1$.
This can be done by substituting the parametrization of Eq.\eqref{par:soft} into the right-hand side of Eqs.\eqref{map:inv_kin_inva1} and~\eqref{map: J1}, obtaining
\begin{equation}
\label{beh:soft_jac}
J_1 \xi_1 \overset{\lambda \rightarrow 0}{\sim} 1 + \mathcal{O}(\lambda) \,\,.
\end{equation}
Inserting Eq.\eqref{beh:soft_dots} into the expressions of the dual counterterms defined in Eqs.\eqref{def:count_qg_comp}, \eqref{def:count_gg_comp}, and~\eqref{def:count_qq_comp}
and using Eqs.\eqref{eqn:phase_space} and~\eqref{beh:soft_jac} to turn the integrals over $\{\Phi_m, q_1\}$ into integrals over $\Phi_{m+1}$, we get the following singular behaviour for the corresponding dual cross sections
\begin{subequations}
\label{eqn:soft_beh_count1}
\begin{align}
\sigma^{D\,(1)}_{qg,b} &\overset{\lambda \rightarrow 0}{\sim} \mu^{2 \varepsilon} \frac{\mathcal{N}_{in}}{S_{\{m\}}} \frac{8\pi\alpha_S}{\lambda^2} \int d \Phi_{m+1} \,_m\langle 1, \dots, m| \mathbf{T}_a \mathbf{T}_b |1, \dots, m \rangle_m  \frac{p'_a \cdot p'_b}{(p'_a \cdot p) (p'_b \cdot p)} + \mathcal{O}(\lambda^{-1}) \\
\sigma^{D\,(1)}_{gg,b}  &\overset{\lambda \rightarrow 0}{\sim} \mu^{2 \varepsilon} \frac{\mathcal{N}_{in}}{S_{\{m\}}} \frac{8\pi\alpha_S}{\lambda^2} \int d \Phi_{m+1} \,_m\langle 1, \dots, m| \mathbf{T}_a \mathbf{T}_b |1, \dots, m \rangle_m  \frac{p'_a \cdot p'_b}{(p'_a \cdot p) (p'_b \cdot p)} + \mathcal{O}(\lambda^{-1})
\end{align}
\end{subequations}
while $\sigma^{D\,(1)}_{q\bar{q},b} \sim \mathcal{O}(\lambda^{-1})$.
Note that we have used $\mathbf{T}_{ac}=\mathbf{T}_a$ because $c$ is a gluon.
In order to obtain the total dual cross section for the soft emission of the gluon $c$, we need to perform a sum over all the possible emitter-spectator pairs and multiply times a factor $1/(m_g+1)$ that turns $S_{\{m\}}$ into $S_{\{m+1\}}$ (see Appendix~\ref{sec:counting}).
The result is the following
\begin{multline}
\label{res:dual_soft_beh}
\frac{1}{m_g+1 }\sum_{a\neq b} \sigma^{D\,(1)}_{ac,b}  \overset{\lambda \rightarrow 0}{\sim} \\
\frac{ \mathcal{N}_{in}}{S_{\{m+1\}}} \frac{4\pi\alpha_S \mu^{2 \varepsilon}}{\lambda^2} \int d \Phi_{m+1} \sum_{a,b} \frac{p'_a \cdot p'_b}{(p'_a \cdot p) (p'_b \cdot p)} \,_m\langle 1, \dots, m| \mathbf{T}_{ac} \mathbf{T}_b |1, \dots, m \rangle_m   + \mathcal{O}(\lambda^{-1}) \,\,.
\end{multline}
Note that, in the limit $\lambda \rightarrow 0$, the matrix element $|1, \dots, m \rangle_m$ can be considered as obtained from the $(m+1)$-parton one by removing $p'_c$, since in this limit we have $(p_i,p_j) \rightarrow (p'_a, p'_b)$.
Eq.\eqref{res:dual_soft_beh} exhibits the same soft behaviour of the $(m+1)$-parton matrix element squared in Eq.\eqref{eqn:soft_beh}.
Therefore, the local cancellation of soft singularities between the dual and the real cross section has been proved.\\
\indent
Let us now move to the collinear limit.
Consider the case of a parton with momentum $p'_c$ emitted collinear to a parton of momentum $p'_a$ and absorbed by a spectator of momentum $p'_b$.
The collinear limit is parametrized in Eqs.\eqref{par:coll} which, once plugged into the right-hand side of Eq.\eqref{map:map1inv}, leads us to
\begin{subequations}
\label{beh:coll_dots}
\begin{align*}
\label{beh:coll_dots1}
q_1 \cdot p_i &\overset{k_\perp \rightarrow 0}{\sim} z p'_a \cdot p'_c \, , & q_1 \cdot p_j &\overset{k_\perp \rightarrow 0}{\sim} (1-z) p \cdot p'_b \numberthis \\
\end{align*}
and
\begin{align*}
\label{beh:coll_dots2}
\,_m\langle 1, \dots, m|  q^\mu_1 q^\nu_1\, |1, \dots, m \rangle_m  &\overset{k_\perp \rightarrow 0}{\sim} k_\perp^\mu k_\perp^\nu \, , & \,_m\langle 1, \dots, m| \,q^\mu_1 p^\nu_j \,|1, \dots, m \rangle_m  &\overset{k_\perp \rightarrow 0}{\sim} k_\perp^\mu {p'_b} ^\nu \\
\,_m\langle 1, \dots, m| \,p^\mu_j q^\nu_1 \, |1, \dots, m \rangle_m &\overset{k_\perp \rightarrow 0}{\sim} {p'_b} ^\mu k_\perp^\nu  \, , &  \,_m\langle 1, \dots, m|\, p_j^\mu p_j^\nu \,|1, \dots, m \rangle_m &\overset{k_\perp \rightarrow 0}{\sim} {p'_b}^\mu {p'_b}^\nu \numberthis \,\,.
\end{align*}
\end{subequations}
In Eq.\eqref{beh:coll_dots2}, we have used the fact that, in the matrix elements, $p_i \rightarrow p$ and, consequently, $\,_m\langle 1, \dots, m| \, p^\mu \rightarrow 0$ as well as $p^\nu \,|1,\dots, m \rangle_m \rightarrow 0$ because of gauge invariance.
Next, we need the behaviour of the product $J_1 \xi_1$ in the collinear limit.
This can be obtained by inserting the collinear limit parametrization \eqref{par:coll} into the right-hand sides of Eqs.\eqref{map:inv_kin_inva1} and \eqref{map: J1}.
We have
\begin{equation}
\label{beh:coll_jac}
J_1 \xi_1 \overset{k_\perp \rightarrow 0}{\sim} z \,\,.
\end{equation}
Substituting the relations of Eq.\eqref{beh:coll_dots} into the dual counterterms of Eqs.\eqref{def:count_qg_comp} \eqref{def:count_gg_comp}, and~\eqref{def:count_qq_comp}
and using Eqs.\eqref{eqn:phase_space} and \eqref{beh:coll_jac} to turn the integrals over $\{\Phi_m, q_1\}$ into integrals over $\Phi_{m+1}$, we get
\begin{align}
\label{beh:coll_countqg1}
\sigma^{D\,(1)}_{qg,b} &\overset{k_\perp \rightarrow 0}{\sim}  \frac{\mathcal{N}_{in}}{S_{\{m\}}} 4\pi\alpha_S \mu^{2 \varepsilon} \int d \Phi_{m+1} \frac{1}{p'_a \cdot p'_c} \nonumber \\
&\hspace{4cm} \times \,_m\langle 1, \dots, m| \mathbf{T}_{ac} \mathbf{T}_b |1, \dots, m \rangle_m \left[ \frac{1+z^2}{1-z} - \varepsilon (1-z)\right] \\
\sigma^{D\,(1)}_{gg,b} &\overset{k_\perp \rightarrow 0}{\sim} \frac{\mathcal{N}_{in}}{S_{\{m\}}} 4\pi\alpha_S \mu^{2 \varepsilon} \int d \Phi_{m+1} \frac{1}{p'_a \cdot p'_c} \nonumber \\
\label{beh:coll_countgg1}
& \times \, _m\langle 1, \dots, m| \mathbf{T}_{ac} \mathbf{T}_b \!  \left[ -2 \left( \frac{z}{1-z} + \frac{1-z}{z}\right)\! g^{\mu \nu}\!\! -4 (1-\varepsilon) z (1-z) \frac{k_\perp^\mu k_\perp^\nu}{k_\perp^2} \right] \!|1, \dots, m \rangle_m \\
\sigma^{D\,(1)}_{q\bar{q},b} &\overset{k_\perp \rightarrow 0}{\sim} \frac{\mathcal{N}_{in}}{S_{\{m\}}} 4\pi\alpha_S \mu^{2 \varepsilon} \int d \Phi_{m+1} \frac{1}{p'_a \cdot p'_c} \nonumber \\
\label{beh:coll_countqq1}
&\hspace{2cm} \times \frac{T_R N_f}{C_A} \, _m\langle 1, \dots, m| \mathbf{T}_{ac} \mathbf{T}_b \!  \left[ g^{\mu \nu}\!\! - 4 z (1-z) \frac{k_\perp^\mu k_\perp^\nu}{k_\perp^2} \right] \!|1, \dots, m \rangle_m \,\,.
\end{align}
The behaviour of the total dual cross section for a given splitting process is obtained by performing a sum over all the possible spectators, multiplying $\sigma^{D\,(1)}_{qg,b}$ and $\sigma^{D\,(1)}_{gg,b}$ times a factor $1/(m_g+1)$ and $\sigma^{D\,(1)}_{q\bar{q},b}$ times a factor $m_g/(m_f+1)(\bar{m}_f+1)/N_f$ as explained in Appendix~\ref{sec:counting}.
Since the only dependence of the integrands of
Eqs.\eqref{beh:coll_countqg1} and \eqref{beh:coll_countgg1}
on the spectator lies in the colour-charge operator $\mathbf{T}_{\text{spec}} \equiv \mathbf{T}_b$, we can easily perform the sum over the spectators by
using the colour conservation relation
\begin{equation}
\label{eqn:colour_cons}
\sum_{\substack{\text{spec},\\ \text{spec} \, \neq \, \text{emit}}} \mathbf{T}_{\text{emit}} \mathbf{T}_{\text{spec}} = - \mathbf{T}_{\text{emit}}^2 \,\,.
\end{equation}
In this way, using $\mathbf{T}_{q}^2=C_F$ and $\mathbf{T}_{g}^2=C_A$,
we get
\begin{subequations}
\label{res:dual_coll1}
\begin{align*}
\label{res:dual_soft_coll1}
&\frac{1}{m_g+1 } \sum_{b \,  \in \,  \{ \text{spectators} \} } \sigma^{D\,(1)}_{qg,b} \overset{k_\perp \rightarrow 0}{\sim} \\
&\hspace{1cm}- \frac{\mathcal{N}_{in}}{S_{\{m+1\}}} 4\pi\alpha_S \mu^{2 \varepsilon} \int d \Phi_{m+1} \, \frac{1}{p'_a \cdot p'_c}  \,\, _m \langle 1, \dots, m + 1| \hat{P}_{q, q} (z, k_{\bot}, \varepsilon) |1, \dots, m + 1 \rangle_m \numberthis \\
\label{res:dual_soft_coll2}
&\frac{1}{m_g+1 }  \sum_{b \,  \in \,  \{ \text{spectators} \} } \sigma^{D\,(1)}_{gg,b}  \overset{k_\perp \rightarrow 0}{\sim} \\
&\hspace{1cm}- \frac{\mathcal{N}_{in}}{S_{\{m+1\}}} 4 \pi \alpha_S \mu^{2 \varepsilon} \int d \Phi_{m+1}\, \frac{1}{p'_a \cdot p'_c} \,\, _m \langle 1, \dots, m + 1| \hat{P}_{g, g} (z, k_{\bot}, \varepsilon) |1, \dots, m + 1 \rangle_m \numberthis \\
&\frac{m_g}{(m_f+1)(\bar{m}_f+1)} \frac{1}{N_f} \sum_{b \,  \in \,  \{ \text{spectators} \} } \sigma^{D\,(1)}_{q\bar{q},b}  \overset{k_\perp \rightarrow 0}{\sim} \\
&\hspace{1cm}- \frac{\mathcal{N}_{in}}{S_{\{m+1\}}} 4 \pi \alpha_S \mu^{2 \varepsilon} \int d \Phi_{m+1}\, \frac{1}{p'_a \cdot p'_c} \,\, _m \langle 1, \dots, m + 1| \hat{P}_{g, q} (z, k_{\bot}, \varepsilon) |1, \dots, m + 1 \rangle_m \numberthis \,\,.
\end{align*}
\end{subequations}
By comparison with Eq.\eqref{eqn:coll_beh}, we see that the integrands of
Eqs.\eqref{res:dual_coll1} share the same local behaviour of the $(m+1)$-parton amplitude squared.
Therefore, we have proved also the local cancellation of collinear singularities between the dual and the real cross section.

\subsubsection*{\textit{$q_1$-$q_2$-cut method}}

In case dual counterterms with $q_2$ on-shell are considered, to study their soft behaviour we need to substitute the parametrization of Eq.\eqref{par:soft} into the mapping of Eq.\eqref{map:map2inv} instead of Eq.\eqref{map:map1inv}.
We have
\begin{align*}
\label{beh:soft_dots2}
q_2 \cdot p_i &\overset{\lambda \rightarrow 0}{\sim} p'_a \cdot p'_b + \lambda\, p'_b \cdot p + \mathcal{O}(\lambda^2) \\
q_2 \cdot p_j &\overset{\lambda \rightarrow 0}{\sim}  \lambda^2 \, \frac{(p'_a \cdot p)(p'_b \cdot p)}{p'_a \cdot p'_b} + \mathcal{O}(\lambda^3) \numberthis
\end{align*}
while, using Eqs.\eqref{map:inv_kin_inva2} and \eqref{map:J2}, we get
\begin{equation}
\label{beh:soft_jac2}
J_2 \xi_2  \overset{\lambda \rightarrow 0}{\sim} \lambda \frac{p'_a \cdot p}{p'_a \cdot p'_b} + \mathcal{O}(\lambda^2) \,\,.
\end{equation}
If we substitute these relations in the corresponding dual cross sections $\sigma^{D\,(2)}_{ac,b}$ of Eqs.\eqref{def:count_qg2} and \eqref{def:count_gg2}, and use Eqs.\eqref{eqn:phase_space2} and \eqref{beh:soft_jac2} to turn the integrals over $\{\Phi_m, q_2\}$ into integrals over $\Phi_{m+1}$, we get
\begin{subequations}
\label{eqn:soft_beh_count2}
\begin{align}
\sigma^{D\,(2)}_{ag,\bar{q}} &\overset{\lambda \rightarrow 0}{\sim} \mu^{2 \varepsilon} \frac{\mathcal{N}_{in}}{S_{\{m\}}} \frac{8\pi\alpha_S}{\lambda^2} \int d \Phi_{m+1} \,_m\langle 1, \dots, m| \mathbf{T}_b \mathbf{T}_a |1, \dots, m \rangle_m  \frac{p'_a \cdot p'_b}{(p'_a \cdot p) (p'_b \cdot p)} \\
\sigma^{D\,(2)}_{ag,g}  &\overset{\lambda \rightarrow 0}{\sim} \mu^{2 \varepsilon} \frac{\mathcal{N}_{in}}{S_{\{m\}}} \frac{8\pi\alpha_S}{\lambda^2} \int d \Phi_{m+1} \,_m\langle 1, \dots, m| \mathbf{T}_b \mathbf{T}_a |1, \dots, m \rangle_m  \frac{p'_a \cdot p'_b}{(p'_a \cdot p) (p'_b \cdot p)}
\end{align}
\end{subequations}
while $\sigma^{D\,(2)}_{a \bar{q},q} \sim \mathcal{O}(\lambda^{-1})$.
We have used $\mathbf{T}_{bc}=\mathbf{T}_b$ because $c$ is a gluon.
The combination of soft contributions given by cutting over $q_1$, Eq.\eqref{eqn:soft_beh_count1}, and $q_2$, Eqs.\eqref{eqn:soft_beh_count2}
reads
\begin{multline}
\label{res:dual_soft_beh2}
\frac{1}{m_g+1 }\sum_{a<b} \left( \tilde{\sigma}^{D\,(1)}_{ac,b} + \tilde{\sigma}^{D\,(2)}_{ac,b} \right) \overset{\lambda \rightarrow 0}{\sim} \\
\frac{ \mathcal{N}_{in}}{S_{\{m+1\}}} \frac{4\pi\alpha_S \mu^{2 \varepsilon}}{\lambda^2} \int d \Phi_{m+1} \sum_{a,b} \frac{p'_a \cdot p'_b}{(p'_a \cdot p) (p'_b \cdot p)} \,_m\langle 1, \dots, m| \mathbf{T}_{ac} \mathbf{T}_b |1, \dots, m \rangle_m   + \mathcal{O}(\lambda^{-1})
\end{multline}
where
\begin{align*}
\label{eqn:tilde_sigma}
\tilde{\sigma}^{D\,(1)}_{ac,b} &\equiv \frac{1}{2} \left( \sigma^{D\,(1)}_{ac,b} + \sigma^{D\,(1)}_{bc,a} \right) \\
\tilde{\sigma}^{D\,(2)}_{ac,b} &\equiv \frac{1}{2} \left( \sigma^{D\,(2)}_{bc,a} + \sigma^{D\,(2)}_{ac,b} \right) \numberthis \,\,.
\end{align*}
More details on the matching among the singularities of the real matrix elements with the dual counterterms are provided in Appendix~\ref{sec:counting}.

If parton $b$ and $c$ become collinear, the singular behaviour of the dual counterterms with $q_2$ on-shell can be obtained by replacing $p'_a \rightarrow p'_b$ in the parametrization of Eqs.\eqref{par:coll} and substituting the result into the momentum mapping of Eq.\eqref{map:map2inv}.
We have
\begin{subequations}
\label{beh:coll2_dots}
\begin{align*}
\label{beh:coll2_dots1}
q_2 \cdot p_i &\overset{k_\perp \rightarrow 0}{\sim} z p \cdot p'_a  \, , & q_2 \cdot p_j &\overset{k_\perp \rightarrow 0}{\sim} (1-z) p'_b \cdot p'_c \numberthis \\
\end{align*}
and
\begin{align*}
\label{beh:coll2_dots2}
\,_m\langle 1, \dots, m|  q^\mu_2 q^\nu_2\, |1, \dots, m \rangle_m  &\overset{k_\perp \rightarrow 0}{\sim} k_\perp^\mu k_\perp^\nu \, , & \,_m\langle 1, \dots, m| \,q^\mu_2 p^\nu_i \,|1, \dots, m \rangle_m  &\overset{k_\perp \rightarrow 0}{\sim} k_\perp^\mu {p'_a} ^\nu \\
\,_m\langle 1, \dots, m| \,p^\mu_i q^\nu_2 \, |1, \dots, m \rangle_m &\overset{k_\perp \rightarrow 0}{\sim} {p'_a} ^\mu k_\perp^\nu  \, , &  \,_m\langle 1, \dots, m|\, p_i^\mu p_i^\nu \,|1, \dots, m \rangle_m &\overset{k_\perp \rightarrow 0}{\sim} {p'_a}^\mu {p'_a}^\nu \numberthis \,\,.
\end{align*}
\end{subequations}
Since $p_j \rightarrow p$,
we have used again that $\,_m\langle 1, \dots, m| \, p^\mu \rightarrow 0$ and $p^\nu \,|1,\dots, m \rangle_m \rightarrow 0$ for gauge invariance.
To obtain the collinear behaviour of the product $J_2 \xi_2$, we substitute the parametrization of Eqs.\eqref{par:coll} into Eqs.\eqref{map:inv_kin_inva2} and \eqref{map:J2}.
We have
\begin{equation}
\label{beh:coll_jac2}
J_2 \xi_2 \overset{k_\perp \rightarrow 0}{\sim} 1 - z \,\,.
\end{equation}
Proceeding in a way analogous to the case of counterterms with $q_1$ on-shell, we obtain
\begin{align}
\label{beh:coll_countqg2}
\sigma^{D\,(2)}_{ag,\bar{q}} &\overset{k_\perp \rightarrow 0}{\sim}  \frac{\mathcal{N}_{in}}{S_{\{m\}}} 4\pi\alpha_S \mu^{2 \varepsilon} \int d \Phi_{m+1} \frac{1}{p'_b \cdot p'_c}  \nonumber \\
& \hspace{4cm}\times \,_m\langle 1, \dots, m| \mathbf{T}_{bc} \mathbf{T}_a |1, \dots, m \rangle_m \left[ \frac{1+z^2}{1-z} - \varepsilon (1-z)\right] \\
\sigma^{D\,(2)}_{ag,g} &\overset{k_\perp \rightarrow 0}{\sim} \frac{\mathcal{N}_{in}}{S_{\{m\}}} 4\pi\alpha_S \mu^{2 \varepsilon} \int d \Phi_{m+1} \frac{1}{p'_b \cdot p'_c} \nonumber \\
\label{beh:coll_countgg2}
& \!\!\!\! \times \, _m\langle 1, \dots, m| \mathbf{T}_{bc} \mathbf{T}_a \!  \left[ -2 \left( \frac{z}{1-z} + \frac{1-z}{z}\right) g^{\mu \nu}\! -4 (1-\varepsilon) z (1-z) \frac{k_\perp^\mu k_\perp^\nu}{k_\perp^2} \right] \! |1, \dots, m \rangle_m \\
\sigma^{D\,(2)}_{a\bar{q},q} &\overset{k_\perp \rightarrow 0}{\sim} \frac{\mathcal{N}_{in}}{S_{\{m\}}} 4\pi\alpha_S \mu^{2 \varepsilon} \int d \Phi_{m+1} \frac{1}{p'_b \cdot p'_c} \nonumber \\
\label{beh:coll_countqq2}
&\hspace{2cm} \times \frac{T_R N_f}{C_A} \, _m\langle 1, \dots, m| \mathbf{T}_{bc} \mathbf{T}_a \!  \left[ g^{\mu \nu}\!\! - 4 z (1-z) \frac{k_\perp^\mu k_\perp^\nu}{k_\perp^2} \right] \!|1, \dots, m \rangle_m \,\,.
\end{align}
The sum of the dual counterterms with $q_1$ and $q_2$ on-shell over all possible spectators, gives
\begin{subequations}
\begin{align*}
&\frac{1}{m_g+1 }  \sum_{a \,  \in \,  \{ \text{spectators} \} } \left( \tilde{\sigma}^{D\,(1)}_{qg,a} +  \tilde{\sigma}^{D\,(2)}_{ag,q} \right) \overset{k_\perp \rightarrow 0}{\sim} \\
&\hspace{1cm}- \frac{\mathcal{N}_{in}}{S_{\{m+1\}}} 4\pi\alpha_S \mu^{2 \varepsilon} \int d \Phi_{m+1} \, \frac{1}{p'_b \cdot p'_c}  \,\, _m \langle 1, \dots, m + 1| \hat{P}_{q, q} (z, k_{\bot}, \varepsilon) |1, \dots, m + 1 \rangle_m \numberthis\\
&\frac{1}{m_g+1 } \sum_{a \,  \in \,  \{ \text{spectators} \} } \left( \tilde{\sigma}^{D\,(1)}_{gg,a} +  \tilde{\sigma}^{D\,(2)}_{ag,g} \right) \overset{k_\perp \rightarrow 0}{\sim} \\
&\hspace{1cm}- \frac{\mathcal{N}_{in}}{S_{\{m+1\}}} 4 \pi \alpha_S \mu^{2 \varepsilon} \int d \Phi_{m+1}\, \frac{1}{p'_b \cdot p'_c} \,\, _m \langle 1, \dots, m + 1| \hat{P}_{g, g} (z, k_{\bot}, \varepsilon) |1, \dots, m + 1 \rangle_m \numberthis \\
&\frac{m_g}{(m_f+1)(\bar{m}_f+1)} \frac{1}{N_f} \sum_{a \,  \in \,  \{ \text{spectators} \} }\left( \tilde{\sigma}^{D\,(1)}_{q\bar{q},a} +  \tilde{\sigma}^{D\,(2)}_{a\bar{q},q} \right)  \overset{k_\perp \rightarrow 0}{\sim} \\
&\hspace{1cm}- \frac{\mathcal{N}_{in}}{S_{\{m+1\}}} 4 \pi \alpha_S \mu^{2 \varepsilon} \int d \Phi_{m+1}\, \frac{1}{p'_b \cdot p'_c} \,\, _m \langle 1, \dots, m + 1| \hat{P}_{g, q} (z, k_{\bot}, \varepsilon) |1, \dots, m + 1 \rangle_m \numberthis \,\,.
\end{align*}
\end{subequations}
(where $\tilde{\sigma}^{D\,(k)}_{ac,b}$ are defined in Eq.\eqref{eqn:tilde_sigma}) so reproducing the correct $p'_b \parallel p'_c$ collinear limit.

\subsection{Integrated Dual Subtractions}
\label{sec:int_dual_sub}

In this section we show how to integrate the dual counterterm over the (constrained) loop momentum.
The result lives on the $m$-particle phase space and, therefore, can be integrated together with the virtual cross section.

Let us start with the case of a quark or an antiquark as the emitter.
In terms of the dimensionless variables $(\xi_1, v_1)$, the subtraction terms in Eq.\eqref{def:count_qg_comp} are given by
\begin{subequations}
\label{eqn:qqg_counterterms_dimensionless}
\begin{align}
V^{(1)}_{qg,b} &= \frac{1}{(4 \pi)^2} \int [d \xi_1 d v_1] \mathcal{R}_1 (\xi_1, v_1) \left[\frac{2}{\xi_1 v_1 (1 - v_1)} - \frac{2}{v_1}  \right] \\
G^{(1)}_{qg,b} &= \frac{1}{(4 \pi)^2} \int [d \xi_1 d v_1] \mathcal{R}_1 (\xi_1, v_1) \,\xi_1 \left( \frac{1 - v_1}{v_1} \right)
\end{align}
\end{subequations}
where the measure $[d \xi_k d v_k]$ is defined as
\begin{equation}
\label{def:measure}
[d \xi_k d v_k] \equiv \left( \frac{\mu^2}{s_{ij}} \right)^{\varepsilon} \frac{\xi_k^{- 2 \varepsilon} [v_k (1 - v_k)]^{- \varepsilon}}{\Gamma (1 - \varepsilon) (4 \pi)^{- \varepsilon}} \,\,.
\end{equation}
The integrands in Eqs. \eqref{eqn:qqg_counterterms_dimensionless} have been expressed in terms of the loop momentum variables $(\xi_1, v_1)$, defined in the center-of-mass frame of $p_i+p_j$ by Eq.\eqref{def:param_loop_momenta},
using
\begin{equation}
\label{eqn:param_loop_momenta_dot}
\begin{split}
q_k \cdot p_i  &=  \frac{s_{ij}}{2} \xi_k v_k \\
q_k \cdot p_j &=  \frac{s_{ij}}{2} \xi_k (1 - v_k)
\end{split}
\end{equation}
and (removing the subscript $k$ for easy of notation)
\begin{equation}
\label{eqn:measure}
\begin{split}
\int_{q} \tilde{\delta} (q) &= \mu^{4 - d} \int \frac{d^{d - 1} q}{(2 \pi)^{d - 1} (2 q^0)} = \frac{\mu^{4 - d}}{(2 \pi)^{d - 1}} \int d \Omega^{d - 2} \int d^{} q^0 \frac{(q^0)^{d - 2}}{2 q^0} \\
&= \left( \frac{\mu^2}{s_{ij}} \right)^{\varepsilon} \frac{s_{ij}}{\Gamma (1 - \varepsilon) (4 \pi)^{2 - \varepsilon}} \int_0^{+ \infty} d \xi \int_0^1 d v \,\xi^{- 2 \varepsilon} [v (1 - v)]^{- \varepsilon} \xi
\end{split}
\end{equation}
the latter being the on-shell loop integration measure in $d=4-2 \varepsilon$ dimensions. The integrals in Eq.\eqref{eqn:qqg_counterterms_dimensionless} can be computed analytically, leading to
\begin{subequations}
\label{eqn:qqg_counterterms_integrated}
\begin{align}
\label{res:count_int_qg_V}
V^{(1)}_{qg,b} &= \frac{(4 \pi)^{\varepsilon - 2}}{\Gamma (1 - \varepsilon)} \left( \frac{\mu^2}{s_{ij}} \right)^{\varepsilon} \left[ \left( \frac{1}{\varepsilon^2} -\frac{\pi^2}{2} \right) + \left(\frac{2}{\varepsilon^{}} +4 + 4\log (2) \right) +\mathcal{O}(\varepsilon) \right] \\
G^{(1)}_{qg,b} &= \frac{(4 \pi)^{\varepsilon - 2}}{\Gamma (1 - \varepsilon)} \left( \frac{\mu^2}{s_{ij}} \right)^{\varepsilon} \left( - \frac{1}{2 \varepsilon}-1 + \mathcal{O}(\varepsilon) \right)
\end{align}
\end{subequations}
where, in the square bracket of Eq.\eqref{res:count_int_qg_V}, we have kept in separated round brackets the contributions of the scalar triangle (former) and bubble (latter) integrals.
By substitution into Eq.\eqref{def:fqg}, we get
\begin{multline}
\label{res:count_qg1_int}
\sigma^{D\,(1)}_{qg,b} 
= \frac{\mathcal{N}_{in}}{S_{\{m\}}} \frac{\alpha_S}{2\pi} \frac{(4 \pi)^{\varepsilon}}{\Gamma (1 - \varepsilon)} \int d \Phi_m \left( \frac{\mu^2}{s_{ij}} \right)^{\varepsilon} \,_m\langle 1, \dots, m| \mathbf{T}_i \mathbf{T}_j |1, \dots, m \rangle_m \\
\times \left[ \frac{1}{\varepsilon^2} + \frac{3}{2\varepsilon} + 3 + 4 \log(2) -\frac{\pi^2}{2} + \mathcal{O}(\varepsilon) \right]
\end{multline}
which has the same poles of the terms in Eq.\eqref{eq:singular} with $\gamma_i = \gamma_q$ and $\mathbf{T}_i =\mathbf{T}_q$, thus confirming that we have extracted in the proper way the IR singular behaviour from the virtual cross section.

Now consider the case of a gluon as the emitter.
The counterterm $V^{(1) \mu\nu}_{gg,b}$ in Eq.\eqref{def:count_gg_comp_V} can be easily integrated by noting that it is very similar to the counterterm $V^{(1)}_{gg,b}$ of Eq.\eqref{def:count_qg_comp} times $-g^{\mu\nu}$.
The only difference lies in the relative coefficient of the scalar bubble with respect to the scalar triangle, which is multiplied by a factor $1/2$ in the case of a gluon as the emitter, as we have already discussed.
Therefore, using Eqs.\eqref{def:measure}, \eqref{eqn:param_loop_momenta_dot} and \eqref{eqn:measure}, we have
\begin{equation}
V^{(1) \mu\nu}_{gg,b}=- \frac{1}{(4 \pi)^2} \int [d \xi_1 d v_1] \mathcal{R}_1 \left[  \frac{2}{\xi_1 v_1 (1 - v_1)} - \frac{1}{v_1} \right] g^{\mu\nu}
\end{equation}
and the result of the integral is
\begin{equation}
\label{res:count_int_gg_V}
V^{(1) \mu\nu}_{gg,b} = - \frac{(4 \pi)^{\varepsilon - 2}}{\Gamma (1 - \varepsilon)} \left( \frac{\mu^2}{s_{ij}} \right)^{\varepsilon} \left[ \left( \frac{1}{\varepsilon^2} -\frac{\pi^2}{2} \right) + \left(\frac{1}{\varepsilon^{}} +2 + 2 \log (2) \right) +\mathcal{O}(\varepsilon) \right] g^{\mu\nu}
\end{equation}
where the remaining tensor $-g^{\mu \nu}$ can be contracted with the spin polarization indices of the parton $p_j$ in the $m$-particle matrix elements, leading to $_m\langle 1, \dots, m| \mathbf{T}_i \mathbf{T}_j |1, \dots, m \rangle_m$ as in Eq.\eqref{eqn:factorization}.

The integration of the non-trivial tensor structure in $G^{(1) \mu\nu}_{gg,b}$ requires particular attention.
Let us focus on the term proportional to $q^{\mu}_1 q^{\nu}_1$ in Eq.\eqref{def:count_gg_comp_G}
\begin{equation}
I_{00}^{\mu \nu} \equiv - \int_{q_1} \tilde{\delta} (q_1) \mathcal{R}_1 \frac{d-2}{(2 q_1 \cdot p_1)^2} \left( \frac{q_1 \cdot p_j}{p_i \cdot p_j} - 1 \right) q^{\mu}_1 q^{\nu}_1 \,\,.
\end{equation}
By Lorentz covariance, the result of the integral must be of the type
\begin{equation}
\label{integrated_tensor_n}
I_{00}^{\mu \nu} = - A_{00} g^{\mu \nu} + A_{11} p_i^\mu p_i^\nu + A_{12} p_i^\mu p_j^\nu + A_{21} p_j^\mu p_i^\nu + A_{22} p_j^\mu p_j^\nu \,\,.
\end{equation}
According to Eq.\eqref{def:fgg}, we have to contract Eq.\eqref{integrated_tensor_n} with the $m$-particle matrix elements.
Because of gauge invariance, Eq.\eqref{eqn:gauge}, the only terms in Eq.\eqref{integrated_tensor_n} that give a non-zero contribution are $- A_{00} g^{\mu \nu} + A_{22} p_j^\mu p_j^\nu$.
The coefficient $A_{00}$ can be extracted by contraction with the tensor
\begin{equation}
\label{projector1_n}
T_{\mu \nu} \equiv - \frac{1}{d-2} \left[ g_{\mu \nu} - \frac{p_{i\,\mu} p_{j\,\nu} + p_{j\,\mu} p_{i\,\nu}}{p_i \cdot p_j} \right]
\end{equation}
since $-T_{\mu\nu} A_{00} g^{\mu\nu} = A_{00}$ and $T_{\mu\nu} p_i^\mu=T_{\mu\nu} p_i^\nu = T_{\mu\nu} p_j^\mu = T_{\mu\nu} p_j^\nu = 0$.
The coefficient  $A_{22}$ can be obtained by contraction with the tensor $p_{i \mu} p_{i \nu}/ (p_i \cdot p_j)^2$.
Therefore we have
\begin{equation}
\label{tensor_n}
I_{00}^{\mu \nu} = - \int_{q_1} \tilde{\delta} (q_1) \mathcal{R}_1 \frac{d-2}{(2 q_1 \cdot p_1)^2} \left( \frac{q_1 \cdot p_j}{p_i \cdot p_j} - 1 \right)
\left(- q^{\alpha}_1 q^{\beta}_1 T_{\alpha \beta}\, g^{\mu \nu} +   \frac{(q_1 \cdot p_i)^2}{(p_i \cdot p_j)^2} p^{\mu}_j p^{\nu}_j \right) \,\,.
\end{equation}
Now consider the integrals
\begin{align*}
I_{01}^{\mu \nu} &\equiv - \int_{q_1} \tilde{\delta} (q_1) \mathcal{R}_1 \frac{d-2}{(2 q_1 \cdot p_i)^2} \left( \frac{q_1 \cdot p_j}{p_i \cdot p_j} - 1 \right) \left( - \frac{q_1 \cdot p_i}{p_i \cdot p_j} \right) q_1^\mu p_j^\nu \\
 I_{10}^{\mu \nu} &\equiv - \int_{q_1} \tilde{\delta} (q_1) \mathcal{R}_1 \frac{d-2}{(2 q_1 \cdot p_i)^2} \left( \frac{q_1 \cdot p_j}{p_i \cdot p_j} - 1 \right) \left( - \frac{q_1 \cdot p_i}{p_i \cdot p_j} \right) p_j^\mu q_1^\nu \numberthis
\end{align*}
which represent the terms in Eq.\eqref{def:count_gg_comp_G} proportional to $q_1^\mu p_j^\nu$ and $p_j^\mu q_1^\nu$, respectively.
By Lorentz covariance, these integrals must be of the type
\begin{align*}
I_{01}^{\mu \nu} &= B_1  p_i^\mu p_j^\nu + B_2 p_j^\mu p_j^\nu \\
I_{10}^{\mu \nu} &= C_1 p_j^\mu p_i^\nu + C_2 p_j^\mu p_j^\nu \numberthis  \,\,.
\end{align*}
Again, because of Eq.\eqref{eqn:gauge}, the only non-vanishing contributions are $B_2 p_j^\mu p_j^\nu$ and $C_2 p_j^\mu p_j^\nu$.
These can be extracted by contraction with $p_i^\mu/p_i \cdot p_j$ and $p_i^\nu/p_i \cdot p_j$, respectively.
Therefore, we can write
\begin{align*}
\label{tensor2_n}
I_{01}^{\mu \nu} & = \int_{q_1} \tilde{\delta} (q_1) \mathcal{R}_1 \frac{d-2}{(2 q_1 \cdot p_i)^2} \left( \frac{q_1 \cdot p_j}{p_i \cdot p_j} - 1 \right) \frac{(q_1 \cdot p_i)^2}{(p_i \cdot p_j)^2} p_j^\mu p_j^\nu \\
 I_{10}^{\mu \nu} & = \int_{q_1} \tilde{\delta} (q_1) \mathcal{R}_1 \frac{d-2}{(2 q_1 \cdot p_i)^2} \left( \frac{q_1 \cdot p_j}{p_i \cdot p_j} - 1 \right) \frac{(q_1 \cdot p_i)^2}{(p_i \cdot p_j)^2} p_j^\mu p_j^\nu \numberthis \,\,.
\end{align*}
Note, by looking at Eqs.\eqref{tensor_n} and \eqref{tensor2_n}, that the term proportional to $p^{\mu}_j p^{\nu}_j$ in $I_{00}^{\mu \nu} + I_{01}^{\mu \nu} + I_{10}^{\mu \nu}$ is opposite to the one proportional to $p^{\mu}_j p^{\nu}_j$ in Eq.\eqref{def:count_gg_comp_G}.
Therefore, we can write
\begin{equation}
G^{(1) \mu\nu}_{gg,b} =  - \int_{q_1} \tilde{\delta} (q_1) \mathcal{R}_1 \frac{1}{(-2 q_1 \cdot p_i)}  \left[ g^{\mu \nu} + \frac{d-2}{2 q_1 \cdot p_i} \left( \frac{q_1 \cdot p_j}{p_i  \cdot p_j} -1 \right) q^{\alpha}_1 q^{\beta}_1 T_{\alpha \beta}\, g^{\mu \nu} \right]
\end{equation}
which, in terms of the dimensionless variables $(\xi_1,v_1)$, becomes
\begin{equation}
G^{(1) \mu\nu}_{gg,b} = \frac{1}{(4 \pi)^2} \int [d \xi_1 d v_1] \mathcal{R}_1 \frac{1}{v_1} \left[1 + \xi_1 (1-v_1) (\xi_1 (1-v_1)-1)  \right] g^{\mu\nu} \,\,.
\end{equation}
The result of the integral is the following
\begin{equation}
\label{res:count_int_gg_G}
G^{(1) \mu\nu}_{gg,b} =- \frac{(4 \pi)^{\varepsilon - 2}}{\Gamma (1 - \varepsilon)} \left( \frac{\mu^2}{s_{ij}} \right)^{\varepsilon} \left(\frac{5}{6 \varepsilon} + \frac{19}{18} + \frac{8}{3} \log(2) + \mathcal{O}(\varepsilon) \right) g^{\mu\nu}
\end{equation}
where, again, the tensor $-g^{\mu \nu}$ can be contracted with the $m$-particle matrix elements, leading to $_m\langle 1, \dots, m| \mathbf{T}_i \mathbf{T}_j |1, \dots, m \rangle_m$.

We now have to insert Eqs.\eqref{res:count_int_gg_V} and \eqref{res:count_int_gg_G} into Eq.\eqref{def:fgg}, contracting $-g^{\mu \nu}$ as in Eq.\eqref{eqn:factorization}.
Note that we do not have to consider the symmetrization $p'_a \leftrightarrow p'_c$, since it is already taken into account in the virtual sector.
We obtain
\begin{multline}
\label{res:int_count_gg1}
\sigma^{D\,(1)}_{gg,b} = \frac{\mathcal{N}_{in}}{S_{\{m\}}} \frac{\alpha_S}{2\pi} \frac{(4 \pi)^{\varepsilon}}{\Gamma (1 - \varepsilon)} \int d \Phi_m \left( \frac{\mu^2}{s_{ij}} \right)^{\varepsilon} \,_m\langle 1, \dots, m| \mathbf{T}_i \mathbf{T}_j |1, \dots, m \rangle_m \,\,  \\
\times \left[ \frac{1}{\varepsilon^2} + \frac{11}{6 \varepsilon} + \frac{55}{18} + \frac{14}{3} \log(2) -\frac{\pi^2}{2} + \mathcal{O}(\varepsilon) \right]
\end{multline}

Following the same reasoning, the integration of the quark contribution to the gluon wave--function renormalization leads to
\begin{equation}
\label{res:count_int_qq_G}
G^{(1) \mu\nu}_{q\bar{q},b} = - \frac{T_R N_f}{C_A} \frac{(4 \pi)^{\varepsilon - 2}}{\Gamma (1 - \varepsilon)} \left( \frac{\mu^2}{s_{ij}} \right)^{\varepsilon} \left( - \frac{2}{3 \varepsilon}  + \frac{2}{9} + \frac{10}{3} \log(2) + \mathcal{O}(\varepsilon) \right) g^{\mu\nu}
\end{equation}
which, inserted into Eq.\eqref{def:count_qq}, gives
\begin{multline}
\label{res:int_count_qq1}
\sigma^{D\,(1)}_{q\bar{q},b} = \frac{\mathcal{N}_{in}}{S_{\{m\}}} \frac{\alpha_S}{2\pi} \frac{(4 \pi)^{\varepsilon}}{\Gamma (1 - \varepsilon)} \int d \Phi_m \left( \frac{\mu^2}{s_{ij}} \right)^{\varepsilon} \,_m\langle 1, \dots, m| \mathbf{T}_i \mathbf{T}_j |1, \dots, m \rangle_m \,\,  \\
\times \frac{T_R N_f}{C_A} \left[ - \frac{2}{3 \varepsilon} + \frac{2}{9} + \frac{10}{3} \log(2) + \mathcal{O}(\varepsilon) \right]
\end{multline}
The sum of the integrated dual cross sections in Eqs.\eqref{res:int_count_gg1} and~\eqref{res:int_count_qq1} has the same poles of the terms in the sum of Eq.\eqref{eq:singular} with $\gamma_i = \gamma_g$ and $\mathbf{T}_i =\mathbf{T}_g$, as we expected.

The dual counterterms with $q_2$ on-shell can be integrated in a way analogous to the case of $q_1$ on-shell.
The results are the following
\begin{align}
V^{(2)}_{ag,\bar{q}} &= \frac{(4 \pi)^{\varepsilon - 2}}{\Gamma (1 - \varepsilon)} \left( \frac{\mu^2}{s_{ij}} \right)^{\varepsilon} \left[ \left( \frac{1}{\varepsilon^2} - \frac{\pi^2}{3} \right) + \left( \frac{2}{\varepsilon^{}} + 4 + 4 \log (2) \right)+\mathcal{O}(\varepsilon) \right] \\
G^{(2)}_{ag,\bar{q}} &= \frac{(4 \pi)^{\varepsilon - 2}}{\Gamma (1 - \varepsilon)} \left( \frac{\mu^2}{s_{ij}} \right)^{\varepsilon} \left( - \frac{1}{2 \varepsilon} - 1 +\mathcal{O}(\varepsilon)\right) \\
V^{(2) \mu\nu}_{ag,g} &= - \frac{(4 \pi)^{\varepsilon - 2}}{\Gamma (1 - \varepsilon)} \left( \frac{\mu^2}{s_{12}} \right)^{\varepsilon} \left[ \left( \frac{1}{\varepsilon^2} - \frac{\pi^2}{3} \right) + \left( \frac{1}{\varepsilon^{}} + 2 + 2 \log (2) \right)+\mathcal{O}(\varepsilon) \right] g^{\mu\nu} \\
G^{(2) \mu\nu}_{ag,g} &=- \frac{(4 \pi)^{\varepsilon - 2}}{\Gamma (1 - \varepsilon)} \left( \frac{\mu^2}{s_{ij}} \right)^{\varepsilon} \left(\frac{5}{6 \varepsilon} + \frac{19}{18} + \frac{8}{3} \log(2) + \mathcal{O}(\varepsilon)\right) g^{\mu\nu} \\
G^{(2) \mu\nu}_{a\bar{q},q} &= - \frac{T_R N_f}{C_A} \frac{(4 \pi)^{\varepsilon - 2}}{\Gamma (1 - \varepsilon)} \left( \frac{\mu^2}{s_{ij}} \right)^{\varepsilon} \left( - \frac{2}{3 \varepsilon} + \frac{2}{9} + \frac{10}{3} \log(2) + \mathcal{O}(\varepsilon) \right) g^{\mu\nu}
\end{align}
leading to
\begin{align}
\sigma^{D\,(2)}_{ag,\bar{q}} &= \frac{\mathcal{N}_{in}}{S_{\{m\}}} \frac{\alpha_S}{2\pi} \frac{(4 \pi)^{\varepsilon}}{\Gamma (1 - \varepsilon)} \int d \Phi_m \left( \frac{\mu^2}{s_{ij}} \right)^{\varepsilon} \,_m\langle 1, \dots, m| \mathbf{T}_i \mathbf{T}_j |1, \dots, m \rangle_m  \nonumber \\
&\hspace{5cm} \times \left[ \frac{1}{\varepsilon^2} + \frac{3}{2\varepsilon} + 3 + 4 \log(2) -\frac{\pi^2}{3} + \mathcal{O}(\varepsilon) \right] \\
\sigma^{D\,(2)}_{ag,g} &= \frac{\mathcal{N}_{in}}{S_{\{m\}}} \frac{\alpha_S}{2\pi} \frac{(4 \pi)^{\varepsilon}}{\Gamma (1 - \varepsilon)} \int d \Phi_m \left( \frac{\mu^2}{s_{ij}} \right)^{\varepsilon} \,_m\langle 1, \dots, m| \mathbf{T}_i \mathbf{T}_j |1, \dots, m \rangle_m \,\, \nonumber \\
&\hspace{5cm}\times \left[ \frac{1}{\varepsilon^2} + \frac{11}{6 \varepsilon} + \frac{55}{18} + \frac{14}{3} \log(2) -\frac{\pi^2}{3} + \mathcal{O}(\varepsilon) \right] \\
\sigma^{D\,(2)}_{a \bar{q},q} &= \frac{\mathcal{N}_{in}}{S_{\{m\}}} \frac{\alpha_S}{2\pi} \frac{(4 \pi)^{\varepsilon}}{\Gamma (1 - \varepsilon)} \int d \Phi_m \left( \frac{\mu^2}{s_{ij}} \right)^{\varepsilon} \,_m\langle 1, \dots, m| \mathbf{T}_i \mathbf{T}_j |1, \dots, m \rangle_m \,\,  \nonumber \\
&\hspace{5cm}\times \frac{T_R N_f}{C_A} \left[ - \frac{2}{3 \varepsilon} + \frac{2}{9} + \frac{10}{3} \log(2) + \mathcal{O}(\varepsilon) \right] \,\,.
\end{align}

It is interesting to note that, if we use only terms with $q_1$ on-shell, we are able to entirely reconstruct the real part of the triangle associated with $p_i$ and $p_j$ in Eq.\eqref{eqn:triangle_Cij}, with no remnant left out.
In fact, as we can appreciate in Eqs.\eqref{res:count_int_qg_V} and \eqref{res:count_int_gg_V}, the dual contributions coming from the scalar triangle in the sum $V^{(1)}_{ag,b}+V^{(1)}_{bg,a}$ lead us to
\begin{equation}
\frac{1}{2 p_i \cdot p_j} \frac{(4 \pi)^{\varepsilon - 2}}{\Gamma (1 - \varepsilon)} \left( \frac{\mu^2}{s_{ij}} \right)^{\varepsilon} \left( \frac{1}{\varepsilon^2} - \frac{\pi^2}{2} \right)
=\text{Re} \, C_0 (p_i, p_j) \,\,.
\end{equation}
It should be noted that this result depends on the function $\mathcal{R}_1$ used to select the loop integration domain which, in turn, is dictated by the mapping.
Therefore, the reconstruction of the whole triangular function may be considered an accident.
However, this could also suggests a connection between the particular mapping of Eq.\eqref{map:map1inv} and the structure of the three-point scalar function.

From now on, we will use only the \textit{$q_1$-method}.

\section{Masses in the final state}
\label{sec:masses}
The presence of massive quarks in the final state does not alter the reasoning of section~\ref{sec:method}.
The difference is that now we have to apply the TLD theorem with massive propagators and that we need a generalization of the mapping in Eq.\eqref{map:map1inv} to take into account the mass of the fermions.
For illustrative purposes, in this section we limit ourselves to the case of a single emitter-spectator pair of fermion-antifermion with the same mass.
This example highlights the main differences with the massless case and can be easily generalized to the cases of pairs with different masses or one massive and one massless particle.

\subsection{Mapping between virtual and real sector}
Let us denote by $M$ the common mass of the quark and the antiquark.
The mapping in Eq.\eqref{map:map1inv} can be generalized as follows
\begin{align*}
\label{map:map1inv_mass}
q_1 &= p'_c \\
p_i &= \frac{2 \beta'_{abc} p'_b +[ \alpha'_{abc}-\gamma'_{abc}- \beta'_{abc} (\alpha'_{abc} + \gamma'_{abc})] (p'_a+p'_b+p'_c)}{2 (\alpha'_{abc}-\gamma'_{abc})} \\
p_j &= - \frac{2 \beta'_{abc} p'_b -[ \alpha'_{abc}-\gamma'_{abc}+ \beta'_{abc} (\alpha'_{abc} + \gamma'_{abc})] (p'_a+p'_b+p'_c)}{2 (\alpha'_{abc}-\gamma'_{abc})} \numberthis
\end{align*}
where $\alpha'_{abc}$, $\beta'_{abc}$ and $\gamma'_{abc}$ are defined by
\begin{align*}
\label{def:al_be_ga}
\alpha'_{abc} &\equiv \frac{s'_{abc}-  2\, p'_a \cdot p'_c -\sqrt{(s'_{abc} - 2\,p'_a \cdot p'_c)^2-4 M^2 s'_{abc}}}{2 s'_{abc}} \\
\beta'_{abc} &\equiv \sqrt{1-\frac{4 M^2}{s'_{abc}}} \qquad\qquad \gamma'_{abc} \equiv \frac{M^2}{s'_{abc} \, \alpha'_{abc}} \numberthis \,\,.
\end{align*}
As in the massless case, the mapping in Eq.\eqref{map:map1inv_mass} automatically verifies momentum conservation $p'_a+p'_b+p'_c=p_i+p_j$ and on-shell conditions ${p_i}^2={p_j}^2=M^2$.
We also choose again the region $ p'_a \cdot p'_c < p'_b \cdot p'_c$ to be the definition domain of the transformation in Eq.\eqref{map:map1inv_mass}.
Note that, in the massless limit ($M \rightarrow 0$), $\alpha'_{abc} \rightarrow 0$, $\beta'_{abc} \rightarrow 1$ and $\gamma'_{abc} \rightarrow 1- 2 (p'_a \cdot p'_c)/s'_{abc}$.
Therefore, in this limit, the mapping in Eq.\eqref{map:map1inv_mass} reduces to the one in Eq.\eqref{map:map1inv} that we have used for the massless case.

Working out the \textit{$q_1$-method}, we set on-shell only the loop momentum flowing into a massless propagator.
For this, we can again use the parametrization of Eq.\eqref{def:param_loop_momenta} to assign $q_1$ in terms of the dimensionless variables $(\xi_1, v_1)$ in the center-of-mass frame of $p_i + p_j$.

In order to obtain the phase space factorization, we may want to express the momentum mapping in Eq.\eqref{map:map1inv_mass} as a relation between the kinematic invariants $(y'_{ac},y'_{bc})$ and the variables $(\xi_1, v_1)$.
The momentum mapping in Eq.\eqref{map:map1inv_mass} implies
\begin{equation}
\label{map:inv_kin_invar1_mass}
\begin{split}
\xi_1 &= y'_{ac} + y'_{bc} \\
v_1 &=  \frac{y'_{ac} -(y'_{ac} + y'_{bc})(y'_{ac}+\alpha'_{abc})}{(y'_{ac} + y'_{bc})(1-y'_{ac}-2\alpha'_{abc})}
\end{split}
\end{equation}
where $\alpha'_{abc}$ can be easily expressed as a function of $(y'_{ac},y'_{bc})$ directly from its definition in Eq.\eqref{def:al_be_ga}.

Phase space factorization reads
\begin{equation}
\label{eqn:phase_space_mass}
\int d \Phi_m \int_{q_1} \tilde{\delta} (q_1) \, \mathcal{R}_1 (\xi_1, v_1)
= \int d \Phi_{m+1}\,
\theta (y'_{bc} - y'_{ac}) J(y'_{ac},y'_{bc}) \xi_1(y'_{ac},y'_{bc}) \beta'_{abc}
\end{equation}
where $J_1(y'_{ac},y'_{bc})$ is the jacobian of the transformation in Eq.\eqref{map:inv_kin_invar1_mass}, given by
\begin{equation}
\label{map: jac1_mass}
J_1 (y'_{ac},y'_{bc})= \sqrt{s'_{abc}}\,\, \frac{s'_{abc}(1-y'_{ac}-y'_{bc})(1-y'_{ac}) - 2M^2(2-y'_{ac}-y'_{bc})}{(y'_{ac}+y'_{bc})\left[ s'_{abc}(1-y'_{ac})^2-4M^2\right]^{3/2}}
\end{equation}
while $\mathcal{R}_1(\xi_1, v_1)$ is the characteristic function that selects the loop integration domain.
To compute its expression in terms of $(\xi_1, v_1)$, we start by considering the inverse of the momentum mapping in Eq.\eqref{map:map1inv_mass}, that is given
by~\cite{Sborlini:2016hat}
\begin{align*}
\label{map:map1_mass}
p'_c &= q_1 \\
p'_a &= (1-\alpha_{ij}) \hat{p}_i+(1-\gamma_{ij}) \hat{p}_j-q_1 \\
p'_b &= \alpha_{ij} \hat{p}_i + \gamma_{ij} \hat{p}_j \numberthis
\end{align*}
where $\hat{p}_i$ and $\hat{p}_j$ are massless momenta related to $p_i$ and $p_j$ by 
\begin{align*}
p_i &=  \frac{1+\beta_{ij}}{2} \,\hat{p}_i + \frac{1-\beta_{ij}}{2} \,\hat{p}_j \\
p_j &=  \frac{1-\beta_{ij}}{2} \,\hat{p}_i + \frac{1+\beta_{ij}}{2} \,\hat{p}_j \numberthis \,\,.
\end{align*}
The expressions of $\alpha_{ij} = \alpha'_{abc}$, $\beta_{ij} = \beta'_{abc}$ and $\gamma_{ij}=\gamma'_{abc}$ in terms of the virtual sector variables are the following
\begin{align*}
\label{eqn:alpha_mass}
\alpha_{ij} &= \frac{1 - \xi_1 - \sqrt{(1-\xi_1)^2- (1-\xi_1+v_1(1-v_1)\xi_1^2)\,4 M^2/s_{ij}}}{2(1-v_1 \xi_1)} \\
\beta_{ij} &\equiv \sqrt{1-\frac{4 M^2}{s_{ij}}} \qquad\qquad
\gamma_{ij} \equiv \frac{M^2}{s_{ij} \, \alpha_{ij}} \numberthis \,\,.
\end{align*}

Inverting \eqref{map:inv_kin_invar1_mass} we get
\begin{align*}
\label{map:map1_kin_inv_mass}
y'_{ac} &= \frac{\xi_1}{1-(1-v_1)\xi_1} \left[ v_1 + \alpha_{ij} (1-2 v_1) \right] \\
y'_{bc} &= \frac{\xi_1}{1-(1-v_1)\xi_1}\left[ (1-v_1) (1-\xi_1) - \alpha_{ij} (1-2 v_1) \right] \numberthis
\end{align*}
that can be used, together with \eqref{eqn:alpha_mass}, to express $\mathcal{R}_1$ in terms of $(\xi_1,v_1)$, obtaining
\begin{equation}
\label{def:R_1_mass}
\theta (y'_{bc} - y'_{ac})\! \equiv\! \mathcal{R}_1(\xi_1, v_1) = \theta (1 - 2 v_1)  \theta\!\!\left[\frac{1-2v_1}{1-v_1}\! \left( 1-\frac{1-\sqrt{1-16v_1(1-v_1)M^2/s_{ij}}}{2v_1}\right) \!-\xi_1  \! \right] \,\,.
\end{equation}
In the region $y'_{bc} <  y'_{ac}$
the prescription of the \textit{$q_1$-cut-method} consists into exchanging both $p'_a \leftrightarrow p'_b$ and $p_i \leftrightarrow p_j$, i.e. we consider the diagram in Fig.\eqref{fig:vertex_ij2} and we set again $q_1$ on-shell.

\subsection{Dual counterterms and singular behaviour}
\label{sec:masses_count}
In the case of a quark-antiquark massive pair as emitter and spectator, the application of LTD to the triangle and the bubble contributions from the virtual cross section leads to the same result as for the massless case.
In fact, according to Eq.\eqref{def:dual propagator}, the relevant dual propagators are given by
\begin{align*}
\label{eqn:dual_prop_mass}
G_D (q_1; q_2)^{-1} &=q_2^2 - M^2 - i 0 (q_2^0 - q_1^0)
= 2 q_1 \cdot p_2 -i0 \\
G_D (q_1; q_3)^{-1} &=q_3^2 - M^2 - i 0 (q_3^0 - q_1^0)
= - 2 q_1 \cdot p_1 +i0 \numberthis
\end{align*}
which do not depend on the mass of the fermions.
The same does not happens if we consider the dual contributions with $q_2$ on-shell (\textit{$q_1$-$q_2$-method}).

With this in mind, we define the dual subtraction term as
\begin{multline}
\label{def:count_mass_qg}
\sigma^{D\,(1)}_{qg,\bar{q}} \equiv 8\pi\alpha_S \frac{\mathcal{N}_{in}}{S_{\{m\}}} \int d \Phi_m \,_m\langle 1, \dots, m| \mathbf{T}_{ac} \mathbf{T}_b |1, \dots, m \rangle_m \\
\times \left[  V^{(1)}_{qg,\bar{q}} (p_i, p_j) + G^{(1)}_{1, qg,\bar{q}} (p_i,p_j) +G^{(1)}_{2, qg,\bar{q}} (p_i,p_j)  \right] 
\end{multline}
where
\begin{subequations}
\label{eqn:count_mass_qg_comp}
\begin{align}
V^{(1)}_{qg,\bar{q}} &\equiv \int_{q_1} \tilde{\delta} (q_1) \mathcal{R}_1 \left[- \frac{4\, p_i \cdot p_j}{(- 2 q_1 \cdot p_i)(2 q_1 \cdot p_j)} + \frac{2}{(-2 q_1 \cdot p_i)} \right] \\
G^{(1)}_{1, qg,\bar{q}} &\equiv - \int_{q_1} \tilde{\delta} (q_1) \mathcal{R}_1  \left[ (1-\varepsilon) \frac{1}{(-2 q_1 \cdot p_i)} \frac{q_1 \cdot p_j}{p_i \cdot p_j} + \frac{2 M^2}{(-2q_1\cdot p_i)^2} \left( 1-\frac{q_1 \cdot p_j}{p_i \cdot p_j} \right) \right] \\
G^{(1)}_{2, qg,\bar{q}} &\equiv \int_{q_1} \tilde{\delta} (q_1) \mathcal{R}_1  \left[ (1-\varepsilon) \frac{1}{(2 q_1 \cdot p_j)} \frac{q_1 \cdot p_i}{p_i \cdot p_j} - \frac{2 M^2}{(2q_1 \cdot p_j)^2} \left( 1 + \frac{q_1 \cdot p_i}{p_i \cdot p_j} \right) \right] \,\,.
\end{align}
\end{subequations}
Here, the first contribution to $V^{(1)}_{qg,\bar{q}}$ represents the triangle scalar function coming from the reduction of the virtual amplitude.
Contrary to the massless case, the coefficient of the bubble integral of the external massive momenta coming from the reduction, is highly non-trivial.
Nevertheless, this scalar integral is IR finite.
However, when the mass of the fermion is very small relative to the other scales of the process, this contribution exhibit a logarithmic enhancement in both the real and the virtual sector.
It is then convenient to keep the bubble in the counterterm $V^{(1)}_{qg,\bar{q}}$ with the coefficient of the massless case.
The dual counterterms $G^{(1)}_{1, qg,\bar{q}}$ and $G^{(1)}_{2, qg,\bar{q}}$ come from the integrand representation of the wave--function renormalization of the emitter and the spectator, respectively, reported in Eq.\eqref{eqn:quark_ren}~\cite{Sborlini:2016hat}.

The procedure to prove that the dual subtraction in Eq.\eqref{def:count_mass_qg} has the same singular behaviour of the real amplitude is algebraically challenging but straightforward.
For this reason, here we limit ourselves to just list the steps of the proof.
As for the quasi-collinear limit, the first thing to do is to express the scalar products appearing in Eq.\eqref{eqn:count_mass_qg_comp} in terms of $p'_a \cdot p'_b$, $p'_a \cdot p'_c$, $p'_b \cdot p'_c$ and $M^2$ by using the momentum mapping in Eq.\eqref{map:map1inv_mass}.
Then, in order to test the collinear behaviour, the invariants $p'_a \cdot p'_b$ and $p'_a \cdot p'_c$ have to be written as a function of $z$, $k_\perp$, $p$, $M^2$ and $n$ via the parametrization in Eqs.\eqref{par:coll_mass}.
At this point, the uniform rescaling in Eq.\eqref{rescaling} can be performed and the dual subtraction in Eq.\eqref{def:count_mass_qg} can be expanded in series of $\lambda$.
What is found is that the leading order perfectly matches the right-hand side of Eq.\eqref{eqn:coll_beh_mass} with $\hat{P}_{(q, q)} (z, k_{\bot}, \{m\}, \varepsilon)$.
In the soft region we need to use Eq.\eqref{par:soft} instead of Eqs.\eqref{par:coll_mass}, and then to consider again the series of $\lambda$.
The result obtained with this procedure is in agreement with the right-hand side of Eq.\eqref{eqn:soft_beh_mass}.

\subsection{Integrated dual counterterms}
\label{sec:masses_int_count}
As in the massless case, we need to integrate the dual counterterms over the loop momentum, in order to obtain a result which lives on the $m$-particle phase space and is ready to cancel the poles of the virtual amplitude.

First of all, the massive momenta $p_i$ and $p_j$ are given, in their center-of-mass frame, by the following expressions
\begin{equation}
\label{eqn:frame_massive}
p_i = \frac{\sqrt{s_{ij}}}{2} (1, 0, 0, \beta_{ij}) \,\, , \qquad p_j = \frac{\sqrt{s_{ij}}}{2} (1, 0, 0, - \beta_{ij}) \,\,.
\end{equation}
The scalar products between the loop momentum $q_1$ parametrized in Eq.\eqref{def:param_loop_momenta} and the external momenta of Eq.\eqref{eqn:frame_massive} turn out to be
\begin{align*}
\label{eqn:param_loop_momenta_dot_mass}
q_1 \cdot p_i  &=  \frac{s_{ij}}{4} \xi_1 [ 1 - \beta_{ij} (1-2 v_1) ] \\
q_1 \cdot p_j &=  \frac{s_{ij}}{4} \xi_1 [ 1 + \beta_{ij} (1-2 v_1) ] \numberthis \,\,.
\end{align*}
By using Eqs.\eqref{eqn:measure}, \eqref{def:measure} and \eqref{eqn:param_loop_momenta_dot_mass}, we can express the dual counterterms of Eq.\eqref{eqn:count_mass_qg_comp} as follows
\begin{align*}
V^{(1)}_{qg, \bar{q}} &= \int \frac{[d \xi_1 d v_1]}{(4 \pi)^2}  \mathcal{R}_1 \left[\frac{8(s_{ij}-2 M^2)}{s_{ij}\xi_1(1-\beta_{ij}^2(1-2 v_1)^2 )} - \frac{4}{1 - \beta_{ij} (1-2 v_1)} \right] \\
G^{(1)}_{1, qg,\bar{q}} &=  \int \frac{[d \xi_1 d v_1]}{(4 \pi)^2} \mathcal{R}_1 \left\{ \frac{ s_{ij} \xi_1 (1-\varepsilon)(1 + \beta_{ij} (1-2 v_1) )}{(s_{ij}-2 M^2)( 1 - \beta_{ij} (1-2 v_1))} \right. \\
&\hspace{4.3cm} \left. - \frac{8 M^2}{s_{ij} \xi_1 ( 1 - \beta_{ij} (1-2 v_1))} \left[ 1 - \frac{\xi s_{ij} ( 1 + \beta_{ij} (1-2 v_1))}{2(s_{ij}-2M^2)} \right] \right\} \\
G^{(1)}_{2, qg,\bar{q}} &=  \int \frac{[d \xi_1 d v_1]}{(4 \pi)^2} \mathcal{R}_1 \left\{ \frac{ s_{ij} \xi_1 (1-\varepsilon)(1 - \beta_{ij} (1-2 v_1) )}{(s_{ij}-2 M^2)( 1 + \beta_{ij} (1-2 v_1))} \right. \\
&\hspace{4.3cm} \left. - \frac{8 M^2}{s_{ij} \xi_1 ( 1 + \beta_{ij} (1-2 v_1))} \left[ 1 - \frac{\xi s_{ij} ( 1 - \beta_{ij} (1-2 v_1))}{2(s_{ij}-2M^2)} \right] \right\} \,\,.
\end{align*}
These integrals can be computed analytically.
Their poles are given by
\begin{align*}
\label{res:poles_mass}
V^{(1)}_{qg, \bar{q}} &= \frac{(4 \pi)^{\varepsilon - 2}}{\Gamma (1 - \varepsilon)} \left( \frac{\mu^2}{s_{ij}} \right)^{\varepsilon} \left[ \left( \frac{1+\beta_{ij}^2}{2 \beta_{ij} \varepsilon} \right) \log \left( \frac{1-\beta_{ij}}{1+\beta_{ij}}\right) + \mathcal{O}(\varepsilon^0) \right] \\
G^{(1)}_{1, qg,\bar{q}} &=  \frac{(4 \pi)^{\varepsilon - 2}}{\Gamma (1 - \varepsilon)} \left( \frac{\mu^2}{s_{ij}} \right)^{\varepsilon} \left[ \frac{1+\beta_{ij}}{2 \varepsilon} + \mathcal{O}(\varepsilon^0) \right] \\
G^{(1)}_{2, qg,\bar{q}} &=  \frac{(4 \pi)^{\varepsilon - 2}}{\Gamma (1 - \varepsilon)} \left( \frac{\mu^2}{s_{ij}} \right)^{\varepsilon} \left[ \frac{1-\beta_{ij}}{2 \varepsilon} + \mathcal{O}(\varepsilon^0) \right] \numberthis
\end{align*}
By inserting Eq.\eqref{res:poles_mass} into Eq.\eqref{def:count_mass_qg} we obtain, for the dual cross section, the following pole
\begin{multline}
\sigma^{D\,(1)}_{qg,\bar{q}} = \frac{\mathcal{N}_{in}}{S_{\{m\}}} \frac{\alpha_S}{2\pi} \frac{(4 \pi)^{\varepsilon}}{\Gamma (1 - \varepsilon)} \int d \Phi_m \left( \frac{\mu^2}{s_{ij}} \right)^{\varepsilon} \,_m\langle 1, \dots, m| \mathbf{T}_i \mathbf{T}_j |1, \dots, m \rangle_m \\
\times \frac{1}{\varepsilon} \left[ 1 + \left( \frac{1+\beta_{ij}^2}{2 \beta_{ij}} \right) \log \left( \frac{1-\beta_{ij}}{1+\beta_{ij}}\right) \right] + \,\,\dots 
\end{multline}
which is doubled when we add $\sigma^{D\,(1)}_{\bar{q}g,q} = \sigma^{D\,(1)}_{qg,\bar{q}}$.
The pole structure just obtained matches the contribution to the one-loop amplitude associated with a massive quark-antiquark emitter-spectator pair~\cite{Catani:2000ef}.

\section{Initial state radiation}
\label{sec:initial_state}

\begin{figure}[t]
    \centering
\begin{subfigure}{.49\textwidth}
    \includegraphics[width=0.8\linewidth]{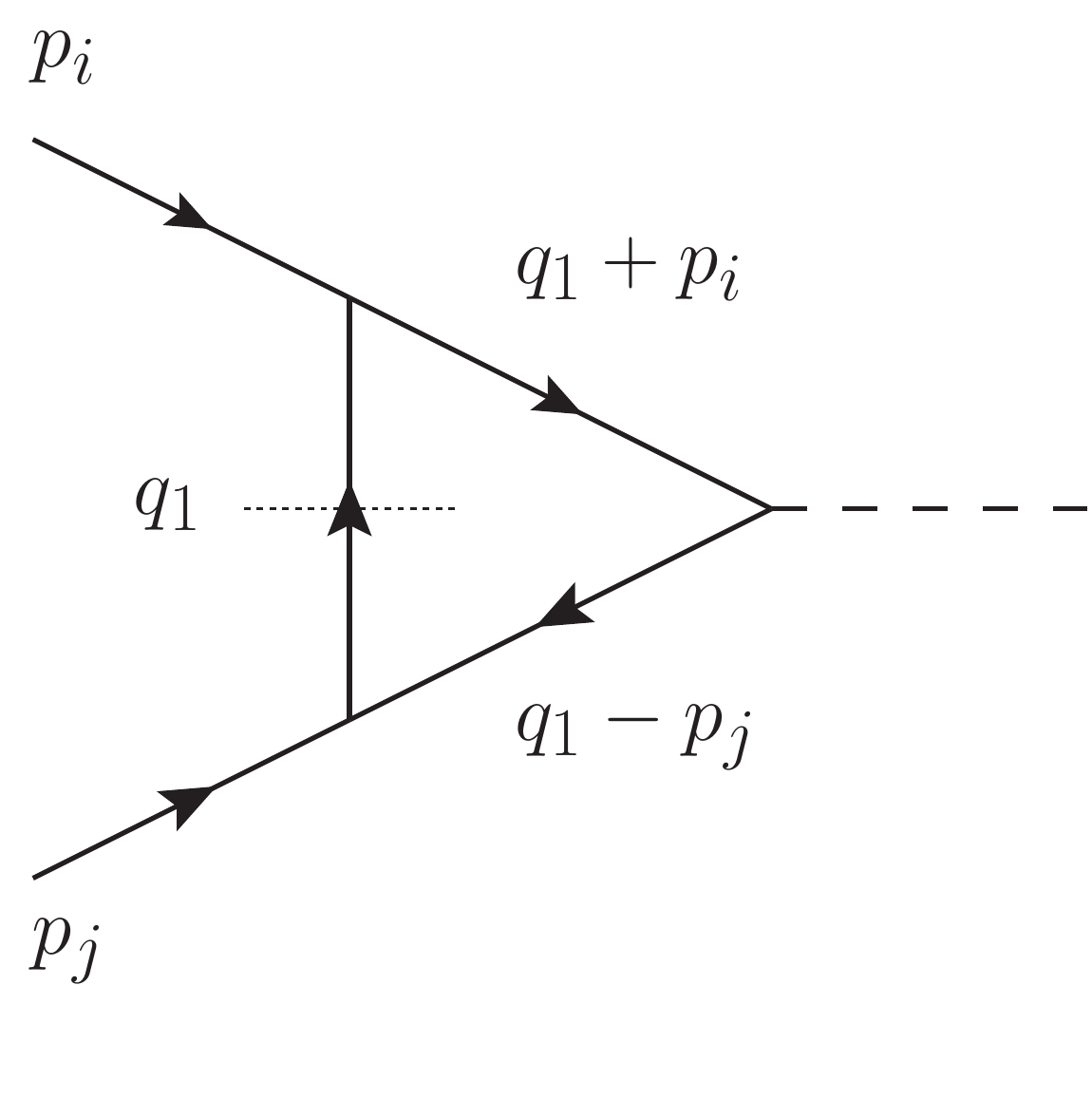}
\end{subfigure}
\begin{subfigure}{.47\textwidth}
\includegraphics[width=0.8\linewidth]{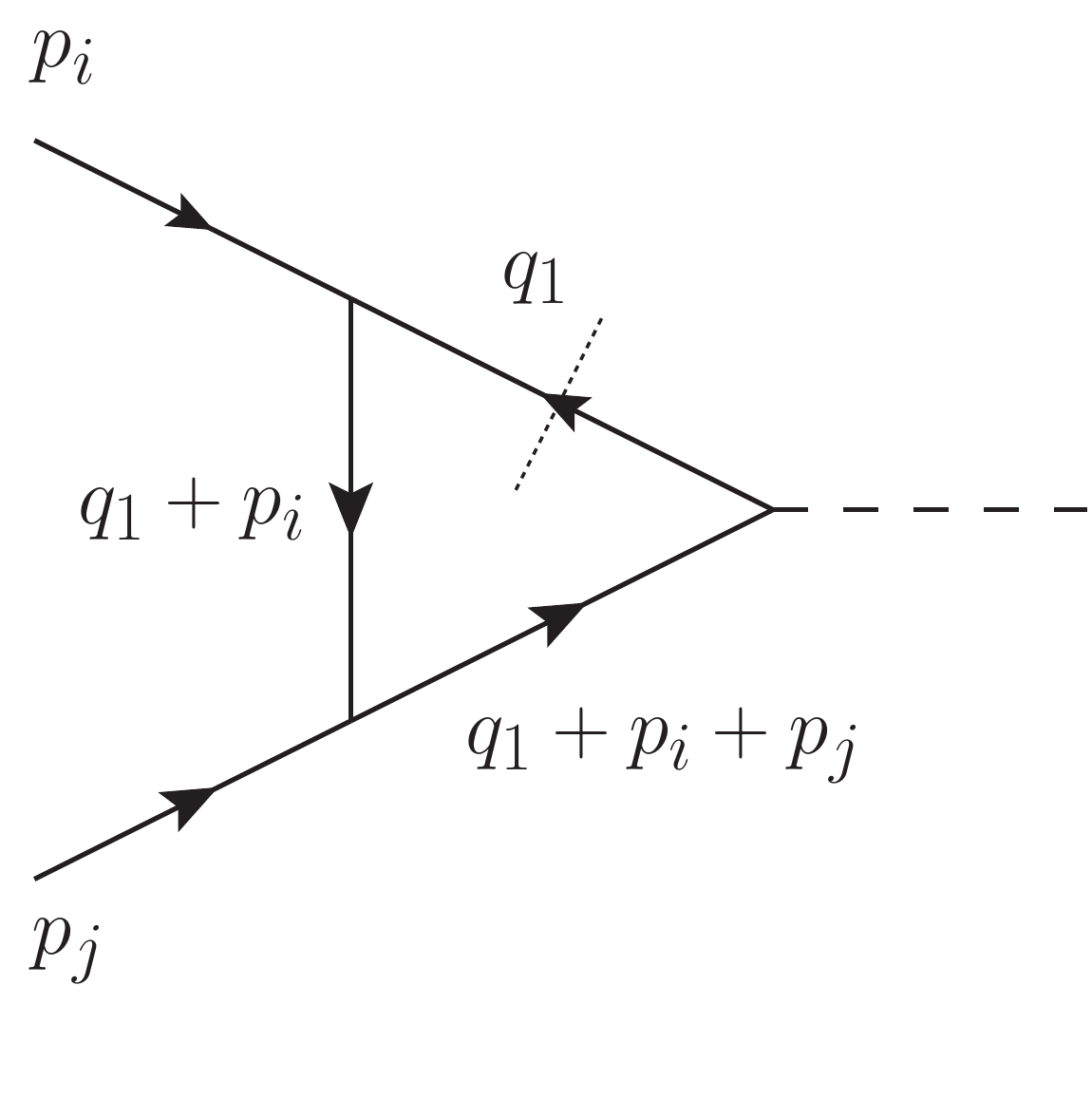}
\end{subfigure}
    \caption{Cut loop contributions giving rise to divergences
    for the case of radiation from the initial state.}
    \label{fig:my_label}
\end{figure}
It is well known~\cite{Altarelli:1977zs} that in presence of partons in the initial state the singularities of the virtual matrix element do not cancel the ones of the real contribution and the proper definition of the NLO cross section requires the subtraction of the remaining collinear initial state singularities.
The reasoning of Section~\ref{sec:method} which allows to reconstruct the universal IR behaviour of the renormalized one-loop amplitude holds also for the case of partons in the initial state.
So that, we will start by the singularities of the virtual matrix element knowing already that they will cancel all but the collinear initial state singularities of the real matrix element. 
We start by considering a quark emitter and a gluon
radiation.
Following the momentum assignment of the left panel in Fig.\eqref{fig:my_label} (note that now momenta $p_i$ and $p_j$ are incoming so that in the expressions below there are different signs with respect to the expressions reported in Section~\ref{sec:algorithm}) and recalling the same steps of Section~\ref{sec:algorithm} the singularities of virtual contribution associated to the emission of a virtual gluon by an hard quark carrying momentum $p_i$ absorbed by the hard parton carrying momentum $p_j$ is given by
\begin{equation}
\label{eqn:count_init_qg}
\sigma^{D\,(1)}_{qg,b} \equiv 8\pi\alpha_S  \int d \Phi_m \frac{\mathcal{N}_{in}}{S_{\{m\}}} \,_m\langle 1, \dots, m| \mathbf{T}_{ac} \mathbf{T}_b |1, \dots, m \rangle_m  \left[  V^{(1)}_{qg,b}(p_i, p_j) + G^{(1)}_{qg,b}(p_i, p_j) \right]
\end{equation}
where
\begin{align*}
\label{eqn:count_init_qg_comp_fake}
V_{qg,b}^{(1)} &\equiv \int_{q_1} \tilde{\delta} (q_1) \,\mathcal{R}_1 \left[- \frac{2 s_{ij}}{(2 \, q_1 \cdot p_i)
(-2\, q_1 \cdot p_j)} +  \frac{2}{(2 \, q_1 \cdot p_i)}  \right] \\
G_{qg,b}^{(1)} &\equiv \frac{(1-\varepsilon)}{s_{ij}} \int_{q_1} \tilde{\delta} (q_1) \,\mathcal{R}_1  \frac{2 q_1 \cdot p_j}{(2 q_1 \cdot p_i)}
\numberthis
\end{align*}
and $\mathcal{R}_1$ is any function that cuts UV divergences, selects the region $q_1 \parallel p_i$ and leaves out the one where $q_1 \parallel p_j$.
Let's focus on the case in which partons $i$ and $j$ are in the initial state.
We will extend the results to the other cases in a moment. When, in the previous sections, we have studied the case of final state radiation, we enforced
the constraint of preserving the invariant mass of the Born system. This time
at Born level we are producing a final state with invariant mass $s_{ij}$ and
so enforcing the conservation of the invariant mass when we cut the scalar loop functions would generate by definition a phase space configuration sitting on
the soft limit for the emitted gluon $q_1$.
We now consider the real phase space volume corresponding to an initial state
configuration with two partons with momenta $p_a$ and $p_b$ such that
$s_{ab} > s_{ij}$. In this case one can exploit the phase space factorization
to map the real phase space onto the Born phase space (with initial momenta 
$p_i$ and $p_j$) and, following the same approach of the previous sections, 
identify the (extended) loop momentum with the momentum of the radiation.
It is a straightforward exercise to show that if parton $a$ is a quark, the
collinear divergences of the real matrix element associated to a gluon 
emission can be described in terms of the above formula multiplied by the 
momentum fraction $x$ of the relation $p_i=x\,p_a$. 
We anticipate that this extra factor of $x$ has to be included for all 
the cases with an initial emitter (i.e. all initial-initial and initial-final cases).
To show that the dual counterterm (including this factor of $x$) cancels the initial state 
collinear singularity of the real matrix element, we adopt the momentum 
mapping for initial-initial singularities of ref.\cite{Catani:1996vz}.
Apart from the boost on the Born system, the mapped momenta are given by
\begin{align*}
\label{map:init1}
p_i &= x p_a & x & \equiv \frac{p_a \cdot p_b - p_c \cdot p_a
- p_c \cdot p_b}{p_a\cdot p_b} \\
p_j &= p_b & v &\equiv \frac{p_a \cdot p_c}{p_a\cdot p_b} \\
q_1 &= p_c \numberthis \,\,.
\end{align*}
where $p_c$ is the momentum of the radiated gluon in the real 
phase space configuration.
We have then
\begin{align}
\label{map:scalarproducts1}
s_{ij} &= x \,2 \,p_a \cdot p_b & q_1 \cdot p_i &= x \,p_c \cdot p_a &
q_1 \cdot p_j &=  p_c \cdot p_b
\end{align}
and expressing all the scalar products in terms of the invariant
built with the momenta of the collinear pair ($p_a\cdot p_c$) and the variables $x$ and $v$, we have
\begin{align}
\label{map:scalarproducts2}
p_a \cdot p_b &= p_a \cdot p_c/v &  p_c \cdot p_b &
= p_a \cdot p_c (1-x-v)/v \,.
\end{align}
Substituting the scalar products in the right hand side of Eq.(\ref{eqn:count_init_qg})
and multiplying by $x$ we find
\begin{eqnarray}
\sigma^{D\,(1)}_{qg,b} &\equiv& 8\pi\alpha_S  \int d \Phi_m \frac{ \mathcal{N}_{in}}{S_{\{m\}}} \,_m\langle 1, \dots, m| \mathbf{T}_{ac} \mathbf{T}_b |1, \dots, m \rangle_m  \, x\,\left[  V^{(1)}_{qg,b}(p_i, p_j) + G^{(1)}_{qg,b}(p_i, p_j) \right] \nonumber \\
&\equiv& 4\pi\alpha_S  \int d \Phi_m 
\frac{\mathcal{N}_{in}}{S_{\{m\}}} \,_m\langle 1, \dots, m| \mathbf{T}_{ac} 
\mathbf{T}_b |1, \dots, m \rangle_m  \\ \nonumber
&&\times \left\{
\frac{1}{x\,p_a \cdot p_c}\left[ \frac{1+x^2-2v +v^2}{1-x-v}
-\varepsilon (1-x-v)
\right]\right\}
\label{def:iqg}
\end{eqnarray}
which in the limit $v \to 0$ coincides with the first of 
Eqs.(\ref{def:splitting_functions_initial}).
This counterterm has to be integrated at fixed $s_{ab}$ and imposing
the condition $p_a \cdot p_c < p_b \cdot p_c$ we constraint the integration
region to $\Omega=\{(v,x)| 0<v<(1-x)/2$, $0<x<1 \}$. As usual, the 
integration has to be performed in the sense of distributions and
gives rise to a distribution in $x$. Finally, to compensate for the different
flux factor of the real contribution ($s_{ab}$) with respect
to the one of the virtual contribution $s_{ij}=x\,s_{ab}$, for the analytic integration
we multiply the counterterm times $x$.
Also the cases of initial-final and final-initial singularities
will be addressed with the corresponding momentum mapping proposed in 
ref.\cite{Catani:1996vz} that enforces one rescaled momentum in the
initial state. For this reason also for these cases the integrated version
of the counterterms is multiplied times this compensating factor of $x$.
A small difference with respect to the case of initial-initial is 
the presence of an extra jacobian factor that is absent (i.e. $1$) for initial-initial,
but is different from 1 for the other two cases, initial-final and 
final-initial.
The result for the integrated version of the dual counterterm in 
Eqs.(\ref{def:splitting_functions_initial}) and for all the others
of the massless case discussed below are collected in 
Appendix~\ref{sec:masslessDS}.

Let us now consider the case of a gluon emitting a gluon in the initial 
state. When we discussed the case of final state emission in Section~\ref{sec:dual_sub} we introduced the dual counterterm
that we report here with incoming momenta for convenience
\begin{equation}
\label{eqn:count_init_gg_second}
\sigma^{D\,(1)}_{gg,b} \equiv 8\pi\alpha_S \frac{\mathcal{N}_{in}}{S_{\{m\}}} \int \! d \Phi_m \,_m\langle 1, \dots, m| \mathbf{T}_{ac} \mathbf{T}_b  \left[  V^{(1) \mu\nu}_{gg,b} (p_i, p_j) + G^{(1) \mu\nu}_{gg,b} (p_i,p_j) \right]  |1, \dots, m \rangle_m
\end{equation}
where
\begin{align*}
\label{eqn:count_init_gg_comp_fake}
V^{(1) \mu\nu}_{gg,b} \!\! &\equiv\! - \!\!\int_{q_1} \tilde{\delta} (q_1) \mathcal{R}_1 \left[- \frac{2 s_{ij}}{( 2\, q_1 \cdot p_i) (-2 \,q_1 \cdot p_j)} + \frac{1}{(2\, q_1 \cdot p_i)} \right] g^{\mu \nu} \\
G^{(1) \mu\nu}_{gg,b} \!\! &\equiv\! - \!\!\int_{q_1} \!\!  \frac{\tilde{\delta} (q_1) \mathcal{R}_1}{(2\, q_1 \cdot p_i)}  \!\left[ g^{\mu \nu}\! - \frac{d-2}{(2\, q_1 \cdot p_i)}\!\! \left( \frac{q_1 \cdot p_j}{p_i \cdot p_j} + 1 \right) \!\!
\left( q_1^{\mu} - \frac{q_1 \cdot p_i}{p_i \cdot p_j} p_j^{\mu} \right)\!\! \left( q_1^{\nu} - \frac{q_1 \cdot p_i}{p_i \cdot p_j} p_j^{\nu} \right) \! \right] \,\,.
\numberthis
\end{align*}
After we introduced the mapping that splits the emitter and spectator momenta
$p_i$ and $p_j$ into the collinear pair $p_a$, $p_c$ and the new spectator 
$p_b$ we proved that the collinear limit of the two partons $p_a$ and $p_c$ 
in real matrix element is reproduced by this virtual contribution by
summing over the two possible final state gluon configurations obtained
by exchanging the momenta $p_a$ and $p_c$ of the collinear pair.
Such symmetric role of the two radiated partons is lost in the case of
radiation from the initial state (note that in the equation above we are not doing the symmetrization), but still of course the above formula correctly reproduces the poles of the virtual matrix element.
Nevertheless, if we multiply this expression times $x$ and consider the extended kinematic range for the loop momentum associated to the possible
radiation in a real phase space kinematic with $s_{ab} > s_{ij}$ this time
we cannot neglect the contribution from the other cut gluon propagator
in the right panel of Fig.\eqref{fig:my_label}. Of course, all three
kinds of integrals, not just the scalar triangle in the figure, but also the scalar bubble and the wave--function
renormalization constant have a second cut contribution, although their
sum has to vanish in the soft limit $x \to 1$. The presence of a square
denominator in the wave--function renormalization is not particularly
problematic {\it per se} thanks to the general results of 
ref.\cite{Sborlini:2016gbr} on the computation of dual integrals with 
denominators with powers greater than~1.
However, to simplify the construction here we include these three extra contributions
expressing them in terms of the same integration variables of the others. To this aim we invert the loop momentum and shift it so that the cut propagator always has momentum $q_1$ (see right panel of Fig.\eqref{fig:my_label}). In this way, the second cut contributions from the scalar bubble
and from the wave--function renormalization constant turns out to be identical
to the ones brought by their first cut (as one could naively expect) while a genuine
new contribution comes from the scalar triangle. The dual subtraction is then given by
\begin{equation}
\label{def:igg}
\sigma^{D\,(1)}_{gg,b} \! \equiv \! 8\pi\alpha_S
\frac{x \mathcal{N}_{in}}{S_{\{m\}}} \! \int \! d \Phi_m \,_m\langle 1, \dots, m| \mathbf{T}_{ac} \mathbf{T}_b \! \left[  V^{(1+2) \mu\nu}_{gg,b} (p_i, p_j)
+ 2\, G^{(1) \mu\nu}_{gg,b} (p_i,p_j) \right] \! |1, \dots, m \rangle_m
\end{equation}
where
\begin{eqnarray}
\label{eqn:count_init_gg_12}
&&V^{(1+2) \mu\nu}_{gg,b} \!\! \equiv \! - \!\!\int_{q_1} \tilde{\delta} (q_1) 
\mathcal{R}_1  \times \\ 
&& \times \left[
- \frac{2\, s_{ij}}{(2\, q_1 \cdot p_i) (-2\, q_1 \cdot p_j)}
- \frac{2 \, s_{ij}}{(2 \, q_1 \cdot p_i)
(2 \,p_i \cdot p_j + 2 \, q_1 \cdot p_j + 2 \, q_1 \cdot p_i)}
  + \frac{2}{(2 \, q_1 \cdot p_i)} \right] g^{\mu \nu} ~~~~ \nonumber
\end{eqnarray}
To study the collinear limit of the expression above we start by analyzing the non trivial tensor part. The vector
\begin{equation}
    N^\mu=q_1^\mu - \frac{q_1 \cdot p_i}{p_i \cdot p_j}p_j^\mu
\end{equation}
is orthogonal to the emitter momentum $p_i$ so that in general it is a superposition of $k_\perp^\mu$ (that is not massless) and $p_i$ (that is massless and orthogonal to $k_\perp$) given by
\begin{equation}
    N^\mu = \alpha k_\perp^\mu + \beta p_i 
\end{equation}
with
\begin{equation}
N^2 \equiv N^\mu N_\mu = -2\,\frac{q_1 \cdot p_i \, q_1 \cdot p_j }{p_i \cdot p_j} = \alpha^2 k_\perp^2 \,.
\qquad    
\end{equation}
The tensorial components proportional to $p_i^\mu$ and $p_i^\nu$ are gauge terms and will not contribute so that multiplying and dividing the tensor part by $N^2$ the counterterm in the square breacket of Eq.(\ref{def:igg}) can be written as
\begin{flalign}
   & \left[  V^{(1+2) \mu\nu}_{gg,b} (p_i, p_j)
+ 2\, G^{(1) \mu\nu}_{gg,b} (p_i,p_j) \right] =  \int_{q_1} \tilde{\delta} (q_1) 
\mathcal{R}_1  \times && \nonumber \\
 & \times \left\{ - \left[  - \frac{2\, s_{ij}}{(2\, q_1 \cdot p_i) (-2\, q_1 \cdot p_j)}
- \frac{2 \, s_{ij}}{(2 \, q_1 \cdot p_i)
(2 \,p_i \cdot p_j + 2 \, q_1 \cdot p_j + 2 \, q_1 \cdot p_i)}
  + \frac{4}{(2 \, q_1 \cdot p_i)} \right] g^{\mu \nu} \right.
 && \nonumber \\
 & \left. \quad\qquad +  2 \, \frac{d-2}{2\, q_1 \cdot p_i}
\left( \frac{q_1 \cdot p_j}{p_i \cdot p_j} + 1 \right)
\left(
-2\,\frac{q_1 \cdot p_i \, q_1 \cdot p_j }{p_i \cdot p_j}
\right)
\frac{k_{\perp}^{\mu} k_{\perp}^{\nu}}{k_\perp^2} \right\} &&
\end{flalign}
Inserting the scalar products in Eqs.(\ref{map:scalarproducts1}) and (\ref{map:scalarproducts2})
and taking the $v \to 0$ limit, the expression above times $x$ coincides with the collinear limit
in Eqs.(\ref{def:splitting_functions_initial}) for a gluon
emitting a gluon.
We now move to the counterterms for the splittings in which the initial parton is different from the hard (emitter) parton.
We first discuss the case of a quark emitting
a quark transforming into an hard gluon for
the underlying Born configuration. In this case
the dual counterterm is built from the quark
contribution to the gluon wave--function renormalization constant and is given by (including the factor of $x$ for the case of an initial emitter)
\begin{equation}
\label{def:iqq}
\sigma^{D\,(1)}_{q\bar{q},b} \equiv 8\pi\alpha_S \frac{\mathcal{N}_{in}}{S_{\{m\}}} \int \! d \Phi_m \,_m\langle 1, \dots, m| \mathbf{T}_{ac} \mathbf{T}_b \,x\, G^{(1) \mu\nu}_{q \bar{q},b} (p_i,p_j) |1, \dots, m \rangle_m
\end{equation}
with
\begin{equation}
\label{def:count_qq_comp}
G^{(1) \mu\nu}_{q\bar{q},b} \!\! \equiv\! \!\!\int_{q_1} \!\!  \frac{\tilde{\delta} (q_1) \mathcal{R}_1 T_R}{C_A (2 \, q_1 \cdot p_i)}  \!\left[ g^{\mu \nu}\! \!-\! \frac{4}{(2 \, q_1 \cdot p_i)}\!\! \left( \frac{q_1 \cdot p_j}{p_i \cdot p_j} + 1 \right) \!\!
\left( q_1^{\mu} - \frac{q_1 \cdot p_i}{p_i \cdot p_j} p_j^{\mu} \right)\!\! \left( q_1^{\nu} - \frac{q_1 \cdot p_i}{p_i \cdot p_j} p_j^{\nu} \right) \! \right].
\end{equation}
Following the same steps as for the case of a gluon emitting a gluon it is easy to show that in the limit $v \to 0$ this counterterm reproduces the functional form of the third
equation of Eqs.(\ref{def:splitting_functions_initial}).
The last case is when an initial gluon radiates an antiquark transforming into a hard quark. In this case we find that the sole contribution
of the quark wave--function renormalization constant (times $x$) is not sufficient to get the correct collinear limit of the real matrix element. We have the correct limit including a contribution given by minus the
second cut of the virtual part associated to a gluon emitter. The result is:
\begin{equation}
\label{def:igq}
\sigma^{D\,(1)}_{g\bar{q},b} \equiv 8\pi\alpha_S \frac{\mathcal{N}_{in}}{S_{\{m\}}} \int \! d \Phi_m \,_m\langle 1, \dots, m| \mathbf{T}_{ac} \mathbf{T}_b \,x\,\left[ -V^{(2)}_{g\bar{q},b}(p_i,p_j) + G^{(1)}_{g\bar{q}} (p_i,p_j)  \right] |1, \dots, m \rangle_m
\end{equation}
with
\begin{eqnarray}
\label{eqn:count_init_gq_2}
V^{(2)}_{g\bar{q},b} &\equiv&  \!\!\int_{q_1} \tilde{\delta} (q_1) 
\mathcal{R}_1  \left[
- \frac{2 \, s_{ij}}{(2 \, q_1 \cdot p_i)
(2 \,p_i \cdot p_j + 2 \, q_1 \cdot p_j + 2 \, q_1 \cdot p_i)}
  + \frac{1}{(2 \, q_1 \cdot p_i)} \right] \\
G_{g\bar{q},b}^{(1)} &\equiv& \frac{(1-\varepsilon)}{s_{ij}} \int_{q_1} \tilde{\delta} (q_1) \,\mathcal{R}_1  \frac{2 \, q_1 \cdot p_j}{(2 \, q_1 \cdot p_i)} \, .
\end{eqnarray}
Once the scalar products are substituted in the formulae above, in the limit $v \to 0$ the resulting expression reproduces proper collinear
behaviour of Eqs.(\ref{def:splitting_functions_initial}).
We stress that the dual dipoles for the final-final case describe also the final-initial case after the corresponding jacobian factor is included to translate the radiation variables into the loop momentum. In the same way the formulae of this section for the initial-initial case describe also the initial-final case once the proper jacobian factor is used.
We have then completed the list of all possible counterterms. A collection of all the relevant
formulas for all the countertems and their integrated version is given in Appendix~\ref{sec:masslessDS}.

\section{Applications}
\label{sec:applications}
In this Section we will show a small selection of examples where we apply
the construction presented in the previous sections.
For illustrative
purposes, in the present section we limit the discussion to a set of relatively simple processes.
In each case, a comparison is made among the results obtained using dual counterterms and the ones obtained using Catani--Seymour dipoles.

\subsection{\texorpdfstring{$\gamma^* \to 2$}{gamma* -> 2 } jets at NLO}
\label{sec:epem2j}
At the lowest order, the two-jet production in $e^+e^-$ collisions has just a quark-antiquark pair in the final state with momenta $p_1$ and $p_2$, respectively.
In the real sector, we have a single sub-process where an additional gluon is radiated with momentum $p'_3$, while the quark and the antiquark have momenta $p'_1$ and $p'_2$, respectively.
Therefore, the only emitter-spectator pair is the one composed by the quark and the antiquark, with the emitter role being played once by the quark and once by the antiquark.
This means that we need two dual counterterms: $\sigma^{D\,(1)}_{qg,\bar{q}}$ for the case where the emitter is the quark and $\sigma^{D\,(1)}_{\bar{q}g,q}$ when that role is played by the antiquark.

In the Monte Carlo implementation, once a real phase space configuration $(p'_1, p'_2, p'_3)$ is generated, the kinematic invariants $s_{13}$ and $s_{23}$ are compared.
If $s_{13} < s_{23}$ ($\mathcal{R}_1$ region), the integrand of the dual cross section $\sigma^{D\,(1)}_{qg,\bar{q}}$ (right-hand side of Eq.\eqref{def:fqg}) is evaluated in the virtual configuration given by the momentum mapping in Eq.\eqref{map:map1inv}.
Otherwise, if $s_{23} < s_{13}$ ($\mathcal{R}_2$ region), the integrand of $\sigma^{D\,(1)}_{\bar{q}g,q}$ is activated.
In any case then, the method utilizes a single counter event for each real phase space configuration.
	
In the virtual sector, we use the integrated dual counterterms of Section~\ref{sec:int_dual_sub}.
In particular, for both counterterms the result of the integration is equal to the right-hand side of Eq.\eqref{res:count_qg1_int}.
Since there are only a quark and an antiquark in the virtual sector, we have $\mathbf{T}_1 \mathbf{T}_2 = - C_F$.
Therefore, we can write
\begin{equation}
\sigma^{D\,(1)}_{q,\bar{q}} = \sigma^{D\,(1)}_{\bar{q},q} = - \frac{\alpha_S}{2\pi} \frac{(4 \pi)^{\varepsilon}}{\Gamma (1 - \varepsilon)} C_F \int d \Phi_2 \left( \frac{\mu^2}{s_{12}} \right)^{\varepsilon} | A^{(0)}_{q \bar{q}} |^2 \left[ \frac{1}{\varepsilon^2} + \frac{3}{2\varepsilon} +3 + 4 \log(2) -\frac{\pi^2}{2}  \right]
\end{equation}
where $| A^{(0)}_{q \bar{q}} |^2$ is the leading-order matrix element squared and we have neglected terms of order $\mathcal{O}(\varepsilon)$.

\begin{figure}[t]
	\centering
	\includegraphics[width=0.7\linewidth]{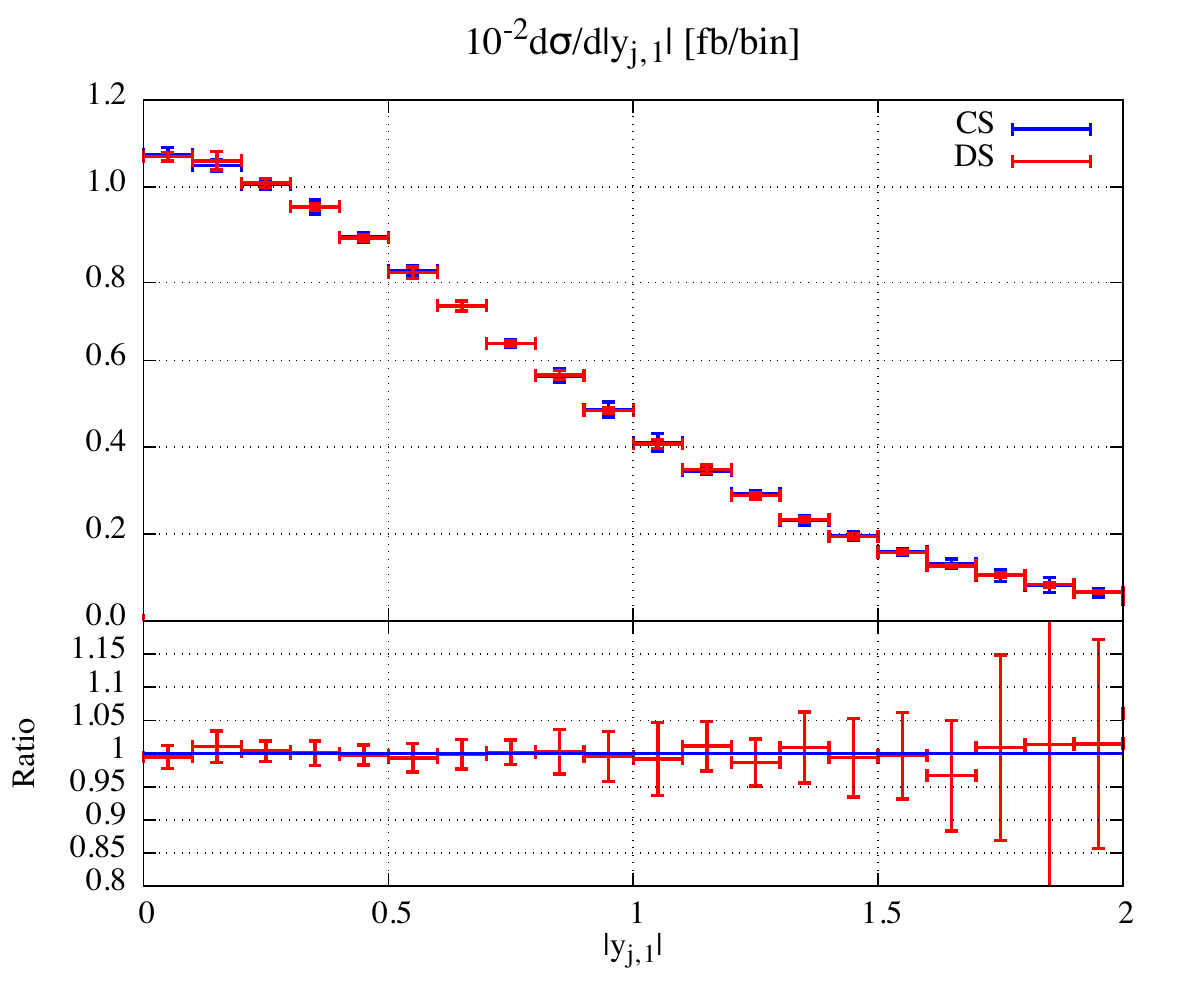}
	\caption{Rapidity distribution of the most energetic jet in $\gamma^* \to 2$ jets at NLO (Durham algorithm, $ y_\text{cut} = 0.1$, $\sqrt{s} = 125\, \text{GeV}$, $\alpha_S (\sqrt{s}) =0.11173799$). The Dual Subtraction (DS) results are shown in red, while the Catani--Seymour (CS) ones in blue.
	The error bars correspond to the statistical error given by the Monte Carlo integration.}
	\label{fig:epem2j_j1_rap}
	\end{figure}

We now show the result for a differential prediction.
We have chosen $\sqrt{s} = 125\, \text{GeV}$ as the energy in the center-of-mass frame.
The corresponding value of the strong coupling constant, $\alpha_S=0.11173799$, has been obtained starting from $\alpha_S(m_Z)=0.117$, $m_Z=91.1876\,$GeV being the mass of the $Z$ boson.
The jet algorithm used in the computation is the Durham
algorithm~\cite{Catani:1992ua} with $y_\text{cut}=0.1$.
Using this setup, in Fig.\eqref{fig:epem2j_j1_rap} we plot the differential cross section with respect to the rapidity of the most energetic jet.
The excellent agreement observed in Fig.\eqref{fig:epem2j_j1_rap} among the two results validates our procedure and is found by plotting any other differential variable.

\subsection{\texorpdfstring{$\gamma^* \to 3$}{gamma* -> 3} jets at NLO}

The basic sub-process contributing to the three-jet production in $e^+e^-$ annihilation into a virtual photon is $\gamma^* \! \rightarrow q(p_1) \, \bar{q}(p_2) \, g(p_3)$, while in the real sector one has the two sub-processes $\gamma^*  \rightarrow q(p'_1)\, \bar{q}(p'_2)\, g(p'_3)\, g(p'_4)$ and $\gamma^* \rightarrow q(p'_1)\, \bar{q}(p'_2)\, q'(p'_3)\, \bar{q}'(p'_4)$, where the two quark-antiquark pairs may have equal or different flavours.
For simplicity we will consider only the gluonic channel, where an additional gluon is radiated with respect of the virtual sector.
The contribution of the other sub-processes can be accounted for by similar considerations.

By inspecting the virtual sector, we find six emitter-spectator pairs for the process $\gamma^* \rightarrow q(p_1) \, \bar{q}(p_2) \, g(p_3)$, that are
$\{ q \bar{q}, qg,\bar{q} g\}$
and the three pairs where the emitter and the spectator switch.
The case of four quark production can be treated along the same lines.
Once a real kinematic configuration $(p'_1, p'_2, p'_3, p'_4)$ is generated, the six kinematic invariants $s_{12}$, $s_{13}$, $s_{14}$, $s_{23}$, $s_{24}$ and $s_{34}$ are analyzed and
six dual counterterms are activated.
To evaluate them, six virtual configurations $(p_1, p_2, p_3, q_1)$ are built by using six times the inverse mapping in Eq.\eqref{map:map1inv}, every time with the proper set of real and virtual momenta.
To each emitter-spectator pair we associate the following dual counterterms
\begin{align*}
\label{scheme:qqgg}
q \bar{q} \quad &\longrightarrow \quad\left\{ \quad
\begin{aligned}
&& g(p'_3)\,\, \text{as radiation:} &&\quad \mathcal{C}^{(1)}_{13,2} \quad \text{if}\,\, s_{13} < s_{23}, &&\quad \mathcal{C}^{(1)}_{23,1} \quad \text{if}\,\, s_{13} > s_{23}\\
&& g(p'_4)\,\, \text{as radiation:} &&\quad\mathcal{C}^{(1)}_{14,2} \quad\text{if}\,\, s_{14} < s_{24}, &&\quad\mathcal{C}^{(1)}_{24,1} \quad \text{if}\,\, s_{14} > s_{24}
\end{aligned} \right. \\
q g \quad &\longrightarrow \quad\left\{ \quad
\begin{aligned}
&& g(p'_3)\,\, \text{as radiation:} &&\quad\mathcal{C}^{(1)}_{13,4} \quad\text{if}\,\, s_{13} < s_{34}, &&\quad\mathcal{C}^{(1)}_{43,1} \quad \text{if}\,\, s_{13} > s_{34}\\
&& g(p'_4)\,\, \text{as radiation:} &&\quad\mathcal{C}^{(1)}_{14,3} \quad\text{if}\,\, s_{14} < s_{34}, &&\quad\mathcal{C}^{(1)}_{34,1}\quad \text{if}\,\, s_{14} > s_{34}
\end{aligned} \right. \\
\bar{q} g \quad &\longrightarrow \quad\left\{ \quad
\begin{aligned}
&& g(p'_3)\,\, \text{as radiation:} &&\quad\mathcal{C}^{(1)}_{23,4} \quad\text{if}\,\, s_{23} < s_{34}, &&\quad\mathcal{C}^{(1)}_{43,2} \quad \text{if}\,\, s_{24} > s_{34}\\
&& g(p'_4)\,\, \text{as radiation:} &&\quad\mathcal{C}^{(1)}_{24,3} \quad \text{if}\,\, s_{24} < s_{34}, &&\quad\mathcal{C}^{(1)}_{34,2} \quad \text{if}\,\, s_{24} > s_{34}
\end{aligned} \right. \numberthis
\end{align*}
where the $\mathcal{C}^{(1)}_{ac,b}$ is defined in Appendix~\ref{sec:counting} and corresponds to the sum of the $V^{(1)}_{ac,b}$ and $G^{(1)}_{ac,b}$ terms of our algorithm.

In the virtual sector, we use the integrated dual counterterms of Section~\ref{sec:int_dual_sub}.
Since there are three partons, a quark, an antiquark and a gluon, the colour algebra factorizes and one has $\mathbf{T}_1 \mathbf{T}_2 = C_A/2- C_F$ and $\mathbf{T}_1 \mathbf{T}_3 = \mathbf{T}_2 \mathbf{T}_3 = - C_A/2$.
Therefore, the integrated dual counterterms are given by
\begin{align*}
\sigma^{D\,(1)}_{q,\bar{q}} &\sim  \frac{\alpha_S}{2\pi} \frac{(4 \pi)^{\varepsilon}}{\Gamma (1 - \varepsilon)}  \left( \frac{C_A}{2} - C_F \right) \int d \Phi_2 \left( \frac{\mu^2}{s_{12}} \right)^{\varepsilon} | A^{(0)}_{q \bar{q} g} |^2 \left[ \frac{1}{\varepsilon^2} + \frac{3}{2\varepsilon} + 3 + 4 \log(2) -\frac{\pi^2}{2}  \right] \\
\sigma^{D\,(1)}_{\bar{q}, q}&\sim  \frac{\alpha_S}{2\pi} \frac{(4 \pi)^{\varepsilon}}{\Gamma (1 - \varepsilon)}  \left( \frac{C_A}{2} - C_F \right) \int d \Phi_2 \left( \frac{\mu^2}{s_{12}} \right)^{\varepsilon} | A^{(0)}_{q \bar{q} g} |^2 \left[ \frac{1}{\varepsilon^2} + \frac{3}{2\varepsilon} + 3 + 4 \log(2) -\frac{\pi^2}{2}  \right] \\
\sigma^{D\,(1)}_{q, g} &\sim - \frac{\alpha_S}{2\pi} \frac{(4 \pi)^{\varepsilon}}{\Gamma (1 - \varepsilon)} \frac{C_A}{2} \int d \Phi_2 \left( \frac{\mu^2}{s_{13}} \right)^{\varepsilon} | A^{(0)}_{q \bar{q} g} |^2 \left[ \frac{1}{\varepsilon^2} + \frac{3}{2\varepsilon} + 3 + 4 \log(2) -\frac{\pi^2}{2}  \right] \\
\sigma^{D\,(1)}_{g,q} &\sim - \frac{\alpha_S}{2\pi} \frac{(4 \pi)^{\varepsilon}}{\Gamma (1 - \varepsilon)} \frac{C_A}{2}  \int d \Phi_2 \left( \frac{\mu^2}{s_{13}} \right)^{\varepsilon} | A^{(0)}_{q \bar{q} g} |^2 \left[ \frac{1}{\varepsilon^2} + \frac{11}{6 \varepsilon} + \frac{55}{18} + \frac{14}{3} \log(2) -\frac{\pi^2}{2} \right] \\
\sigma^{D\,(1)}_{\bar{q},g} &\sim - \frac{\alpha_S}{2\pi} \frac{(4 \pi)^{\varepsilon}}{\Gamma (1 - \varepsilon)} \frac{C_A}{2}  \int d \Phi_2 \left( \frac{\mu^2}{s_{23}} \right)^{\varepsilon} | A^{(0)}_{q \bar{q} g} |^2 \left[ \frac{1}{\varepsilon^2} + \frac{3}{2\varepsilon} + 3 + 4 \log(2) -\frac{\pi^2}{2}  \right] \\
\sigma^{D\,(1)}_{g, \bar{q}} &\sim - \frac{\alpha_S}{2\pi} \frac{(4 \pi)^{\varepsilon}}{\Gamma (1 - \varepsilon)} \frac{C_A}{2}  \int d \Phi_2 \left( \frac{\mu^2}{s_{23}} \right)^{\varepsilon} | A^{(0)}_{q \bar{q} g} |^2 \left[ \frac{1}{\varepsilon^2} + \frac{11}{6 \varepsilon} + \frac{55}{18} + \frac{14}{3} \log(2) -\frac{\pi^2}{2}  \right] \numberthis
\end{align*}
where $| A^{(0)}_{q \bar{q} g} |^2$ is the Born amplitude for the process $\gamma^* \rightarrow q \bar{q} g$ and we have neglected terms of order $\mathcal{O}(\varepsilon)$.

\begin{figure}[t]
	\centering
	\includegraphics[width=0.7\linewidth]{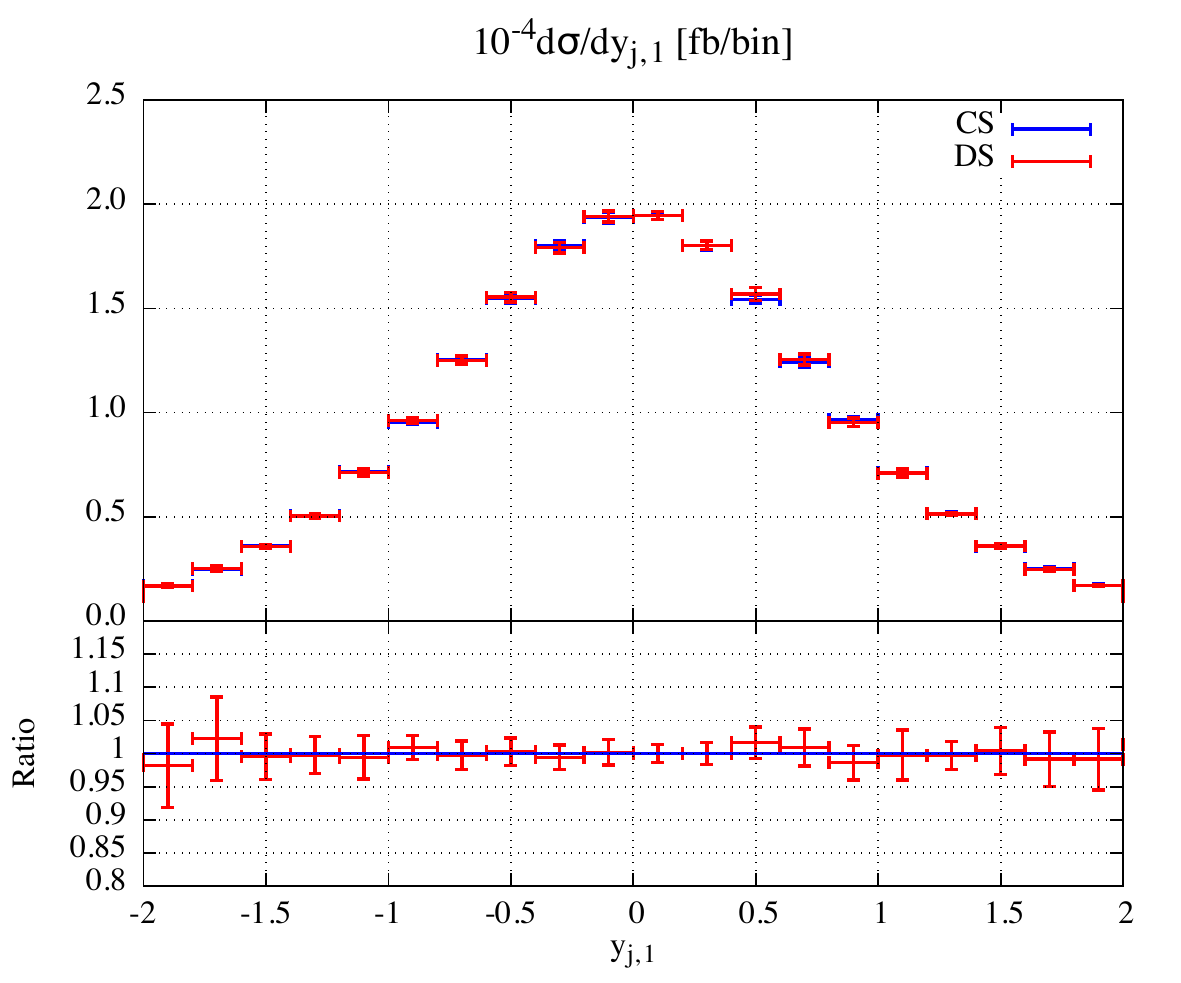}
	\caption{Rapidity distribution of the most energetic jet in $\gamma^* \to 3$ jets  (Durham algorithm, $ y_\text{cut} = 0.05$, $\sqrt{s} = 125\, \text{GeV}$, $\alpha_S(\sqrt{s})=0.11173799$). Only the NLO corrections are plotted. The Dual Subtraction (DS) results are shown in red, while the Catani--Seymour (CS) ones in blue.
	The error bars correspond to the statistical error given by the Monte Carlo integration.}
	\label{fig:epem3j_cor_j1_rap}
	\end{figure}
In Fig.(\ref{fig:epem3j_cor_j1_rap}) we show the differential cross section with respect to the rapidity of the most energetic jet using again $\sqrt{s} = 125\, \text{GeV}$ as the energy in the center-of-mass frame, $\alpha_S(\sqrt{s})=0.11173799$ (starting from $\alpha_S(m_Z)=0.117$) and the Durham jet algorithm with $y_\text{cut}=0.05$.
Also in this case, we observe a perfect agreement with the same computation performed using Catani--Seymour dipoles.
We refrain to include other plots because the same level of agreement is observed for any other differential distribution.

\subsection{\texorpdfstring{$H \to b\,\bar{b}$}{H -> b b-bar} at NLO}
In the present Section, we will make of use the formulas for the massive case derived in Section~\ref{sec:masses} to study the Higgs boson decay into bottom quarks.
The virtual sector process is $H \rightarrow b(p_1)\, \bar{b} (p_2)$, while in the real sector an addition gluon is emitted, $H \rightarrow b(p'_1)\, \bar{b} (p'_2)\,g(p'_3)$.
Therefore, the counting of the dual counterterms is the same as for the process $\gamma^* \rightarrow 2$ jets analyzed in Section~\ref{sec:epem2j}.

\begin{figure}[t]
	\centering
	\includegraphics[width=0.7\linewidth]{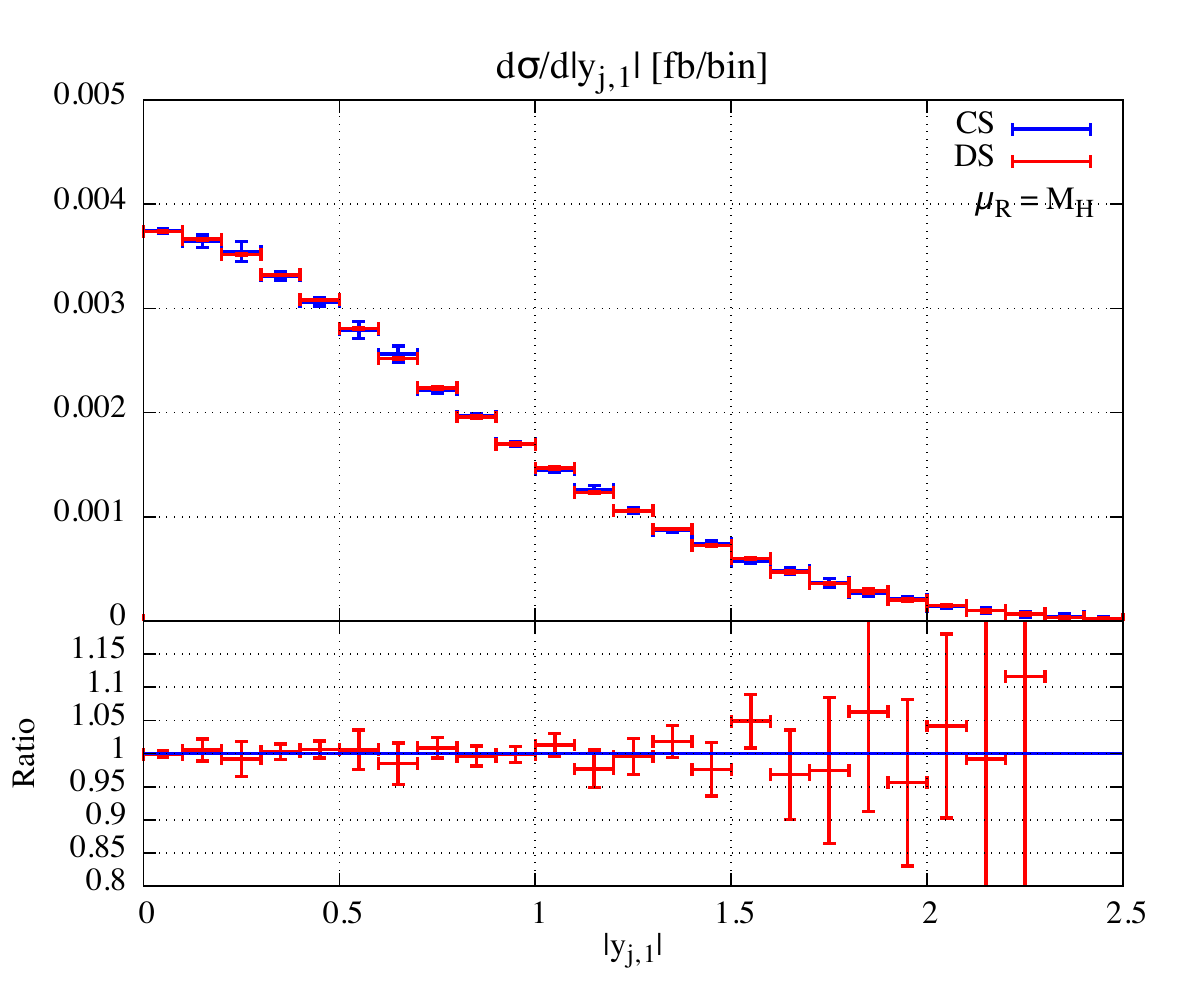}
	\caption{Rapidity distribution of the most energetic jet in $H \to b\,\bar{b}$ at NLO. (Durham algorithm, $ y_\text{cut} = 0.01$, $m_b=4.78  \, \text{GeV}$, $m_H= 125.09 \, \text{GeV}$, $\alpha_S(m_H)=0.11263619$). The Dual Subtraction (DS) results are shown in red, while the Catani--Seymour (CS) ones in blue.
	The error bars correspond to the statistical error given by the Monte Carlo integration.}
	\label{fig:Hbb_j1_rap}
	\end{figure}
Once a real phase space configuration is generated, the proper counterterm in Eq.\eqref{def:count_mass_qg} is selected comparing the $s_{13}$ and $s_{23}$ invariants.
As for the virtual contribution, we have renormalized the mass of the wave--function of the bottom quark on-shell.
The virtual singularities are canceled by the integrated dual counterterms of Section~\ref{sec:masses_int_count} with $\mathbf{T}_1 \mathbf{T}_2 = - C_F$.

The differential distribution of the rapidity of the most energetic jet is plotted in Fig.\eqref{fig:Hbb_j1_rap}.
We have used $m_H= 125.09 \, \text{GeV}$ for the Higgs boson mass and $\alpha_S(m_H)=0.11263619$ corresponding to $\alpha_S(m_Z)=0.118$.
The on-shell mass of the bottom quark has been set to $m_b=4.78  \, \text{GeV}$.
The jet algorithm used in the computation is the Durham algorithm with $y_\text{cut}=0.01$.
A fairly good agreement with the computation performed using Catani--Seymour dipoles is observed.

\subsection{Drell-Yan pair production plus 0 or 1 jet at NLO}
In this Section we consider proton-proton collisions where a $W^+$ boson is produced. We start by considering the Born process with no further radiation.
In the virtual sector the basic sub-process is $q (p_1)\, \bar{q} (p_2) \rightarrow W^+ (p_3)$, while in the real sector we have $q (p'_1)\, \bar{q} (p'_2) \rightarrow  W^+ (p'_3) \, g(p'_4)$.
Since there are two partons in the initial state, we apply the method shown in Section~\ref{sec:initial_state}.
For the sake of comparison and for simplicity, we limit here to the quark-antiquark initiated sub-processes.

\begin{figure}[t]
	\centering
	\includegraphics[width=0.7\linewidth]{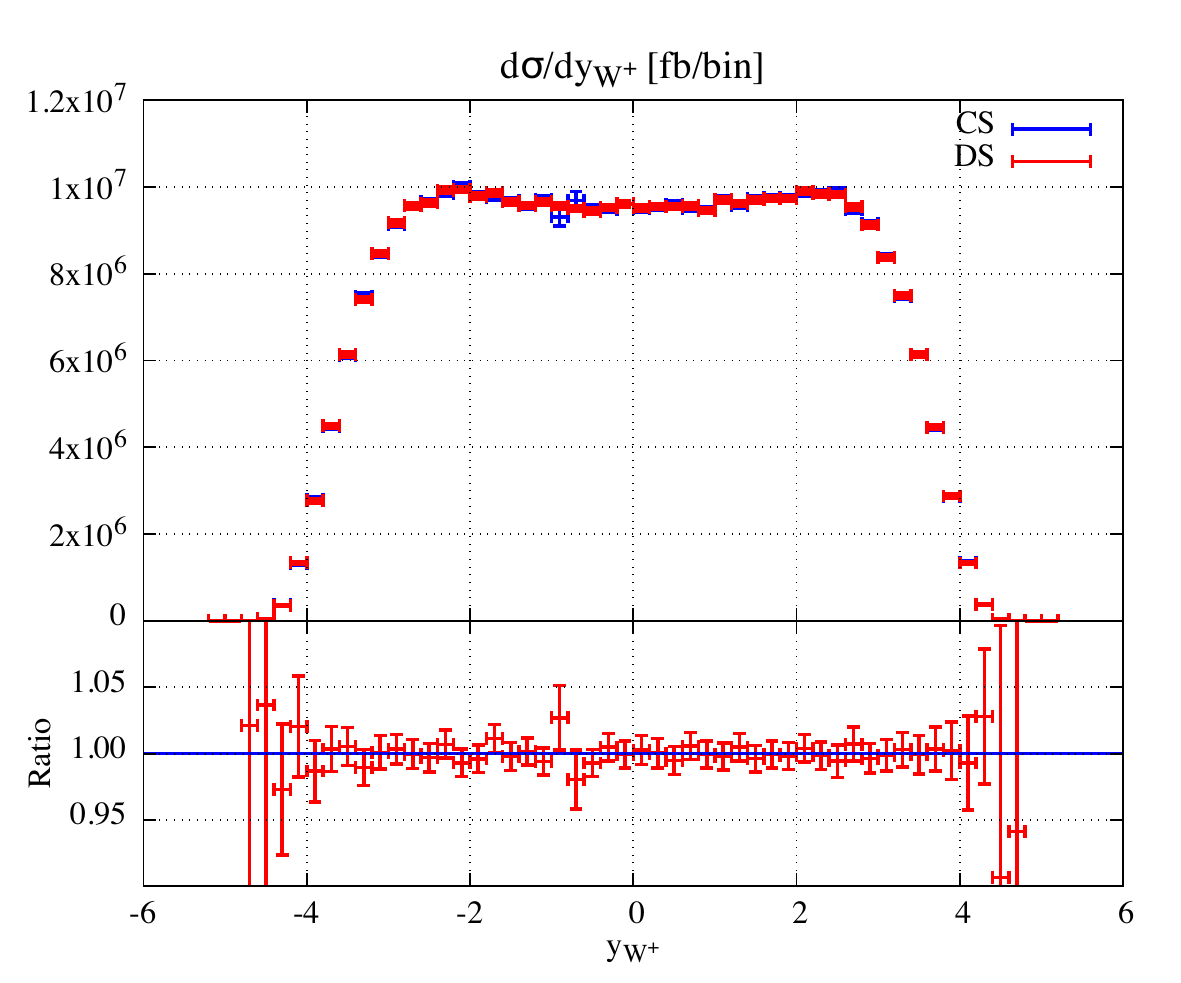}
	\caption{Rapidity distribution of the $W^+$ boson in proton-proton collisions at $8 \,\text{TeV}$ center-of-mass energy ($q\bar{q}$-channel only). The renormalization and the factorization scale are both equal to $\mu=80 \, \text{GeV}$, while $\alpha_S(\mu)=0.12264887$, $m_W= 80.398 \, \text{GeV}$ and $G_F= 1.16639 \cdot 10^-5 \, \text{GeV}^{-2}$.
	The Dual Subtraction (DS) results are shown in red, while the Catani--Seymour (CS) ones in blue.
	The error bars correspond to the statistical error given by the Monte Carlo integration.}
	\label{fig:pp_Wp_rap}
	\end{figure}
In the Monte Carlo implementation, we generate a real phase space configuration $(p'_3,p'_4)$ and compare the kinematic invariants $s_{14}$ and $s_{24}$.
If $s_{14} < s_{24}$, we apply the mapping in Eq.\eqref{map:init1}, with $(p'_a, p'_b, p'_c) = (p'_1, p'_2, p'_3)$, to obtain a virtual configuration $(p_1, p_2, q_1)$.
Then, we evaluate the integrand of the dual counterterm $\sigma^{D\,(1)}_{qg,\bar{q}}$ in Eq.\eqref{def:iqg} and we add it to the real cross section.
If $s_{14} > s_{24}$, we use the mapping in Eq.\eqref{map:init1} with $(p'_a, p'_b, p'_c) = (p'_2, p'_1, p'_3)$, 
instead.
In the virtual sector we use the integrated dual counterterms of Section~\ref{sec:initial_state}, with $\mathbf{T}_1 \mathbf{T}_2 = - C_F$.

In Fig.(\ref{fig:pp_Wp_rap}) we show the differential cross section with respect to the rapidity of the $W^+$ boson.
We consider proton-proton collisions at $8 \,\text{TeV}$ in the center-of-mass frame and we have used the MSTW2008NLO set of parton distribution functions with $\alpha_S(m_Z)=0.12018$.
The renormalization and the factorization scale have been set equal to $\mu_R = \mu_F = \mu = 80 \, \text{GeV}$ and we have used $m_W= 80.398 \, \text{GeV}$ for the $W^+$ boson mass, $G_F= 1.16639 \cdot 10^-5 \, \text{GeV}^{-2}$ for the Fermi constant and $\alpha_S(\mu)=0.12264887$ for the strong coupling constant. 
Also in this case, we observe an excellent agreement among the computation performed using dual subtractions and the one obtained with Catani--Seymour dipoles.

\begin{figure}[t]
	\centering
	\includegraphics[width=0.49\linewidth]{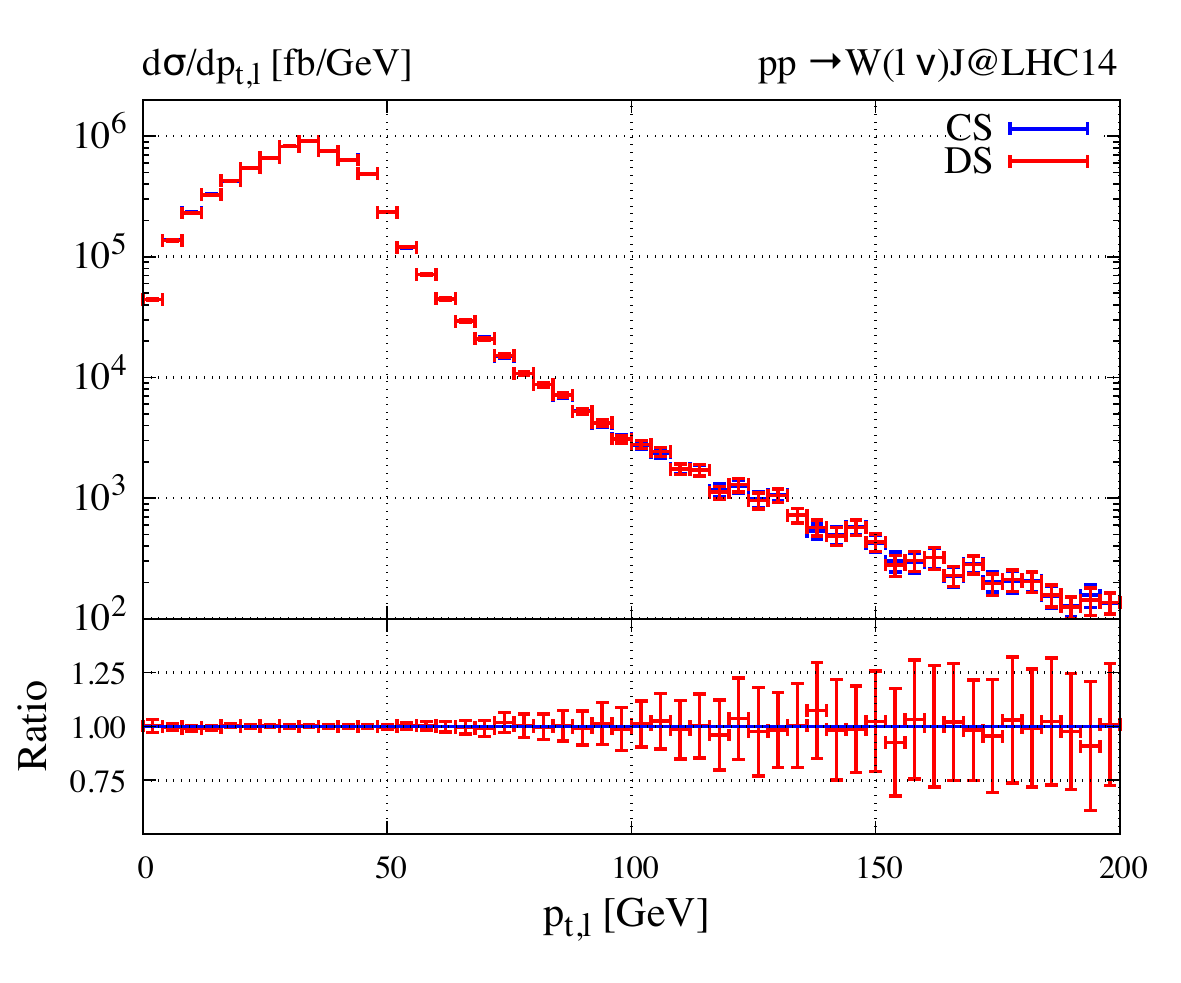}
    \includegraphics[width=0.49\linewidth]{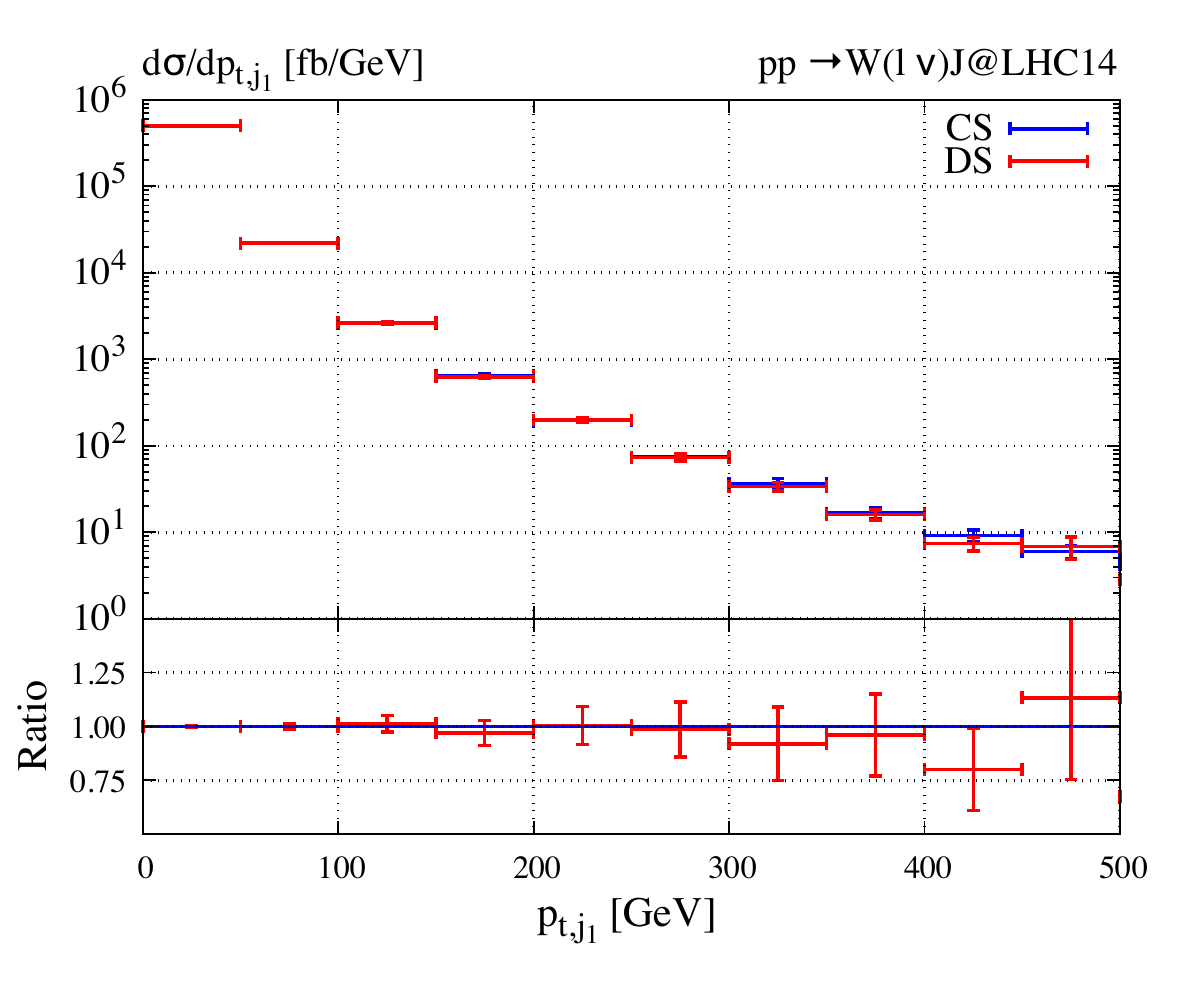}
	\caption{Charged lepton (left) and hardest jet (right) transverse momentum distributions for $W^+$ plus 1 jet production in proton-proton collisions at $14 \,\text{TeV}$ center-of-mass energy ($q\bar{q}$-channel only). The renormalization and the factorization scale are both equal to $\mu=80 \, \text{GeV}$.
	The Dual Subtraction (DS) results are shown in red, while the Catani--Seymour (CS) ones in blue.
	The error bars correspond to the statistical error given by the Monte Carlo integration.}
	\label{fig:w1jplots}
	\end{figure}
The case in which there is radiation also at Born level, beyond the initial-initial counterterms, requires the computation also of initial-final and final inital counterterms. For each possible emitter spectator pair the two kinematic invariants are computed and compared. The counterterm to subtract correspond to the smaller of the two invariants.
In Fig.(\ref{fig:w1jplots}) we show the lepton and
leading jet transverse momentum distributions respectively. Jets are formed using the antikt algorithm with resolution paramenter $R=0.5$.
Only the $q\bar{q}$ channel is included.

\subsection{Higgs boson production in gluon fusion plus 0 or 1 jet at NLO}
We consider the production of an Higgs boson
in proton-proton collision in the framework of the effective field theory where the heavy top quark degree of freedom that mediates the interaction among the Higgs boson and gluons has been integrated out.
At the Born and virtual sector the basic sub-process is then $g (p_1)\, g (p_2) \rightarrow H (p_3)$, while in the real sector we have $g (p'_1)\, g (p'_2) \rightarrow  H (p'_3) \, g(p'_4)$.
For simplicity, we limit here to a theory with only gluons, so that these are the only two partonic subprocesses for a NLO calculation.
Since the Born process has a colourless final state and presents two partons in the initial state, we apply the method shown in Section~\ref{sec:initial_state}.
\begin{figure}[t]
	\centering
	\includegraphics[width=0.7\linewidth]{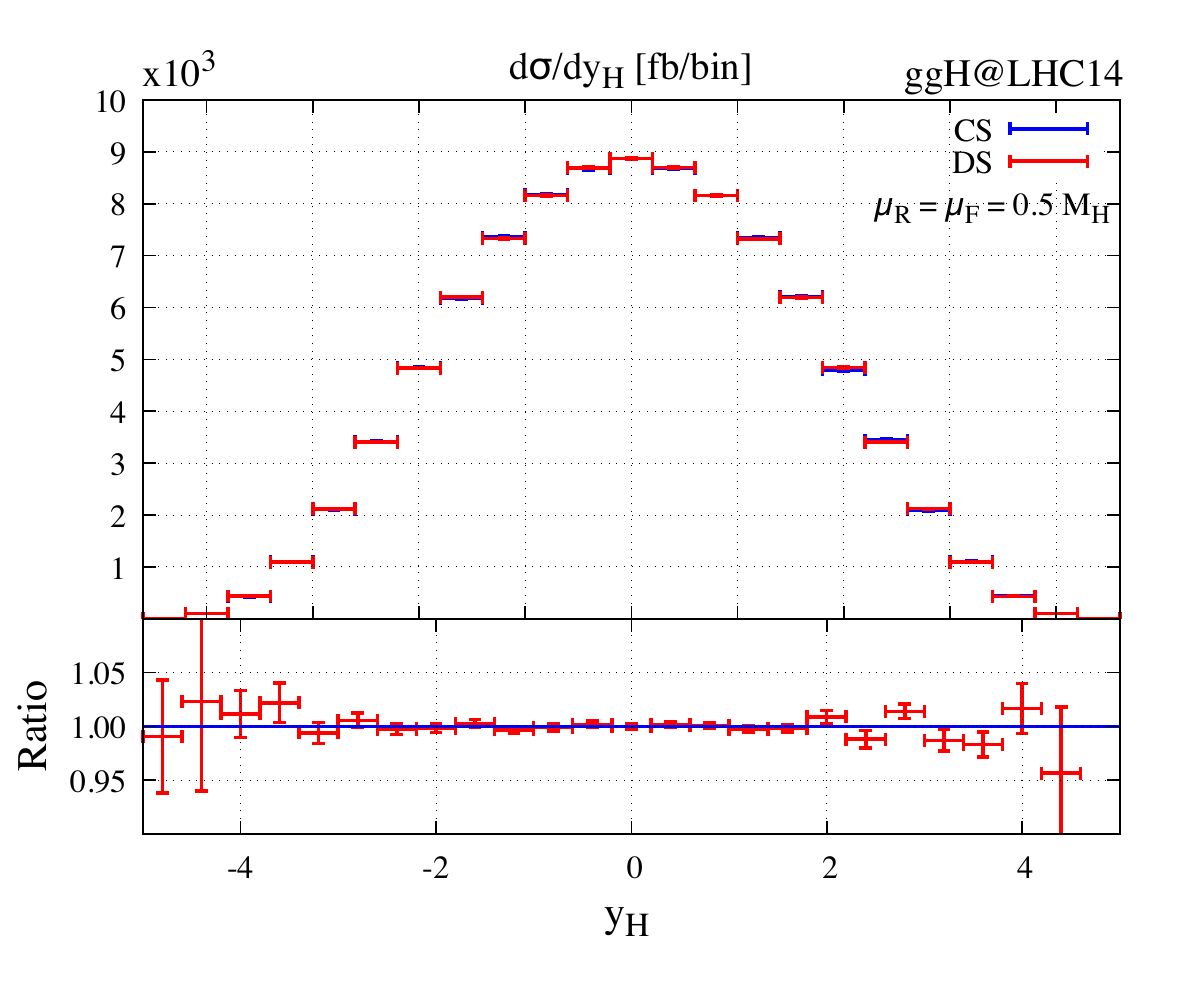}
	\caption{Rapidity distribution of the Higgs boson in proton-proton collisions at $14 \,\text{TeV}$ center-of-mass energy ($gg$-channel only). The renormalization and the factorization scale are both equal to $\mu=m_H/2=62.5 \, \text{GeV}$.
	The Dual Subtraction (DS) results are shown in red, while the Catani--Seymour (CS) ones in blue.
	The error bars correspond to the statistical error given by the Monte Carlo integration performed with the same setup.}
	\label{fig:pp_H_rap}
	\end{figure}
The mapping and the rest of the procedure is precisely the one already discussed for the case of the W boson production with the only difference that for the present case we use the formulas for a gluon emitting a gluon.	
In Fig.(\ref{fig:pp_H_rap}) we show the rapidity distribution of the Higgs boson produced at LHC14. We have set $m_H=125\,$GeV for the Higgs boson mass and used the ${\rm NNPDF31\_nnlo\_as\_0118\_luxqed}$ set of parton distribution functions with $\alpha_S(m_Z)=0.11800216$. 
The renormalization and the factorization scale have been set equal to $\mu_R = \mu_F = m_H/2$. 
Once again, we observe excellent agreement among the computation performed using dual subtractions and the one obtained with Catani--Seymour dipoles.

\begin{figure}[t]
	\centering
	\includegraphics[width=0.49\linewidth]{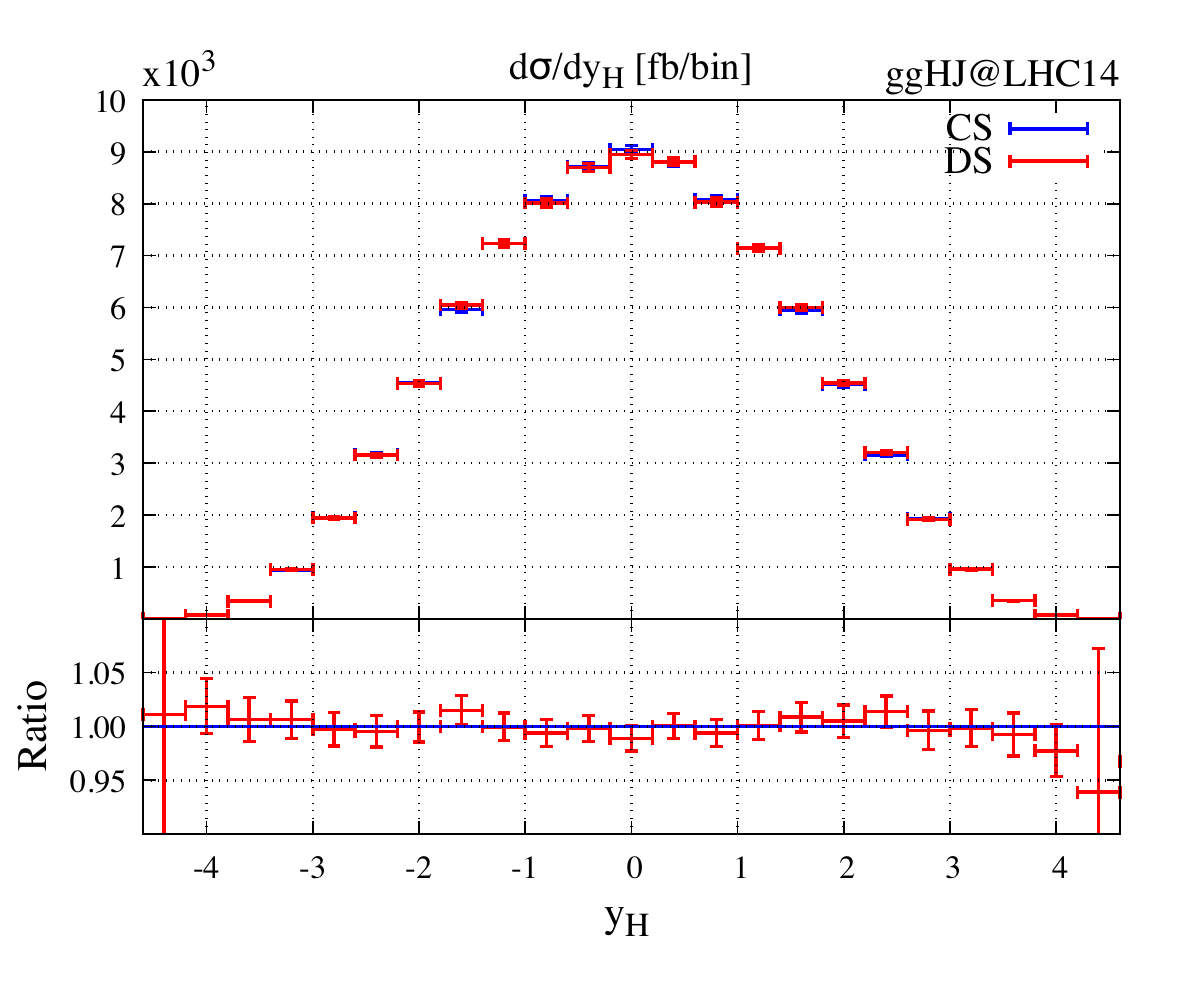}
    \includegraphics[width=0.49\linewidth]{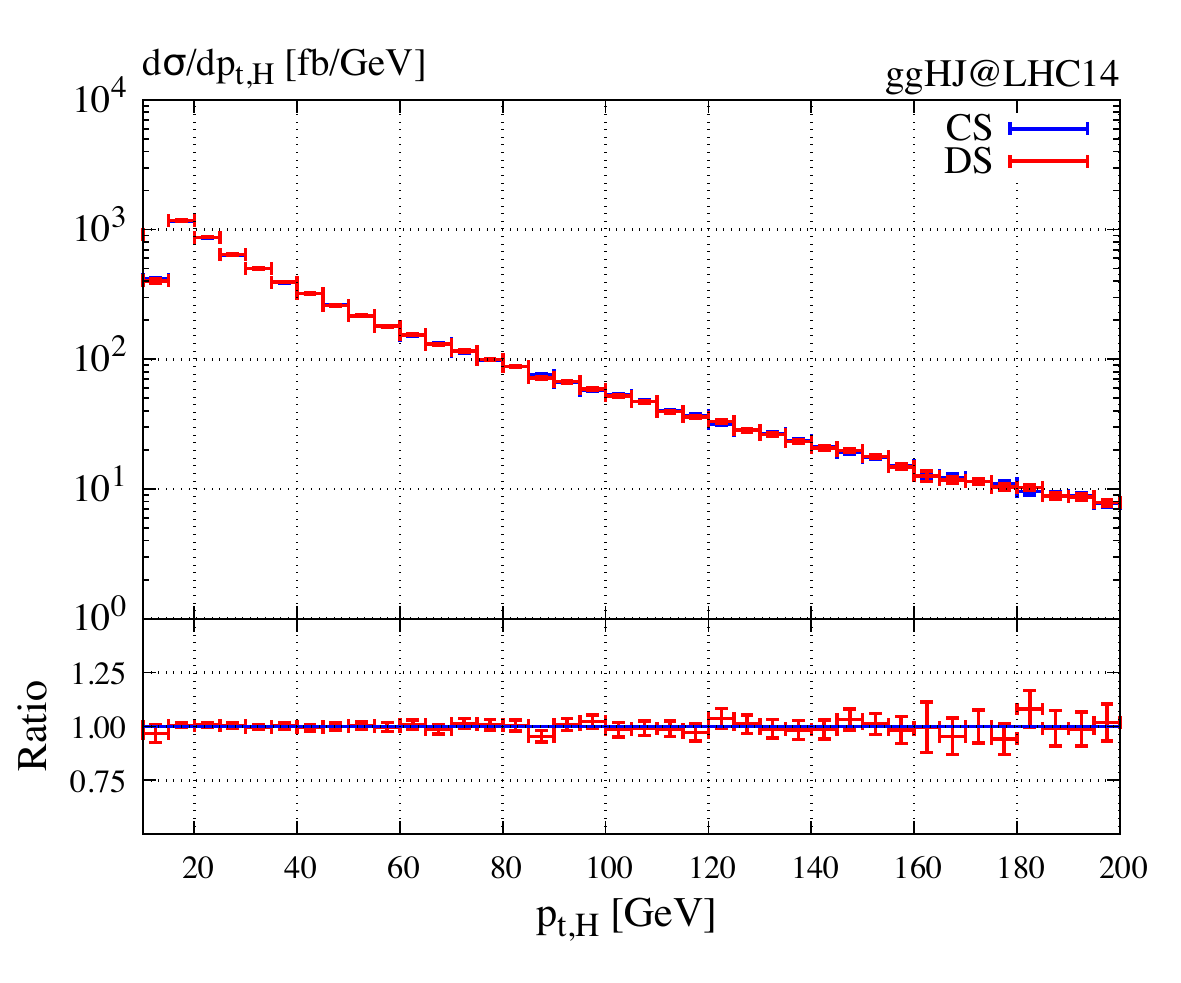}
	\caption{Higgs boson rapidity (left) and transverse momentum (right) distributions for $H$ plus 1 jet production in proton-proton collisions at $14 \,\text{TeV}$ center-of-mass energy ($gg$-channel only).
	The Dual Subtraction (DS) results are shown in red, while the Catani--Seymour (CS) ones in blue.
	The error bars correspond to the statistical error given by the Monte Carlo integration.}
	\label{fig:h1jplots}
	\end{figure}
Finally, we consider the process in which one hard radiation is present already at Born level. In this case the basic subprocess is $g (p_1)\, g (p_2) \rightarrow H (p_3) g(p_4)$ and the real subprocess
contains one further radiated gluon. Using the
formulas for initial-initial, initial-final and final-initial gluon radiation from a gluon we cancel locally all the singularities obtaining fully differential predictions.
In Fig.(\ref{fig:h1jplots}) we show the Higgs boson
rapidity (left) and transverse momentum distributions (right) respectively. Again we observe a perfect agreement with the same computation performed using Catani--Seymour dipoles.

\section{Conclusion}
\label{sec:conclusions}
Starting from the analysis of the divergences of one-loop amplitudes reduced to scalar integrals, in the present paper we have built a subtraction formalism exploiting the LTD theorem.
Beyond the singularities of the virtual diagrams, we have also to take into account the ones brought by the wave--function renormalization.
An integrand--level representation for these counterterms suitable for massless and massive fermions can be found in~\cite{Sborlini:2016gbr} and~\cite{Sborlini:2016hat}, respectively.
In the present work we provide an integrand--level representation of the gluon wave--function renormalization constant.
Furthermore, our construction is naturally extended to the case of initial state singularities.
Starting from the dual representation of the divergent part of virtual contribution, we cancelled the singularities of the real contribution by simply feeding the former with the mapped momenta of the latter and multiplying times the momentum fraction $x$.

We now expand a bit more on this last point.
Let us consider a process in which a colourless 
final state is produced and, to further simplify the discussion,
let's take the production of a pair or leptons
from $q\bar{q}$ annihilation into a virtual photon, so ignoring the contribution of the $Z$ boson.
Normally, when calculating higher--order 
contributions in a proton-proton collision
one starts from a leading order process
producing a final state with a given
invariant mass $s_{ij}$. This might seems a
complication in our case, but for the sake
of the discussion, let's consider that our
experiment is capable to see pairs
of leptons only in a well defined narrow
window of the invariant mass of the pair
at the time. Once a value for the invariant
mass is chosen we compute the Born, the virtual
and the real contribution to the scattering
process. It is well known that requiring a fixed
final state invariant mass for the lepton pair
the singularities will not cancel. Living
on different phase spaces, $s_{ij}$ for the
virtual and $s_{ab}>s_{ij}$ for the real, they cannot match.
Thanks to the remarkable property of 
factorization, this has been solved through a 
redefinition of the initial state in the form of
a renormalization of the parton distribution
functions.
Let us now take a slightly different point of 
view. We could define the corrections, both
real and virtual, to be initiated by the same
center of mass energy $s_{ab}$. The real part
is precisely as before, but the virtual part
is not contributing at all if $s_{ab}$ is
greater then $s_{ij}$.
By the way, the 
divergences cancel if we consider 
$s_{ab}=s_{ij}$.
The only difference now is that, if we have
$s_{ab}>s_{ij}$ and cut the loop amplitude in such a way to produce a
gluon that is not soft, then the invariant
mass of the leptons in the Born system will be lower then
$s_{ab}$ and could be tuned to be the value $s_{ij}$ of 
the Born process. By doing so (and also
multiplying times $x$) we observe
the cancellation of the singularities.
Of course, this way of cutting the loops
produces a result that does not correspond to 
the original loop integral living on $s_{ab}$,
but it represents a generalization of the
loop integral living on $s_{ij}$ for $x<1$.
In the present work we have exploited this
feature that integrating such cut
virtual loops, originally living on $s_{ab}$, suffices
to cancel both the loop divergences at $s_{ij}$ 
and the collinear initial state singularities.
This fact provides a general strategy to use
Loop--Tree Duality to compute cross sections
with hadrons in the initial state.
In fact our procedure can be applied also to the
case of full amplitudes that are not reduced to 
scalar integrals. In practice, for the cases studied in the previous section,
such as Drell-Yan and Higgs production in gluon fusion, the divergent part of the virtual
contribution that we have extracted, differs very little (by finite terms and just one IR finite 
bubble) from the whole renormalized virtual contribution.\\
\indent Acting such generalized virtual as a counterterm that cancels the IR divergences
of the real part, there seems to be no room for 
an explicit presence of the factorization scale.
Of course, we want to take benefit of parton 
evolution so the question is what is the 
factorization scale to be used in PDF's if we use a dual version of the (extended and cut) virtual contribution in the sense discussed above.
This issue will be discussed elsewhere but we anticipate here a simple analytic strategy to answer to this question. It consists in deducing
the corresponding factorization scale
through the comparison among results obtained
by performing the NLO computation in
the two ways, using the cut extended loops on
one side and the dipole subtractions plus
the universal collinear counterterms (which bring the
explicit presence of the factorization scale in the hard
cross section) on the other.\\
\indent On a technical ground, the structure of the dual
counterterms is very close to the Catani--Seymour
dipoles.
In fact, even if we count a lower number of 
counter events per real phase space point with 
respect to the Catani--Seymour construction, in 
the latter it is possible to reduce the number 
of active dipoles by restricting the phase space
region where they are computed to the 
neighbourhood of the singularities.
On a more formal ground, the dual subtraction 
builds a direct link for the cancellation among 
virtual and real singularities.\\
\indent We conclude by adding a few comments.
The extension of the formalism presented here to the case of NNLO corrections
is an interesting path that will be addressed elsewhere.
We note that many ingredients are already available~\cite{Bierenbaum:2012th, Aguilera-Verdugo:2019kbz, Aguilera-Verdugo:2020nrp,TorresBobadilla:2020ekr}, including examples of full NNLO computation using LTD, momentum mappings and master integrals.
Further applications of the work of the present paper would be the implementation of novel shower and matching schemes based on the dual subtractions discussed here.
These subjects are beyond the aim of the present work and are left for future investigations.

\appendix
\section{Counting of the dual counterterms}
\label{sec:counting}

The computational scheme proposed in the current work exploits the dual representation of the loop amplitude with $m$ partons in the final state in order to cancel soft and collinear singularities of the real matrix element living on the $(m+1)$-parton phase space.
The subtractions are extracted from the virtual sector on the base of the number of emitter-spectator pairs in its $m$-parton phase space.
Our starting point is represented by a set of dual counterterms $ \mathcal{C}_{i,j}$ (with $i,j=1,\dots,m$ and $i \neq j$), each associated with two functions $V_{i,j}$ and $G_{i,j}$ that depend on the emitter-spectator pair $(p_i, p_j)$.
These are constructed by application of LTD to the scalar three- and two-point functions, $C_0 (p_i, p_j)$
and $B_0(p_i)$, respectively, and the wave--function renormalization counterterm $\Delta Z (p_i)$.
In particular, $\mathcal{C}_{ij}$ has the general structure in terms of the dual $V$ and $G$
\begin{equation}
\mathcal{C}_{ij} (p_i,p_j) \equiv \,_m\langle 1, \dots, m| \mathbf{T}_{i} \mathbf{T}_j  \left(  V_{i,j} (p_i, p_j) + G_{i,j} (p_i,p_j) \right)  |1, \dots, m \rangle_m
\end{equation}
The dual cross section is obtained by summing over all the possible emitter-spectator pairs
\begin{equation}
\label{def:dual_sub2}
\sigma^{D} \equiv \mathcal{N}_{in} \sum_{\{ m\} } \int d \Phi_{m} \, \frac{1}{S_{\{m\}}} \sum_{i} \sum_{j \neq i} \mathcal{C}_{i,j}(p_1, \dots, p_m)
\end{equation}
where $\mathcal{N}_{in}$ includes all the non-QCD factors, $\sum_{\{ m\} }$ indicates a sum over all the $m$-parton configurations and $S_{\{m\}}$ is the Bose symmetry factor for identical partons.
The sum over the different emitters on the right-hand side of Eq.\eqref{def:dual_sub2} can be split into sums over different flavours
\begin{equation}
\label{eqn:sum_flav}
\sum_i \sum_{j \neq i} \dots  = \sum_{i=q}  \sum_{j \neq i} \dots+ \sum_{i=\bar{q}}  \sum_{j \neq i} \dots+ \sum_{i=g}  \sum_{j \neq i}\dots \,\,.
\end{equation}
Then, we need to map the emitter-spectator pairs of the virtual sector into the dipoles of the real sector.
Each pair $(i,j)$ can be mapped in more than a single dipole $(a,c,b)$, where parton $ac$ is the emitter, $c$ is the radiation and $b$ is the spectator.
Therefore, to distinguish the case in which the pair $(i,j)$ is linked to a specific dipole $(a,c,b)$, we use the following notation for the momentum mappings of Section~\ref{sec:maps}
\begin{align*}
\label{map:singular_regions}
\Phi_m &= \mathbf{M}^{i,j}_{a,c,b}(\Phi_{m+1}) \\
q_k &= \mathbf{M}^k_{a,c,b}(\Phi_{m+1}) \,\,. \numberthis
\end{align*}
The application of Eq.\eqref{map:singular_regions} in the corresponding dual counterterm leads to
\begin{equation}
\label{eqn:map_appl}
\mathcal{C}_{ac,b} \equiv \mathcal{C}_{i,j} (\mathbf{M}^{i,j}_{a,c,b}(\Phi_{m+1})) \,\,.
\end{equation}
We now want to use Eq.\eqref{eqn:map_appl} and count the number of counterterms through the number of dipoles in the real sector, by transforming each sum on the right-hand side of Eq.\eqref{eqn:sum_flav} in the following way
\begin{equation}
\label{eqn:sum_change}
\sum_{i=f} \sum_{j \neq i}  = \frac{\#(f)_m}{\#\{a,c\}_{m+1}} \sum_{a<c} \sum_{b \neq a,c}
\end{equation}
where $\#(f)_m$ denotes the number of partons with flavour $f$ in the $m$-parton configuration and $\#\{a,c\}_{m+1}$ the number of pairs $\{a,c\}$ in the $(m+1)$-parton configuration such that the emitter parton $ac$ has flavour $f$.
Given an $m$-parton configuration, all the possible $(m+1)$-parton processes can be obtained either by increasing by one the number $m_g$ of gluons, or by decreasing by one $m_g$ and adding a quark-antiquark pair.
Let us focus on the first case, which includes all the dual counterterms beside the quark contributions to the gluon wave--function renormalization.
The latters are required in the second class of $(m+1)$-parton processes, treated at the end of this Appendix.
Therefore, for the moment we assume that the counterterms $\mathcal{C}_{g,j}$ only include gluon and ghost contribution to the gluon renormalization. We have
\begin{align*}
\label{eqn:pairs}
\frac{\#(q_f)_m}{\#\{q_f,g\}_{m+1}} &= \frac{m_f}{m_f (m_g+1)} = \frac{1}{(m_g+1)} \\
\frac{\#(\bar{q}_f)_m}{\#\{\bar{q}_f,g\}_{m+1}} &= \frac{\bar{m}_f}{\bar{m}_f (m_g+1)} = \frac{1}{(m_g+1)} \\
\frac{\#(g)_m}{\#\{g,g\}_{m+1}} &= \frac{m_g}{m_g (m_g+1)/2} = \frac{2}{(m_g+1)} \numberthis
\end{align*}
$m_q$ and $m_{\bar{q}}$ being the number of quark and antiquarks of a given flavour, respectively.
The counting in Eq.\eqref{eqn:sum_change} is possible only if the integral of $\mathcal{C}_{i,j}$ over the $m$-particle phase space does not change when we exchange parton $i$ with a parton of the same flavour.
This condition is certainly fulfilled if we use the \textit{$q_1$-cut method} and so always set on-shell the loop momentum propagator connecting the external partons ($q_1$).
However, particular attention must be payed to this point if we want to use the \textit{$q_1$-$q_2$-method}.
In fact, if $a$ and $a'$ had the same flavour and we used $\sigma^{D\,(1)}_{ac,b}$ to describe the radiation collinear to $a$ and $\sigma^{D\,(2)}_{bc,a'}$ for the emission collinear to $a'$, then the dual subtractions would differ among each other.
As a solution to this problem we can perform, for each dual subtraction, the following symmetrization 
\begin{align*}
\label{eqn:symmetrization}
\sigma^{D\,(1)}_{ac,b} &\rightarrow \frac{1}{2} \left( \sigma^{D\,(1)}_{ac,b} + \sigma^{D\,(2)}_{bc,a} \right) \\
\sigma^{D\,(2)}_{ac,b} &\rightarrow \frac{1}{2} \left( \sigma^{D\,(1)}_{bc,a} + \sigma^{D\,(2)}_{ac,b} \right) \numberthis
\end{align*}
which ensures that the dual subtractions depend only on the flavours of the partons in the associated dipoles.
Eq.\eqref{eqn:symmetrization} may be derived from a different prescription for the application of LTD, which can be summarized as follows: when we consider the generic contribution $\mathcal{A}$ (be it a triangle $C_0$, a bubble $B_0$ or a wave--function renormalization counterterm $\Delta Z$) from a given emitter-spectator pair $(i,j)$ of the virtual sector, we first use the identity $\mathcal{A} = \mathcal{A}/2 + \mathcal{A}/2$; then, for the first term, we apply LTD as usual while, for the second term, we denote the internal momenta e we apply LTD as if we had considered the pair $(j,i)$ instead of $(i,j)$.
From a diagrammatic point of view, this means that we associate the triangle in Fig.\eqref{fig:vertex_ij1} to the first $\mathcal{A}$ term  and the triangle in Fig.\eqref{fig:vertex_ij2} to the second one.

Using Eqs.\eqref{eqn:sum_change} and \eqref{eqn:pairs} we can turn Eq.\eqref{eqn:sum_flav} into
\begin{equation}
\label{eqn:sum_flav2}
\sum_i \sum_{k \neq i} \dots  = \frac{1}{(m_g+1)} \left(\,\, \sum_{\substack{\text{pairs} \\ a,c=q,g}}  \sum_{b \neq a,c} \dots+ \sum_{\substack{\text{pairs} \\ a,c=\bar{q},g}}  \sum_{b \neq a,c} \dots+ 2 \sum_{\substack{\text{pairs} \\ a,c=g,g}}  \sum_{b \neq a,c}\dots \right)
\end{equation}
so that, reminding the notation in Eq.\eqref{eqn:map_appl}, the dual cross section in Eq.\eqref{def:dual_sub2} can be rewritten as
\begin{multline}
\label{def:dual_sub3}
\sigma^{D} = \mathcal{N}_{in} \sum_{\{ m\} } \int d \phi_{m} \, \frac{1}{S_{\{m\}}} \frac{1}{(m_g+1)} \\
\times \left(\,\, \sum_{\substack{\text{pairs} \\ a,c=q,g}}  \sum_{b \neq a,c} \mathcal{C}_{ac,b}+ \sum_{\substack{\text{pairs} \\ a,c=\bar{q},g}}  \sum_{b \neq a,c} \mathcal{C}_{ac,b} + 2 \sum_{\substack{\text{pairs} \\ a,c=g,g}}  \sum_{b \neq a,c} \mathcal{C}_{ac,b} \right) \,\,.
\end{multline}
The presence of a factor $2$ in front of the sum over gluon-gluon pairs on the right-hand side of Eq.\eqref{def:dual_sub3} has a clear interpretation: given a certain pair of gluons in the real sector, their roles can be exchanged and, therefore, we need to take twice the dual counterterm.
In this way, indeed, in one counterterm we can apply the mapping where one of the two gluons is the radiation and, in the other counterterm, we can use the mapping where the role of radiation is played by the other gluon.
Furthermore, using the relation
\begin{equation}
\label{eqn:bose}
\frac{S_{\{m\}}}{S_{\{m+1\}}} = \frac{\dots m_g !}{ \dots (m_g+1)!} = \frac{1}{m_g+1}
\end{equation}
we can turn Eq.\eqref{def:dual_sub3} into
\begin{multline}
\label{def:dual_sub4}
\sigma^{D} = \mathcal{N}_{in} \sum_{\{ m\} } \int d \phi_{m} \, \frac{1}{S_{\{m+1\}}}  \\
\times \left[\,\, \sum_{\substack{\text{pairs} \\ a,c=q,g}}  \sum_{b \neq a,c} \mathcal{C}_{ac,b}+ \sum_{\substack{\text{pairs} \\ a,c=\bar{q},g}}  \sum_{b \neq a,c} \mathcal{C}_{ac,b} + \sum_{\substack{\text{pairs} \\ a,c=g,g}}  \sum_{b \neq a,c} \left( \mathcal{C}_{ac,b} + \mathcal{C}_{ca,b} \right) \right] \,\,.
\end{multline}
The sums in Eq.\eqref{def:dual_sub4} present one dual counterterm for each dipole of the real phase space.
Therefore, Eq.\eqref{def:dual_sub4} tells us how to associate every single counterterm of the dual subtraction scheme with the proper singular configuration of the real phase space.

Let us now move to processes with a quark-antiquark pair replacing a gluon.
These involve the counterterms $\mathcal{C}_{g,j}$, that we assume to include only the quark contribution to the gluon renormalization, since gluon and ghost contributions are used in processes with one additional gluon in the final state.
Before to apply Eq.\eqref{eqn:sum_change}, we first multiply and divide the counterterms by $N_f$ and then use the $N_f$ in the numerator to introduce a sum over the different quark flavours
\begin{equation}
\sum_{i=g} \sum_{j\neq i} \mathcal{C}_{i,j}=
\frac{1}{N_f} \sum_{i=g} \Biggl(\,\, \underbrace{\sum_{j\neq i} \mathcal{C}_{i,j} + \dots + \sum_{j\neq i} \mathcal{C}_{i,j}}_{N_f \,\text{times}} \,\,\Biggr) = \frac{1}{N_f} \sum_{i=g} \sum_f \sum_{j\neq i} \mathcal{C}_{i,j}\,\,.
\end{equation}
Then, we apply Eq.\eqref{eqn:sum_change} ending up with
\begin{equation}
\frac{1}{N_f} \sum_{i=g} \sum_f \sum_{j\neq i} \mathcal{C}_{i,j} = \frac{1}{N_f} \sum_f \frac{\#(g)_m}{\#\{q_f,\bar{q}_f\}_{m+1}} \sum_{\substack{\text{pairs} \\ a,c=q_f,\bar{q}_f}} \sum_{b \neq a,c} \mathcal{C}_{ac,b}
\end{equation}
where
\begin{equation}
\frac{\#(g)_m}{\#\{q_f,\bar{q}_f\}_{m+1}} = \frac{m_g}{(m_f+1)(\bar{m}_f+1)}
\end{equation}
turns the Bose symmetry factor $S_{\{m\}}$ into $S_{\{m+1\}}$, since
\begin{equation}
\label{eqn:bose2}
\frac{S_{\{m\}}}{S_{\{m+1\}}} = \frac{\dots m_f! \bar{m}_f! m_g !}{ \dots (m_f+1)!(\bar{m}_f+1)!(m_g-1)!} = \frac{m_g}{(m_f+1)(\bar{m}_f+1)}\,\,.
\end{equation}
Therefore, the dual cross section for processes with a quark-antiquark pair in place of a final state gluon can be written as
\begin{align}
\label{def:dual_sub4.2}
\sigma^{D} &= \mathcal{N}_{in} \sum_{\{ m\} } \int d \phi_{m} \, \frac{1}{S_{\{m\}}} \frac{m_g}{(m_f+1)(\bar{m}_f+1)} \frac{1}{N_f} \sum_f \sum_{\substack{\text{pairs} \\ a,c=q_f,\bar{q}_f}} \sum_{b \neq a,c} \mathcal{C}_{ac,b} \nonumber \\
&= \mathcal{N}_{in} \sum_{\{ m\} } \int d \phi_{m} \, \frac{1}{S_{\{m+1\}}} \frac{1}{N_f} \sum_f \sum_{\substack{\text{pairs} \\ a,c=q_f,\bar{q}_f}} \sum_{b \neq a,c} \mathcal{C}_{ac,b} \,\,.
\end{align}
Eq.\eqref{def:dual_sub4.2} associates each dual counterterm with a collinear quark-antiquark configuration of the real phase space.

\section{Dual Counterterms for the massless case}
\label{sec:masslessDS}
In this section we collect all the formulas to be used for a numerical implementation of the dual subtractions. To regulate the divergences of the real matrix elements, we have cut the relevant loop integrals and collected their divergences. Then, in principle, we could organize the integrated counterpart in the form of finite remnants
of the loop integrals. This could be useful for implementations of our formalism in general purposed programs that perform the full reduction of the virtual
amplitude.
On the other hand, to make the implementation easier, in this appendix we will continue to treat our cut loop integrals as normal countertems and proceed to present the list of their integrated version.
The formulation of our method is given as follows: in Table~\ref{tab:dualsub} we list the equation number of the dual subtraction along with the name of the corresponding jacobian factor that converts the radiation variables to that of the loop momentum
as explained in Section~\ref{sec:algorithm}. We label with $p_a$ and $p_b$ the partons in the initial state and $p_i$, $p_j$, $p_k$ and so on for the ones in the final state. Moreover, we remember here that the formulae for the dual dipoles are written following the convention of in-going and out-going momenta, $p_a + p_b = p_i + p_j+p_k+...\,$. Furthermore, the names of the counterterms are written using the same conventions as in the presentation of the Altarelli--Parisi splitting functions in Section~\ref{sec:real_beh}. The prefixes specify the emitter and spectator respectively. The 
suffixes have to be understood as $(ac)c$ for a final state emitter $(ac)$
radiating the parton $c$ and it is $a(ac)$ for an initial state
emitter $(ac)$ (the hard parton) that is the result of the splitting of the
incoming parent parton $a$. So for example: $if\_qq$ stands for the counterterm for an initial
state quark ($i$ in the first part of the prefix and $q$ in the first part of the suffix) emitting
a gluon and becoming the hard quark of the underlying event (the second part of the suffix $q$), and this gluon is connected to
a final state spectator (as indicated by the $f$ in the second part of the prefix); $ff\_qg$ stands instead for a virtual quark $(q)$ splitting into a $qg$ pair with the gluon $(g)$ being the radiated parton connected to another
final state parton. The dual dipoles formulae do not depend from the location of the spectator momentum.
\begin{table}[]
    \centering
    \begin{tabular}{|c|l||c|l|}
        \hline
        Counterterm & Equation $\times$ jac &
        Counterterm & Equation $\times$ jac\\
        \hline
        ff\_qg  & Eq.(\ref{def:fqg})$\times$ jac(ff) &
        fi\_qg  & Eq.(\ref{def:fqg})$\times$ jac(fi) \\
        ff\_gq  & Eq.(\ref{def:fgq})$\times$ jac(ff) &
        fi\_gq  & Eq.(\ref{def:fgq})$\times$ jac(fi) \\
        ff\_gg  & Eq.(\ref{def:fgg})$\times$ jac(ff) &
        fi\_gg  & Eq.(\ref{def:fgg})$\times$ jac(fi) \\
        \hline
        if\_qq  & Eq.(\ref{def:iqq})$\times$ jac(if) &
        ii\_qq  & Eq.(\ref{def:iqq})$\times$ jac(ii) \\
        if\_qg  & Eq.(\ref{def:iqg})$\times$ jac(if) &
        ii\_qg  & Eq.(\ref{def:iqg})$\times$ jac(ii) \\
        if\_gq  & Eq.(\ref{def:igq})$\times$ jac(if) &
        ii\_gq  & Eq.(\ref{def:igq})$\times$ jac(ii) \\
        if\_gg  & Eq.(\ref{def:igg})$\times$ jac(if) &
        ii\_gg  & Eq.(\ref{def:igg})$\times$ jac(ii) \\
        \hline
    \end{tabular}
    \caption{Counterterms definition as product of the cut loop integrals
    in the equations times the corresponding jacobian factor.}
\label{tab:dualsub}
\end{table}
The formulae for the dipole subtractions do not depend from the specific momentum mappings nevertheless for their numerical implementation one choice has to be made.
Here we adopt the momentum mappings reported in ref.\cite{Catani:1996vz} that from now on we dub CS and will refer to the equation x.y in that paper with CS(x.y)
for easy of reference\footnote{Note that sometimes we have relabelled the suffixes of the momenta with respect to the equations of ref.\cite{Catani:1996vz}.}.

\subsection{final-final}
For the case of final state emitter and
spectator we will report in detail all the formulas that we take from CS. For the
other cases we will just point to the other
formulas in the same paper.
The momentum mapping to transform the three momentum $\{i,j,k\}$ into $\{\tilde{ij},\tilde{k}\}$ is defined in  the Equation CS(5.3)
\begin{align}
\label{eq:ffmap1}
    \tilde{p}_k^\mu&=\frac{1}{1-y}p_k^\mu \\
    \tilde{p}_{ij}^\mu&=p_i^\mu+p_j^\mu-
    \frac{y}{1-y}p_k^\mu \nonumber
\end{align}
in terms of the variables
\begin{align}
\label{eq:ffmap2}
    y&=\frac{p_i p_j}{p_ip_j+p_jp_k+p_kp_i}\\
    z_i&=\frac{p_ip_k}{p_ip_k+p_jp_k} \nonumber \\
    z_j&=1-z_i\, . \nonumber
\end{align}
The factorization of the phase space in Eq. CS(5.17-20) is then
\begin{equation}
    d\phi(p_i,p_j,p_k;Q)=
    d\phi(\tilde{p}_{ij},\tilde{p}_k;Q)
    [dp_j(\tilde{p}_{ij},\tilde{p}_k)]
\end{equation}
where
\begin{flalign}
    [dp_j(\tilde{p}_{ij},\tilde{p}_k)]&=
    \frac{d^dp_j}{(2\,\pi)^{d-1}}\delta_+(p_j^2)\mathcal{I}(p_j;\tilde{p}_{ij},\tilde{p}_k)
\end{flalign}
with
\begin{align}
    \mathcal{I}(p_j;\tilde{p}_{ij},\tilde{p}_k)&=
    \Theta(1-z_j)\Theta(1-y)\frac{(1-y)^{d-3}}{1-z_j} \,.
\end{align}
In terms of the radiation variables introduced above Equation CS(5.20) reads
\begin{align}
\label{def:ff_measure}
    [dp_j(\tilde{p}_{ij},\tilde{p}_k)]=&
    \frac{(2\,\tilde{p}_{ij}\tilde{p}_k)^{1-\varepsilon}}{16\,\pi^2}
    \frac{d\Omega^{d-3}}{(2\,\pi)^{1-2\varepsilon}}
    dz_i dy \Theta(z_i(1-z_i))\Theta(y(1-y)) \nonumber \\
    &\times (z_i(1-z_i))^{-\varepsilon}
    (1-y)^{1-2\,\varepsilon}y^{-\varepsilon} \,.
\end{align}
As explained in Section~\ref{sec:algorithm}, the local cancellation of the 
singularities among virtual and real contribution
is obtained only if one takes into account also the jacobian factor
associated to the mapping. In particular we have to divide 
the formulae for the counterterms in Tab.\ref{tab:dualsub}
by the factor $\mathcal{I}(p_j;\tilde{p}_{ij},\tilde{p}_k)$ that for
this case amount to multiply them by
\begin{align}
\label{def:ff_jac}
    {\rm jac(ff)}=\frac{z_i}{(1-y)^{1-2\varepsilon}}
\end{align}
Furthermore, instead of the whole region $0<z_i,y<1$ implied by the $\Theta$ functions in Eq.(\ref{def:ff_measure}) the restriction of the phase space induced by the $R_1$ function
corresponds to limit the integration region to the configurations with $p_ip_j < p_jp_k$ or in terms of the radiation variables to the restriction
\begin{align}
\label{def:ff_region}
0<y<\frac{1-z_i}{2-z_i} \qquad \qquad 0<z_i<1 \,.
\end{align}
We introduce the normalization factor
\begin{align}
    N_\varepsilon&=\frac{(4\,\pi)^\varepsilon}{\Gamma(1-\varepsilon)}
\end{align}
and the short hand notation $l_2\equiv \log(2)$. We integrate the counterterms built by the product of the dual integrands multiplied times the corresponding jacobian factor (jac(ff) in this case), using the measure in Eq.(\ref{def:ff_measure}) and restricting the integration
to the region in Eq.(\ref{def:ff_region}),
 obtaining
\begin{align}
    {\rm ff\_qg\_int}&=\frac{\alpha_s}{2\,\pi}N_\varepsilon \left(\frac{\mu^2}{2\,\tilde{p}_{ij}\tilde{p}_k}\right)^\varepsilon
    \left(\frac{1}{\varepsilon^2}+\frac{3}{2\,\varepsilon}
    -\frac{\pi^2}{2}+4\,l_2+3
    \right)
\\
    {\rm ff\_gq\_int}&=\frac{\alpha_s}{2\,\pi}N_\varepsilon
    \frac{N_f}{C_A} \left(\frac{\mu^2}{2\,\tilde{p}_{ij}\tilde{p}_k}\right)^\varepsilon
    \left(-\frac{2}{3\,\varepsilon}
    -\frac{10}{3}\,l_2+\frac{2}{9}
    \right)
\\
    {\rm ff\_gg\_int}&=\frac{\alpha_s}{2\,\pi}N_\varepsilon \left(\frac{\mu^2}{2\,\tilde{p}_{ij}\tilde{p}_k}\right)^\varepsilon
    \left(\frac{2}{\varepsilon^2}+\frac{11}{3\,\varepsilon}
    -\pi^2+\frac{28}{3}\,l_2+\frac{55}{9}
    \right)
\end{align}
Note that we adopt the convention to have in the prefactor the kinematic invariant built with the momenta of the emitter and spectator partons. The same convention is adopted also for the other cases~\cite{Campbell:2004ch}. Following the
approach of combining the integrated countertems
with the virtual matrix element it is more convenient to do so because this is the invariant which is kept fixed during the $x$ integration. In the above equations the case for
${\rm ff\_g\bar{q}\_int}$ is included in
${\rm ff\_gq\_int}$ as a factor of 2 compensated with
the $T_R$ factor. Furthermore, the two parts corresponding to
the simmetrization in Eq.(\ref{def:fqg}) gives identical
integrated contribution. We point out that the momentum mapping in Eqs.(\ref{eq:ffmap1},\ref{eq:ffmap2})
is symmetric for the exchange of $p_i$ and $p_j$ and so there
is just one counter event although its weight is determined
by the phase space restrictions that event by event
are different for the two contributions.
The same considerations also apply to the next
case that is the final-initial.

\subsection{final-initial}

The momentum mapping to transform the three momentum 
$\{i,j,a\}$ into $\{\tilde{ij},\tilde{a}\}$ is defined 
in  the Equation CS(5.37)
\begin{align}
    \tilde{p}_a^\mu&= x' p_a^\mu \\
    \tilde{p}_{ij}^\mu&=p_i^\mu+p_j^\mu-
    (1-x')p_a^\mu \nonumber
\end{align}
in terms of the variables
\begin{align}
    x'&=\frac{p_ip_a+p_jp_a-p_ip_j}{p_ip_a+p_jp_a}\\
    z_i&=\frac{p_ip_a}{p_ip_a+p_jp_a} \nonumber \\
    z_j&=1-z_i\, . \nonumber
\end{align}
The factorization of the three parton phase space in Eq. CS(5.45-48) is then
\begin{equation}
    d\phi(p_i,p_j;Q+p_a)=
    \int_0^1 dx d\phi(\tilde{p}_{ij};Q+xp_a)
    [dp_j(\tilde{p}_{ij};p_a,x)]
\end{equation}
where
\begin{flalign}
    \label{def:jac_fi}
    [dp_j(\tilde{p}_{ij};p_a,x)]&=
    \frac{d^dp_j}{(2\,\pi)^{d-1}}\delta_+(p_j^2)
    \Theta(x(1-x))\delta(x-x')
    \frac{1}{1-z_j}
\end{flalign}
witch in terms of the radiation variables introduced above reads
\begin{align}
\label{def:fi_measure}
    [dp_j(\tilde{p}_{ij};p_a,x)]=&
    \frac{(2\,\tilde{p}_{ij}p_a)^{1-\varepsilon}}{16\,\pi^2}
    \frac{d\Omega^{d-3}}{(2\,\pi)^{1-2\varepsilon}}
    dz_i \Theta(z_i(1-z_i))
    \Theta(x(1-x))\delta(x-x') \nonumber \\
    &\times (z_i(1-z_i))^{-\varepsilon}
    (1-x)^{-\varepsilon} \,.
\end{align}
From Eq.(\ref{def:jac_fi}) the jacobian factor
needed to transform the radiated momentum into
the loop momentum is given by
\begin{align}
\label{def:fi_jac}
    {\rm jac(fi)}=z_i
\end{align}
Furthermore, the integration region corresponding
to $p_ip_j < p_jp_a$ in terms of the radiation variables is given by
\begin{align}
\label{def:fi_region}
0<z_i<x \qquad \qquad 0<x<1 \,.
\end{align}
We integrate the counterterms built by the product of the dual integrands multiplied times the corresponding jacobian factor in Eq.(\ref{def:fi_jac}), using the measure in Eq.(\ref{def:fi_measure}) and restricting the integration to the region in Eq.(\ref{def:fi_region}),
 obtaining
 \begin{align}
    {\rm fi\_qg\_int}&=\frac{\alpha_s}{2\,\pi}N_\varepsilon \left(\frac{\mu^2}{2\,\tilde{p}_{ij}\tilde{p}_a}\right)^\varepsilon
    \left[
    \delta(1-x)\left(
    \frac{1}{\varepsilon^2}+\frac{3}{2\,\varepsilon}
    -\frac{\pi^2}{3}+\frac{7}{2}
    \right) 
    -\frac{3}{2}\frac{1}{(1-x)}_+  \right. \nonumber \\
    & \left. -\left(\frac{2\log(1-x)}{1-x}\right)_+
    +2\log(1-x)+\frac{1}{2}(3+3x+x^2)
    \right]
\\
    {\rm fi\_gq\_int}&=\frac{\alpha_s}{2\,\pi}N_\varepsilon
    \frac{N_f}{C_A} \left(\frac{\mu^2}{2\,\tilde{p}_{ij}\tilde{p}_a}\right)^\varepsilon
    \left[
    \delta(1-x)\left(
    -\frac{2}{3\,\varepsilon}
    -\frac{10}{9}
    \right) 
    +\frac{2}{3}\frac{1}{(1-x)}_+ \right. \nonumber \\
    & \left. -\frac{1}{3}(2+2x-x^2+2x^3)
    \right]
\\
    {\rm fi\_gg\_int}&=\frac{\alpha_s}{2\,\pi}N_\varepsilon \left(\frac{\mu^2}{2\,\tilde{p}_{ij}\tilde{p}_a}\right)^\varepsilon
    \left[
    \delta(1-x)\left(
    \frac{2}{\varepsilon^2}+\frac{11}{3\,\varepsilon}
    -\frac{2\,\pi^2}{3}+\frac{67}{9}
    \right) 
    -\frac{11}{3}\frac{1}{(1-x)}_+ \right. \nonumber \\
    &\left. -\left(\frac{4\log(1-x)}{1-x}\right)_+
    +4\log(1-x)+\frac{1}{3}(11+11x-x^2+2x^3)
    \right]
\end{align}

\subsection{initial-final}

The momentum mapping to transform the three momentum 
$\{a,j,i\}$ into $\{\tilde{aj},\tilde{i}\}$ is defined 
in  the Equation CS(5.62-64)
\begin{align}
    \tilde{p}_{aj}^\mu&= x' p_a^\mu \\
    \tilde{p}_{i}^\mu&=p_i^\mu+p_j^\mu-
    (1-x)p_a^\mu. \nonumber
\end{align}
in terms of the variables
\begin{align}
    x'&=\frac{p_ip_a+p_jp_a-p_jp_i}{p_ip_a+p_jp_a}\\
    u_j&=\frac{p_jp_a}{p_jp_a+p_ip_a} = z_j\,. \nonumber
\end{align}
This momentum mapping coincides with the previous one.
This is particularly useful for our
approach because such a map connects a set of radiation variables and a Born configuration with an initial(final) state emitter and a final(initial) state spectator to a unique real configuration
and vivecersa. The factorization of the three parton phase space in Eq. CS(5.70-73) is the same as in the previous
case and the jacobian factor needed to transform the radiated momentum into the loop momentum is again
given by
\begin{align}
\label{def:if_jac}
    {\rm jac(if)}=1-u_j = z_i = {\rm jac(fi)}
\end{align}
The integration region corresponding
to $p_ap_j < p_ip_j$ in terms of the radiation variables is now given by
\begin{align}
\label{def:if_region}
0<u_j<1-x \qquad \qquad 0<x<1 \,.
\end{align}
We integrate the counterterms built by the product of the dual integrands multiplied times the corresponding jacobian factor in Eq.(\ref{def:if_jac}), using the measure in Eq.(\ref{def:fi_measure}) and restricting the integration
to the region in Eq.(\ref{def:if_region}),
 obtaining
\begin{align}
    {\rm if\_qq\_int}&=\frac{\alpha_s}{2\,\pi}N_\varepsilon \left(\frac{\mu^2}{2\,\tilde{p}_{aj}\tilde{p}_{i}}\right)^\varepsilon
    \left[
    \delta(1-x)
    \frac{1}{\varepsilon^2}
    -\frac{2}{\varepsilon}\frac{1}{(1-x)}_+ 
     +\left(\frac{4\log(1-x)}{1-x}\right)_+ \right. \nonumber \\
    & \left.
    +(1+x)\left( \frac{1}{\varepsilon}-2\log(1-x) \right)
    -\frac{1+x^2}{1-x}\log(x)
    -x(x^2+x+3)+1
    \right]
\\  {\rm if\_qg\_int}&=\frac{\alpha_s}{2\,\pi}N_\varepsilon
    \frac{T_R}{C_A}
    \left(\frac{\mu^2}{2\,\tilde{p}_{aj}\tilde{p}_{i}}\right)^\varepsilon
    \left[
    -\frac{1+(1-x)^2}{x}
    \left(\frac{1}{\varepsilon}-2\log(1-x)+2\log(x)+1
    \right) \right. \nonumber \\
    &\left. +(2-x)\log(x)+x^2
    \right]
\\
    {\rm if\_gq\_int}&=\frac{\alpha_s}{2\,\pi}N_\varepsilon \left(\frac{\mu^2}{2\,\tilde{p}_{aj}\tilde{p}_{i}}\right)^\varepsilon
    \left[
    -(1-2x+2x^2)\left(\frac{1}{\varepsilon}-2\log(1-x) +\log(x) \right)
    \right. \nonumber \\
    & \left. -\left(\frac{1}{2}-2x+2x^2\right)\log(1-2x+2x^2)
    +1-x
    \right]
\\
    {\rm if\_gg\_int}&=\frac{\alpha_s}{2\,\pi}N_\varepsilon \left(\frac{\mu^2}{2\,\tilde{p}_{aj}\tilde{p}_{i}}\right)^\varepsilon
    \left[
    \delta(1-x)
    \frac{1}{\varepsilon^2}
    -\frac{2}{\varepsilon}\frac{1}{(1-x)}_+ 
     +\left(\frac{4\log(1-x)}{1-x}\right)_+ \right. \nonumber \\
    & \left. -2\left(
    \frac{1-x}{x}+x(1-x)-1
    \right)\left(
    \frac{1}{\varepsilon}-2\log(1-x) +\log(x)
    \right)
    -2\frac{1-2x+x^2}{x}\log(x) \right. \nonumber \\
    &\left. -\frac{2}{1-x}\log(x)+\left(
    \frac{1}{2}-2x+2x^2\right)\log(1-2x+2x^2)
    -x(x^2+3)
    \right]
\end{align}
Note that in $if\_qg\_int$ there is no $N_f$ factor because
it refers to the contribution from a single kind of identified initial quark (or antiquark). The same will happen also
for the case initial-initial.

\subsection{initial-initial}

The momentum mapping to transform the three momenta 
$\{a,j,b\}$ into $\{\tilde{aj},\tilde{b}\}$ is defined 
in  the Equation CS(5.137-140)
\begin{align}
    \tilde{p}_{aj}^\mu&= x' p_a^\mu \\
    \tilde{p}_{b}^\mu&=p_b^\mu \nonumber
\end{align}
in terms of the variables
\begin{align}
    x'&=\frac{p_ap_b-p_jp_a-p_jp_b}{p_ap_b}\\
    v_j&=\frac{p_ap_j}{p_ap_b} \,. \nonumber
\end{align}
Note that in this case the momenta of all the other final state particles $k_i$ are transformed as follows (CS(5.139-140))
\begin{align}
\tilde{k}_i^\mu&=k_i^\mu
-\frac{2\,k_i\cdot (K+\tilde{K})}{(K+\tilde{K})^2}(K+\tilde{K})^\mu
+\frac{2\,k_i\cdot K}{K^2}\tilde{K}^\mu \\
K^\mu&=p_a^\mu+p_b^\mu-p_j^\mu  \nonumber \\
\tilde{K}^\mu&=\tilde{p}_{aj}^\mu+p_b^\mu \, . \nonumber
\end{align}
The factorization of the three parton phase space in Eq. CS(5.149-151) is then
\begin{equation}
    d\phi(p_k,k_1,\dots;p_a+p_b)=
    \int_0^1 dx d\phi(\tilde{k}_1,\dots;\tilde{p}_a+\tilde{p}_b)
    [dp_j(p_a,p_b,x)]
\end{equation}
where
\begin{flalign}
    \label{def:jac_ii}
    [dp_j(p_a,p_b,x)]&=
    \frac{d^dp_j}{(2\,\pi)^{d-1}}\delta_+(p_j^2)
    \Theta(x(1-x))\delta(x-x')
\end{flalign}
witch in terms of the radiation variables introduced above reads
\begin{align}
\label{def:ii_measure}
    [dp_j(p_a,p_b,x)]=&
    \frac{(2\,p_ap_b)^{1-\varepsilon}}{16\,\pi^2}
    \frac{d\Omega^{d-3}}{(2\,\pi)^{1-2\varepsilon}}
    dv_j \Theta(v_j)\Theta\left(1-\frac{v_j}{1-x}\right)\Theta(x(1-x))\delta(x-x') \nonumber \\
    &\times (v_j(1-x-v_j))^{-\varepsilon} \,.
\end{align}
From Eq.(\ref{def:jac_ii}) the jacobian factor
needed to transform the radiated momentum into
the loop momentum is given by
\begin{align}
\label{def:ii_jac}
    {\rm jac(ii)}=1
\end{align}
Furthermore, the integration region corresponding
to $p_ap_j < p_bp_j$ in terms of the radiation variables is given by
\begin{align}
\label{def:ii_region}
0<v_j<\frac{1-x}{2} \qquad \qquad 0<x<1 \,.
\end{align}
We integrate the dual counterterms using the measure in Eq.(\ref{def:ii_measure}) and restricting the integration
to the region in Eq.(\ref{def:ii_region}),
 obtaining
\begin{align}
    {\rm ii\_qq\_int}&=\frac{\alpha_s}{2\,\pi}N_\varepsilon \left(\frac{\mu^2}{2\,\tilde{p}_{aj}\tilde{p}_{b}}\right)^\varepsilon
    \left[
    \delta(1-x)
    \left(
    \frac{1}{\varepsilon^2}
    -\frac{\pi^2}{6}
    \right)
    -\frac{2}{\varepsilon}\frac{1}{(1-x)}_+ 
     +\left(\frac{4\log(1-x)}{1-x}\right)_+ \right. \nonumber \\
    & \left.
    +(1+x)\left( \frac{1}{\varepsilon}-2\log(1-x) -l_2 \right)
    -\frac{1+x^2}{1-x}\log(x)+\frac{1-x}{2}
    \right]
\\  {\rm ii\_qg\_int}&=\frac{\alpha_s}{2\,\pi}N_\varepsilon 
    \frac{T_R}{C_A}
    \left(\frac{\mu^2}{2\,\tilde{p}_{aj}\tilde{p}_{b}}\right)^\varepsilon
    \left[
    -\frac{1+(1-x)^2}{x}
    \left(\frac{1}{\varepsilon}-2\log(1-x)+\log(x)+l_2
    \right) \right. \nonumber \\
    &\left. 
    -\frac{3}{4}\frac{(1-x)^2}{x}-3\frac{1-x}{x}
    \right]
\\
    {\rm ii\_gq\_int}&=\frac{\alpha_s}{2\,\pi}N_\varepsilon
    \left(\frac{\mu^2}{2\,\tilde{p}_{aj}\tilde{p}_{b}}\right)^\varepsilon
    \left[
    -(1-2\,x+2\,x^2)\left(\frac{1}{\varepsilon}-2\log(1-x) +\log(x) 
    +l_2\right) \right. \nonumber \\
    & \left. +2x^2(l_2 - \log(1 + 2\,x - x^2))
    +\frac{1-x}{2}
    \right]
\\
    {\rm ii\_gg\_int}&=\frac{\alpha_s}{2\,\pi}N_\varepsilon 
    \left(\frac{\mu^2}{2\,\tilde{p}_{aj}\tilde{p}_{b}}\right)^\varepsilon
    \left[
    \delta(1-x)\left(
    \frac{1}{\varepsilon^2}
    -\frac{\pi^2}{6}
    \right) 
    -\frac{2}{\varepsilon}\frac{1}{(1-x)}_+ 
    +\left(\frac{4\log(1-x)}{1-x}\right)_+ \right. \nonumber \\
    &\left. 
    -2\left(
    \frac{1-x}{x}+x(1-x)-1
    \right)\left(
    \frac{1}{\varepsilon}-2\log(1-x)+\log(x)+l_2
    \right)
    -\frac{2}{1-x}\log(x) \right. \nonumber \\
    &\left. 
    +2\,x^2\log(1+2\,x-x^2) -2\,(1+x+x^2)\,l_2 
    -\frac{3}{4}\frac{(1-x)^2}{x}-\frac{1-x}{x}
    \right]
\end{align}

\acknowledgments
The authors gratefully acknowledge Germ\'an Rodrigo for a critical reading of the manuscript, G\'abor Somogyi for useful communications and also Jonathan Ronca, German Sborlini and William J. Torres Bobadilla for useful discussions. Our work is supported by INFN.

\bibliographystyle{JHEP}
\bibliography{main}

\end{document}